\documentclass[twocolumn,tighten]{aastex631}
\usepackage{natbib, twoopt}
\usepackage{float}
\usepackage{graphicx}
\usepackage{epstopdf}
\usepackage{enumitem}
\usepackage[caption=false]{subfig}
\usepackage{color}\definecolor{AliceBlue}{rgb}{0.94,0.97,1.00}
\definecolor{AntiqueWhite1}{rgb}{1.00,0.94,0.86}
\definecolor{AntiqueWhite2}{rgb}{0.93,0.87,0.80}
\definecolor{AntiqueWhite3}{rgb}{0.80,0.75,0.69}
\definecolor{AntiqueWhite4}{rgb}{0.55,0.51,0.47}
\definecolor{AntiqueWhite}{rgb}{0.98,0.92,0.84}
\definecolor{BlanchedAlmond}{rgb}{1.00,0.92,0.80}
\definecolor{BlueViolet}{rgb}{0.54,0.17,0.89}
\definecolor{CadetBlue1}{rgb}{0.60,0.96,1.00}
\definecolor{CadetBlue2}{rgb}{0.56,0.90,0.93}
\definecolor{CadetBlue3}{rgb}{0.48,0.77,0.80}
\definecolor{CadetBlue4}{rgb}{0.33,0.53,0.55}
\definecolor{CadetBlue}{rgb}{0.37,0.62,0.63}
\definecolor{CornflowerBlue}{rgb}{0.39,0.58,0.93}
\definecolor{DarkBlue}{rgb}{0.00,0.00,0.55}
\definecolor{DarkCyan}{rgb}{0.00,0.55,0.55}
\definecolor{DarkGoldenrod1}{rgb}{1.00,0.73,0.06}
\definecolor{DarkGoldenrod2}{rgb}{0.93,0.68,0.05}
\definecolor{DarkGoldenrod3}{rgb}{0.80,0.58,0.05}
\definecolor{DarkGoldenrod4}{rgb}{0.55,0.40,0.03}
\definecolor{DarkGoldenrod}{rgb}{0.72,0.53,0.04}
\definecolor{DarkGray}{rgb}{0.66,0.66,0.66}
\definecolor{DarkGreen}{rgb}{0.00,0.39,0.00}
\definecolor{DarkGrey}{rgb}{0.66,0.66,0.66}
\definecolor{DarkKhaki}{rgb}{0.74,0.72,0.42}
\definecolor{DarkMagenta}{rgb}{0.55,0.00,0.55}
\definecolor{DarkOliveGreen1}{rgb}{0.79,1.00,0.44}
\definecolor{DarkOliveGreen2}{rgb}{0.74,0.93,0.41}
\definecolor{DarkOliveGreen3}{rgb}{0.64,0.80,0.35}
\definecolor{DarkOliveGreen4}{rgb}{0.43,0.55,0.24}
\definecolor{DarkOliveGreen}{rgb}{0.33,0.42,0.18}
\definecolor{DarkOrange1}{rgb}{1.00,0.50,0.00}
\definecolor{DarkOrange2}{rgb}{0.93,0.46,0.00}
\definecolor{DarkOrange3}{rgb}{0.80,0.40,0.00}
\definecolor{DarkOrange4}{rgb}{0.55,0.27,0.00}
\definecolor{DarkOrange}{rgb}{1.00,0.55,0.00}
\definecolor{DarkOrchid1}{rgb}{0.75,0.24,1.00}
\definecolor{DarkOrchid2}{rgb}{0.70,0.23,0.93}
\definecolor{DarkOrchid3}{rgb}{0.60,0.20,0.80}
\definecolor{DarkOrchid4}{rgb}{0.41,0.13,0.55}
\definecolor{DarkOrchid}{rgb}{0.60,0.20,0.80}
\definecolor{DarkRed}{rgb}{0.55,0.00,0.00}
\definecolor{DarkSalmon}{rgb}{0.91,0.59,0.48}
\definecolor{DarkSeaGreen1}{rgb}{0.76,1.00,0.76}
\definecolor{DarkSeaGreen2}{rgb}{0.71,0.93,0.71}
\definecolor{DarkSeaGreen3}{rgb}{0.61,0.80,0.61}
\definecolor{DarkSeaGreen4}{rgb}{0.41,0.55,0.41}
\definecolor{DarkSeaGreen}{rgb}{0.56,0.74,0.56}
\definecolor{DarkSlateBlue}{rgb}{0.28,0.24,0.55}
\definecolor{DarkSlateGray1}{rgb}{0.59,1.00,1.00}
\definecolor{DarkSlateGray2}{rgb}{0.55,0.93,0.93}
\definecolor{DarkSlateGray3}{rgb}{0.47,0.80,0.80}
\definecolor{DarkSlateGray4}{rgb}{0.32,0.55,0.55}
\definecolor{DarkSlateGray}{rgb}{0.18,0.31,0.31}
\definecolor{DarkSlateGrey}{rgb}{0.18,0.31,0.31}
\definecolor{DarkTurquoise}{rgb}{0.00,0.81,0.82}
\definecolor{DarkViolet}{rgb}{0.58,0.00,0.83}
\definecolor{DeepPink1}{rgb}{1.00,0.08,0.58}
\definecolor{DeepPink2}{rgb}{0.93,0.07,0.54}
\definecolor{DeepPink3}{rgb}{0.80,0.06,0.46}
\definecolor{DeepPink4}{rgb}{0.55,0.04,0.31}
\definecolor{DeepPink}{rgb}{1.00,0.08,0.58}
\definecolor{DeepSkyBlue1}{rgb}{0.00,0.75,1.00}
\definecolor{DeepSkyBlue2}{rgb}{0.00,0.70,0.93}
\definecolor{DeepSkyBlue3}{rgb}{0.00,0.60,0.80}
\definecolor{DeepSkyBlue4}{rgb}{0.00,0.41,0.55}
\definecolor{DeepSkyBlue}{rgb}{0.00,0.75,1.00}
\definecolor{DimGray}{rgb}{0.41,0.41,0.41}
\definecolor{DimGrey}{rgb}{0.41,0.41,0.41}
\definecolor{DodgerBlue1}{rgb}{0.12,0.56,1.00}
\definecolor{DodgerBlue2}{rgb}{0.11,0.53,0.93}
\definecolor{DodgerBlue3}{rgb}{0.09,0.45,0.80}
\definecolor{DodgerBlue4}{rgb}{0.06,0.31,0.55}
\definecolor{DodgerBlue}{rgb}{0.12,0.56,1.00}
\definecolor{FloralWhite}{rgb}{1.00,0.98,0.94}
\definecolor{ForestGreen}{rgb}{0.13,0.55,0.13}
\definecolor{GhostWhite}{rgb}{0.97,0.97,1.00}
\definecolor{GreenYellow}{rgb}{0.68,1.00,0.18}
\definecolor{HotPink1}{rgb}{1.00,0.43,0.71}
\definecolor{HotPink2}{rgb}{0.93,0.42,0.65}
\definecolor{HotPink3}{rgb}{0.80,0.38,0.56}
\definecolor{HotPink4}{rgb}{0.55,0.23,0.38}
\definecolor{HotPink}{rgb}{1.00,0.41,0.71}
\definecolor{IndianRed1}{rgb}{1.00,0.42,0.42}
\definecolor{IndianRed2}{rgb}{0.93,0.39,0.39}
\definecolor{IndianRed3}{rgb}{0.80,0.33,0.33}
\definecolor{IndianRed4}{rgb}{0.55,0.23,0.23}
\definecolor{IndianRed}{rgb}{0.80,0.36,0.36}
\definecolor{LavenderBlush1}{rgb}{1.00,0.94,0.96}
\definecolor{LavenderBlush2}{rgb}{0.93,0.88,0.90}
\definecolor{LavenderBlush3}{rgb}{0.80,0.76,0.77}
\definecolor{LavenderBlush4}{rgb}{0.55,0.51,0.53}
\definecolor{LavenderBlush}{rgb}{1.00,0.94,0.96}
\definecolor{LawnGreen}{rgb}{0.49,0.99,0.00}
\definecolor{LemonChiffon1}{rgb}{1.00,0.98,0.80}
\definecolor{LemonChiffon2}{rgb}{0.93,0.91,0.75}
\definecolor{LemonChiffon3}{rgb}{0.80,0.79,0.65}
\definecolor{LemonChiffon4}{rgb}{0.55,0.54,0.44}
\definecolor{LemonChiffon}{rgb}{1.00,0.98,0.80}
\definecolor{LightBlue1}{rgb}{0.75,0.94,1.00}
\definecolor{LightBlue2}{rgb}{0.70,0.87,0.93}
\definecolor{LightBlue3}{rgb}{0.60,0.75,0.80}
\definecolor{LightBlue4}{rgb}{0.41,0.51,0.55}
\definecolor{LightBlue}{rgb}{0.68,0.85,0.90}
\definecolor{LightCoral}{rgb}{0.94,0.50,0.50}
\definecolor{LightCyan1}{rgb}{0.88,1.00,1.00}
\definecolor{LightCyan2}{rgb}{0.82,0.93,0.93}
\definecolor{LightCyan3}{rgb}{0.71,0.80,0.80}
\definecolor{LightCyan4}{rgb}{0.48,0.55,0.55}
\definecolor{LightCyan}{rgb}{0.88,1.00,1.00}
\definecolor{LightGoldenrod1}{rgb}{1.00,0.93,0.55}
\definecolor{LightGoldenrod2}{rgb}{0.93,0.86,0.51}
\definecolor{LightGoldenrod3}{rgb}{0.80,0.75,0.44}
\definecolor{LightGoldenrod4}{rgb}{0.55,0.51,0.30}
\definecolor{LightGoldenrodYellow}{rgb}{0.98,0.98,0.82}
\definecolor{LightGoldenrod}{rgb}{0.93,0.87,0.51}
\definecolor{LightGray}{rgb}{0.83,0.83,0.83}
\definecolor{LightGreen}{rgb}{0.56,0.93,0.56}
\definecolor{LightGrey}{rgb}{0.83,0.83,0.83}
\definecolor{LightPink1}{rgb}{1.00,0.68,0.73}
\definecolor{LightPink2}{rgb}{0.93,0.64,0.68}
\definecolor{LightPink3}{rgb}{0.80,0.55,0.58}
\definecolor{LightPink4}{rgb}{0.55,0.37,0.40}
\definecolor{LightPink}{rgb}{1.00,0.71,0.76}
\definecolor{LightSalmon1}{rgb}{1.00,0.63,0.48}
\definecolor{LightSalmon2}{rgb}{0.93,0.58,0.45}
\definecolor{LightSalmon3}{rgb}{0.80,0.51,0.38}
\definecolor{LightSalmon4}{rgb}{0.55,0.34,0.26}
\definecolor{LightSalmon}{rgb}{1.00,0.63,0.48}
\definecolor{LightSeaGreen}{rgb}{0.13,0.70,0.67}
\definecolor{LightSkyBlue1}{rgb}{0.69,0.89,1.00}
\definecolor{LightSkyBlue2}{rgb}{0.64,0.83,0.93}
\definecolor{LightSkyBlue3}{rgb}{0.55,0.71,0.80}
\definecolor{LightSkyBlue4}{rgb}{0.38,0.48,0.55}
\definecolor{LightSkyBlue}{rgb}{0.53,0.81,0.98}
\definecolor{LightSlateBlue}{rgb}{0.52,0.44,1.00}
\definecolor{LightSlateGray}{rgb}{0.47,0.53,0.60}
\definecolor{LightSlateGrey}{rgb}{0.47,0.53,0.60}
\definecolor{LightSteelBlue1}{rgb}{0.79,0.88,1.00}
\definecolor{LightSteelBlue2}{rgb}{0.74,0.82,0.93}
\definecolor{LightSteelBlue3}{rgb}{0.64,0.71,0.80}
\definecolor{LightSteelBlue4}{rgb}{0.43,0.48,0.55}
\definecolor{LightSteelBlue}{rgb}{0.69,0.77,0.87}
\definecolor{LightYellow1}{rgb}{1.00,1.00,0.88}
\definecolor{LightYellow2}{rgb}{0.93,0.93,0.82}
\definecolor{LightYellow3}{rgb}{0.80,0.80,0.71}
\definecolor{LightYellow4}{rgb}{0.55,0.55,0.48}
\definecolor{LightYellow}{rgb}{1.00,1.00,0.88}
\definecolor{LimeGreen}{rgb}{0.20,0.80,0.20}
\definecolor{MediumAquamarine}{rgb}{0.40,0.80,0.67}
\definecolor{MediumBlue}{rgb}{0.00,0.00,0.80}
\definecolor{MediumOrchid1}{rgb}{0.88,0.40,1.00}
\definecolor{MediumOrchid2}{rgb}{0.82,0.37,0.93}
\definecolor{MediumOrchid3}{rgb}{0.71,0.32,0.80}
\definecolor{MediumOrchid4}{rgb}{0.48,0.22,0.55}
\definecolor{MediumOrchid}{rgb}{0.73,0.33,0.83}
\definecolor{MediumPurple1}{rgb}{0.67,0.51,1.00}
\definecolor{MediumPurple2}{rgb}{0.62,0.47,0.93}
\definecolor{MediumPurple3}{rgb}{0.54,0.41,0.80}
\definecolor{MediumPurple4}{rgb}{0.36,0.28,0.55}
\definecolor{MediumPurple}{rgb}{0.58,0.44,0.86}
\definecolor{MediumSeaGreen}{rgb}{0.24,0.70,0.44}
\definecolor{MediumSlateBlue}{rgb}{0.48,0.41,0.93}
\definecolor{MediumSpringGreen}{rgb}{0.00,0.98,0.60}
\definecolor{MediumTurquoise}{rgb}{0.28,0.82,0.80}
\definecolor{MediumVioletRed}{rgb}{0.78,0.08,0.52}
\definecolor{MidnightBlue}{rgb}{0.10,0.10,0.44}
\definecolor{MintCream}{rgb}{0.96,1.00,0.98}
\definecolor{MistyRose1}{rgb}{1.00,0.89,0.88}
\definecolor{MistyRose2}{rgb}{0.93,0.84,0.82}
\definecolor{MistyRose3}{rgb}{0.80,0.72,0.71}
\definecolor{MistyRose4}{rgb}{0.55,0.49,0.48}
\definecolor{MistyRose}{rgb}{1.00,0.89,0.88}
\definecolor{NavajoWhite1}{rgb}{1.00,0.87,0.68}
\definecolor{NavajoWhite2}{rgb}{0.93,0.81,0.63}
\definecolor{NavajoWhite3}{rgb}{0.80,0.70,0.55}
\definecolor{NavajoWhite4}{rgb}{0.55,0.47,0.37}
\definecolor{NavajoWhite}{rgb}{1.00,0.87,0.68}
\definecolor{NavyBlue}{rgb}{0.00,0.00,0.50}
\definecolor{OldLace}{rgb}{0.99,0.96,0.90}
\definecolor{OliveDrab1}{rgb}{0.75,1.00,0.24}
\definecolor{OliveDrab2}{rgb}{0.70,0.93,0.23}
\definecolor{OliveDrab3}{rgb}{0.60,0.80,0.20}
\definecolor{OliveDrab4}{rgb}{0.41,0.55,0.13}
\definecolor{OliveDrab}{rgb}{0.42,0.56,0.14}
\definecolor{OrangeRed1}{rgb}{1.00,0.27,0.00}
\definecolor{OrangeRed2}{rgb}{0.93,0.25,0.00}
\definecolor{OrangeRed3}{rgb}{0.80,0.22,0.00}
\definecolor{OrangeRed4}{rgb}{0.55,0.15,0.00}
\definecolor{OrangeRed}{rgb}{1.00,0.27,0.00}
\definecolor{PaleGoldenrod}{rgb}{0.93,0.91,0.67}
\definecolor{PaleGreen1}{rgb}{0.60,1.00,0.60}
\definecolor{PaleGreen2}{rgb}{0.56,0.93,0.56}
\definecolor{PaleGreen3}{rgb}{0.49,0.80,0.49}
\definecolor{PaleGreen4}{rgb}{0.33,0.55,0.33}
\definecolor{PaleGreen}{rgb}{0.60,0.98,0.60}
\definecolor{PaleTurquoise1}{rgb}{0.73,1.00,1.00}
\definecolor{PaleTurquoise2}{rgb}{0.68,0.93,0.93}
\definecolor{PaleTurquoise3}{rgb}{0.59,0.80,0.80}
\definecolor{PaleTurquoise4}{rgb}{0.40,0.55,0.55}
\definecolor{PaleTurquoise}{rgb}{0.69,0.93,0.93}
\definecolor{PaleVioletRed1}{rgb}{1.00,0.51,0.67}
\definecolor{PaleVioletRed2}{rgb}{0.93,0.47,0.62}
\definecolor{PaleVioletRed3}{rgb}{0.80,0.41,0.54}
\definecolor{PaleVioletRed4}{rgb}{0.55,0.28,0.36}
\definecolor{PaleVioletRed}{rgb}{0.86,0.44,0.58}
\definecolor{PapayaWhip}{rgb}{1.00,0.94,0.84}
\definecolor{PeachPuff1}{rgb}{1.00,0.85,0.73}
\definecolor{PeachPuff2}{rgb}{0.93,0.80,0.68}
\definecolor{PeachPuff3}{rgb}{0.80,0.69,0.58}
\definecolor{PeachPuff4}{rgb}{0.55,0.47,0.40}
\definecolor{PeachPuff}{rgb}{1.00,0.85,0.73}
\definecolor{PowderBlue}{rgb}{0.69,0.88,0.90}
\definecolor{RosyBrown1}{rgb}{1.00,0.76,0.76}
\definecolor{RosyBrown2}{rgb}{0.93,0.71,0.71}
\definecolor{RosyBrown3}{rgb}{0.80,0.61,0.61}
\definecolor{RosyBrown4}{rgb}{0.55,0.41,0.41}
\definecolor{RosyBrown}{rgb}{0.74,0.56,0.56}
\definecolor{RoyalBlue1}{rgb}{0.28,0.46,1.00}
\definecolor{RoyalBlue2}{rgb}{0.26,0.43,0.93}
\definecolor{RoyalBlue3}{rgb}{0.23,0.37,0.80}
\definecolor{RoyalBlue4}{rgb}{0.15,0.25,0.55}
\definecolor{RoyalBlue}{rgb}{0.25,0.41,0.88}
\definecolor{SaddleBrown}{rgb}{0.55,0.27,0.07}
\definecolor{SandyBrown}{rgb}{0.96,0.64,0.38}
\definecolor{SeaGreen1}{rgb}{0.33,1.00,0.62}
\definecolor{SeaGreen2}{rgb}{0.31,0.93,0.58}
\definecolor{SeaGreen3}{rgb}{0.26,0.80,0.50}
\definecolor{SeaGreen4}{rgb}{0.18,0.55,0.34}
\definecolor{SeaGreen}{rgb}{0.18,0.55,0.34}
\definecolor{SkyBlue1}{rgb}{0.53,0.81,1.00}
\definecolor{SkyBlue2}{rgb}{0.49,0.75,0.93}
\definecolor{SkyBlue3}{rgb}{0.42,0.65,0.80}
\definecolor{SkyBlue4}{rgb}{0.29,0.44,0.55}
\definecolor{SkyBlue}{rgb}{0.53,0.81,0.92}
\definecolor{SlateBlue1}{rgb}{0.51,0.44,1.00}
\definecolor{SlateBlue2}{rgb}{0.48,0.40,0.93}
\definecolor{SlateBlue3}{rgb}{0.41,0.35,0.80}
\definecolor{SlateBlue4}{rgb}{0.28,0.24,0.55}
\definecolor{SlateBlue}{rgb}{0.42,0.35,0.80}
\definecolor{SlateGray1}{rgb}{0.78,0.89,1.00}
\definecolor{SlateGray2}{rgb}{0.73,0.83,0.93}
\definecolor{SlateGray3}{rgb}{0.62,0.71,0.80}
\definecolor{SlateGray4}{rgb}{0.42,0.48,0.55}
\definecolor{SlateGray}{rgb}{0.44,0.50,0.56}
\definecolor{SlateGrey}{rgb}{0.44,0.50,0.56}
\definecolor{SpringGreen1}{rgb}{0.00,1.00,0.50}
\definecolor{SpringGreen2}{rgb}{0.00,0.93,0.46}
\definecolor{SpringGreen3}{rgb}{0.00,0.80,0.40}
\definecolor{SpringGreen4}{rgb}{0.00,0.55,0.27}
\definecolor{SpringGreen}{rgb}{0.00,1.00,0.50}
\definecolor{SteelBlue1}{rgb}{0.39,0.72,1.00}
\definecolor{SteelBlue2}{rgb}{0.36,0.67,0.93}
\definecolor{SteelBlue3}{rgb}{0.31,0.58,0.80}
\definecolor{SteelBlue4}{rgb}{0.21,0.39,0.55}
\definecolor{SteelBlue}{rgb}{0.27,0.51,0.71}
\definecolor{VioletRed1}{rgb}{1.00,0.24,0.59}
\definecolor{VioletRed2}{rgb}{0.93,0.23,0.55}
\definecolor{VioletRed3}{rgb}{0.80,0.20,0.47}
\definecolor{VioletRed4}{rgb}{0.55,0.13,0.32}
\definecolor{VioletRed}{rgb}{0.82,0.13,0.56}
\definecolor{WhiteSmoke}{rgb}{0.96,0.96,0.96}
\definecolor{YellowGreen}{rgb}{0.60,0.80,0.20}
\definecolor{aliceblue}{rgb}{0.94,0.97,1.00}
\definecolor{antiquewhite}{rgb}{0.98,0.92,0.84}
\definecolor{aquamarine1}{rgb}{0.50,1.00,0.83}
\definecolor{aquamarine2}{rgb}{0.46,0.93,0.78}
\definecolor{aquamarine3}{rgb}{0.40,0.80,0.67}
\definecolor{aquamarine4}{rgb}{0.27,0.55,0.45}
\definecolor{aquamarine}{rgb}{0.50,1.00,0.83}
\definecolor{azure1}{rgb}{0.94,1.00,1.00}
\definecolor{azure2}{rgb}{0.88,0.93,0.93}
\definecolor{azure3}{rgb}{0.76,0.80,0.80}
\definecolor{azure4}{rgb}{0.51,0.55,0.55}
\definecolor{azure}{rgb}{0.94,1.00,1.00}
\definecolor{beige}{rgb}{0.96,0.96,0.86}
\definecolor{bisque1}{rgb}{1.00,0.89,0.77}
\definecolor{bisque2}{rgb}{0.93,0.84,0.72}
\definecolor{bisque3}{rgb}{0.80,0.72,0.62}
\definecolor{bisque4}{rgb}{0.55,0.49,0.42}
\definecolor{bisque}{rgb}{1.00,0.89,0.77}
\definecolor{black}{rgb}{0.00,0.00,0.00}
\definecolor{blanchedalmond}{rgb}{1.00,0.92,0.80}
\definecolor{blue1}{rgb}{0.00,0.00,1.00}
\definecolor{blue2}{rgb}{0.00,0.00,0.93}
\definecolor{blue3}{rgb}{0.00,0.00,0.80}
\definecolor{blue4}{rgb}{0.00,0.00,0.55}
\definecolor{blueviolet}{rgb}{0.54,0.17,0.89}
\definecolor{blue}{rgb}{0.00,0.00,1.00}
\definecolor{brown1}{rgb}{1.00,0.25,0.25}
\definecolor{brown2}{rgb}{0.93,0.23,0.23}
\definecolor{brown3}{rgb}{0.80,0.20,0.20}
\definecolor{brown4}{rgb}{0.55,0.14,0.14}
\definecolor{brown}{rgb}{0.65,0.16,0.16}
\definecolor{burlywood1}{rgb}{1.00,0.83,0.61}
\definecolor{burlywood2}{rgb}{0.93,0.77,0.57}
\definecolor{burlywood3}{rgb}{0.80,0.67,0.49}
\definecolor{burlywood4}{rgb}{0.55,0.45,0.33}
\definecolor{burlywood}{rgb}{0.87,0.72,0.53}
\definecolor{cadetblue}{rgb}{0.37,0.62,0.63}
\definecolor{chartreuse1}{rgb}{0.50,1.00,0.00}
\definecolor{chartreuse2}{rgb}{0.46,0.93,0.00}
\definecolor{chartreuse3}{rgb}{0.40,0.80,0.00}
\definecolor{chartreuse4}{rgb}{0.27,0.55,0.00}
\definecolor{chartreuse}{rgb}{0.50,1.00,0.00}
\definecolor{chocolate1}{rgb}{1.00,0.50,0.14}
\definecolor{chocolate2}{rgb}{0.93,0.46,0.13}
\definecolor{chocolate3}{rgb}{0.80,0.40,0.11}
\definecolor{chocolate4}{rgb}{0.55,0.27,0.07}
\definecolor{chocolate}{rgb}{0.82,0.41,0.12}
\definecolor{coral1}{rgb}{1.00,0.45,0.34}
\definecolor{coral2}{rgb}{0.93,0.42,0.31}
\definecolor{coral3}{rgb}{0.80,0.36,0.27}
\definecolor{coral4}{rgb}{0.55,0.24,0.18}
\definecolor{coral}{rgb}{1.00,0.50,0.31}
\definecolor{cornflowerblue}{rgb}{0.39,0.58,0.93}
\definecolor{cornsilk1}{rgb}{1.00,0.97,0.86}
\definecolor{cornsilk2}{rgb}{0.93,0.91,0.80}
\definecolor{cornsilk3}{rgb}{0.80,0.78,0.69}
\definecolor{cornsilk4}{rgb}{0.55,0.53,0.47}
\definecolor{cornsilk}{rgb}{1.00,0.97,0.86}
\definecolor{cyan1}{rgb}{0.00,1.00,1.00}
\definecolor{cyan2}{rgb}{0.00,0.93,0.93}
\definecolor{cyan3}{rgb}{0.00,0.80,0.80}
\definecolor{cyan4}{rgb}{0.00,0.55,0.55}
\definecolor{cyan}{rgb}{0.00,1.00,1.00}
\definecolor{darkblue}{rgb}{0.00,0.00,0.55}
\definecolor{darkcyan}{rgb}{0.00,0.55,0.55}
\definecolor{darkgoldenrod}{rgb}{0.72,0.53,0.04}
\definecolor{darkgray}{rgb}{0.66,0.66,0.66}
\definecolor{darkgreen}{rgb}{0.00,0.39,0.00}
\definecolor{darkgrey}{rgb}{0.66,0.66,0.66}
\definecolor{darkkhaki}{rgb}{0.74,0.72,0.42}
\definecolor{darkmagenta}{rgb}{0.55,0.00,0.55}
\definecolor{darkolive}{rgb}{0.33,0.42,0.18}
\definecolor{darkorange}{rgb}{1.00,0.55,0.00}
\definecolor{darkorchid}{rgb}{0.60,0.20,0.80}
\definecolor{darkred}{rgb}{0.55,0.00,0.00}
\definecolor{darksalmon}{rgb}{0.91,0.59,0.48}
\definecolor{darksea}{rgb}{0.56,0.74,0.56}
\definecolor{darkslate}{rgb}{0.18,0.31,0.31}
\definecolor{darkslate}{rgb}{0.18,0.31,0.31}
\definecolor{darkslate}{rgb}{0.28,0.24,0.55}
\definecolor{darkturquoise}{rgb}{0.00,0.81,0.82}
\definecolor{darkviolet}{rgb}{0.58,0.00,0.83}
\definecolor{deeppink}{rgb}{1.00,0.08,0.58}
\definecolor{deepsky}{rgb}{0.00,0.75,1.00}
\definecolor{dimgray}{rgb}{0.41,0.41,0.41}
\definecolor{dimgrey}{rgb}{0.41,0.41,0.41}
\definecolor{dodgerblue}{rgb}{0.12,0.56,1.00}
\definecolor{firebrick1}{rgb}{1.00,0.19,0.19}
\definecolor{firebrick2}{rgb}{0.93,0.17,0.17}
\definecolor{firebrick3}{rgb}{0.80,0.15,0.15}
\definecolor{firebrick4}{rgb}{0.55,0.10,0.10}
\definecolor{firebrick}{rgb}{0.70,0.13,0.13}
\definecolor{floralwhite}{rgb}{1.00,0.98,0.94}
\definecolor{forestgreen}{rgb}{0.13,0.55,0.13}
\definecolor{gainsboro}{rgb}{0.86,0.86,0.86}
\definecolor{ghostwhite}{rgb}{0.97,0.97,1.00}
\definecolor{gold1}{rgb}{1.00,0.84,0.00}
\definecolor{gold2}{rgb}{0.93,0.79,0.00}
\definecolor{gold3}{rgb}{0.80,0.68,0.00}
\definecolor{gold4}{rgb}{0.55,0.46,0.00}
\definecolor{goldenrod1}{rgb}{1.00,0.76,0.15}
\definecolor{goldenrod2}{rgb}{0.93,0.71,0.13}
\definecolor{goldenrod3}{rgb}{0.80,0.61,0.11}
\definecolor{goldenrod4}{rgb}{0.55,0.41,0.08}
\definecolor{goldenrod}{rgb}{0.85,0.65,0.13}
\definecolor{gold}{rgb}{1.00,0.84,0.00}
\definecolor{gray0}{rgb}{0.00,0.00,0.00}
\definecolor{gray100}{rgb}{1.00,1.00,1.00}
\definecolor{gray10}{rgb}{0.10,0.10,0.10}
\definecolor{gray11}{rgb}{0.11,0.11,0.11}
\definecolor{gray12}{rgb}{0.12,0.12,0.12}
\definecolor{gray13}{rgb}{0.13,0.13,0.13}
\definecolor{gray14}{rgb}{0.14,0.14,0.14}
\definecolor{gray15}{rgb}{0.15,0.15,0.15}
\definecolor{gray16}{rgb}{0.16,0.16,0.16}
\definecolor{gray17}{rgb}{0.17,0.17,0.17}
\definecolor{gray18}{rgb}{0.18,0.18,0.18}
\definecolor{gray19}{rgb}{0.19,0.19,0.19}
\definecolor{gray1}{rgb}{0.01,0.01,0.01}
\definecolor{gray20}{rgb}{0.20,0.20,0.20}
\definecolor{gray21}{rgb}{0.21,0.21,0.21}
\definecolor{gray22}{rgb}{0.22,0.22,0.22}
\definecolor{gray23}{rgb}{0.23,0.23,0.23}
\definecolor{gray24}{rgb}{0.24,0.24,0.24}
\definecolor{gray25}{rgb}{0.25,0.25,0.25}
\definecolor{gray26}{rgb}{0.26,0.26,0.26}
\definecolor{gray27}{rgb}{0.27,0.27,0.27}
\definecolor{gray28}{rgb}{0.28,0.28,0.28}
\definecolor{gray29}{rgb}{0.29,0.29,0.29}
\definecolor{gray2}{rgb}{0.02,0.02,0.02}
\definecolor{gray30}{rgb}{0.30,0.30,0.30}
\definecolor{gray31}{rgb}{0.31,0.31,0.31}
\definecolor{gray32}{rgb}{0.32,0.32,0.32}
\definecolor{gray33}{rgb}{0.33,0.33,0.33}
\definecolor{gray34}{rgb}{0.34,0.34,0.34}
\definecolor{gray35}{rgb}{0.35,0.35,0.35}
\definecolor{gray36}{rgb}{0.36,0.36,0.36}
\definecolor{gray37}{rgb}{0.37,0.37,0.37}
\definecolor{gray38}{rgb}{0.38,0.38,0.38}
\definecolor{gray39}{rgb}{0.39,0.39,0.39}
\definecolor{gray3}{rgb}{0.03,0.03,0.03}
\definecolor{gray40}{rgb}{0.40,0.40,0.40}
\definecolor{gray41}{rgb}{0.41,0.41,0.41}
\definecolor{gray42}{rgb}{0.42,0.42,0.42}
\definecolor{gray43}{rgb}{0.43,0.43,0.43}
\definecolor{gray44}{rgb}{0.44,0.44,0.44}
\definecolor{gray45}{rgb}{0.45,0.45,0.45}
\definecolor{gray46}{rgb}{0.46,0.46,0.46}
\definecolor{gray47}{rgb}{0.47,0.47,0.47}
\definecolor{gray48}{rgb}{0.48,0.48,0.48}
\definecolor{gray49}{rgb}{0.49,0.49,0.49}
\definecolor{gray4}{rgb}{0.04,0.04,0.04}
\definecolor{gray50}{rgb}{0.50,0.50,0.50}
\definecolor{gray51}{rgb}{0.51,0.51,0.51}
\definecolor{gray52}{rgb}{0.52,0.52,0.52}
\definecolor{gray53}{rgb}{0.53,0.53,0.53}
\definecolor{gray54}{rgb}{0.54,0.54,0.54}
\definecolor{gray55}{rgb}{0.55,0.55,0.55}
\definecolor{gray56}{rgb}{0.56,0.56,0.56}
\definecolor{gray57}{rgb}{0.57,0.57,0.57}
\definecolor{gray58}{rgb}{0.58,0.58,0.58}
\definecolor{gray59}{rgb}{0.59,0.59,0.59}
\definecolor{gray5}{rgb}{0.05,0.05,0.05}
\definecolor{gray60}{rgb}{0.60,0.60,0.60}
\definecolor{gray61}{rgb}{0.61,0.61,0.61}
\definecolor{gray62}{rgb}{0.62,0.62,0.62}
\definecolor{gray63}{rgb}{0.63,0.63,0.63}
\definecolor{gray64}{rgb}{0.64,0.64,0.64}
\definecolor{gray65}{rgb}{0.65,0.65,0.65}
\definecolor{gray66}{rgb}{0.66,0.66,0.66}
\definecolor{gray67}{rgb}{0.67,0.67,0.67}
\definecolor{gray68}{rgb}{0.68,0.68,0.68}
\definecolor{gray69}{rgb}{0.69,0.69,0.69}
\definecolor{gray6}{rgb}{0.06,0.06,0.06}
\definecolor{gray70}{rgb}{0.70,0.70,0.70}
\definecolor{gray71}{rgb}{0.71,0.71,0.71}
\definecolor{gray72}{rgb}{0.72,0.72,0.72}
\definecolor{gray73}{rgb}{0.73,0.73,0.73}
\definecolor{gray74}{rgb}{0.74,0.74,0.74}
\definecolor{gray75}{rgb}{0.75,0.75,0.75}
\definecolor{gray76}{rgb}{0.76,0.76,0.76}
\definecolor{gray77}{rgb}{0.77,0.77,0.77}
\definecolor{gray78}{rgb}{0.78,0.78,0.78}
\definecolor{gray79}{rgb}{0.79,0.79,0.79}
\definecolor{gray7}{rgb}{0.07,0.07,0.07}
\definecolor{gray80}{rgb}{0.80,0.80,0.80}
\definecolor{gray81}{rgb}{0.81,0.81,0.81}
\definecolor{gray82}{rgb}{0.82,0.82,0.82}
\definecolor{gray83}{rgb}{0.83,0.83,0.83}
\definecolor{gray84}{rgb}{0.84,0.84,0.84}
\definecolor{gray85}{rgb}{0.85,0.85,0.85}
\definecolor{gray86}{rgb}{0.86,0.86,0.86}
\definecolor{gray87}{rgb}{0.87,0.87,0.87}
\definecolor{gray88}{rgb}{0.88,0.88,0.88}
\definecolor{gray89}{rgb}{0.89,0.89,0.89}
\definecolor{gray8}{rgb}{0.08,0.08,0.08}
\definecolor{gray90}{rgb}{0.90,0.90,0.90}
\definecolor{gray91}{rgb}{0.91,0.91,0.91}
\definecolor{gray92}{rgb}{0.92,0.92,0.92}
\definecolor{gray93}{rgb}{0.93,0.93,0.93}
\definecolor{gray94}{rgb}{0.94,0.94,0.94}
\definecolor{gray95}{rgb}{0.95,0.95,0.95}
\definecolor{gray96}{rgb}{0.96,0.96,0.96}
\definecolor{gray97}{rgb}{0.97,0.97,0.97}
\definecolor{gray98}{rgb}{0.98,0.98,0.98}
\definecolor{gray99}{rgb}{0.99,0.99,0.99}
\definecolor{gray9}{rgb}{0.09,0.09,0.09}
\definecolor{gray}{rgb}{0.75,0.75,0.75}
\definecolor{green1}{rgb}{0.00,1.00,0.00}
\definecolor{green2}{rgb}{0.00,0.93,0.00}
\definecolor{green3}{rgb}{0.00,0.80,0.00}
\definecolor{green4}{rgb}{0.00,0.55,0.00}
\definecolor{greenyellow}{rgb}{0.68,1.00,0.18}
\definecolor{green}{rgb}{0.00,1.00,0.00}
\definecolor{grey0}{rgb}{0.00,0.00,0.00}
\definecolor{grey100}{rgb}{1.00,1.00,1.00}
\definecolor{grey10}{rgb}{0.10,0.10,0.10}
\definecolor{grey11}{rgb}{0.11,0.11,0.11}
\definecolor{grey12}{rgb}{0.12,0.12,0.12}
\definecolor{grey13}{rgb}{0.13,0.13,0.13}
\definecolor{grey14}{rgb}{0.14,0.14,0.14}
\definecolor{grey15}{rgb}{0.15,0.15,0.15}
\definecolor{grey16}{rgb}{0.16,0.16,0.16}
\definecolor{grey17}{rgb}{0.17,0.17,0.17}
\definecolor{grey18}{rgb}{0.18,0.18,0.18}
\definecolor{grey19}{rgb}{0.19,0.19,0.19}
\definecolor{grey1}{rgb}{0.01,0.01,0.01}
\definecolor{grey20}{rgb}{0.20,0.20,0.20}
\definecolor{grey21}{rgb}{0.21,0.21,0.21}
\definecolor{grey22}{rgb}{0.22,0.22,0.22}
\definecolor{grey23}{rgb}{0.23,0.23,0.23}
\definecolor{grey24}{rgb}{0.24,0.24,0.24}
\definecolor{grey25}{rgb}{0.25,0.25,0.25}
\definecolor{grey26}{rgb}{0.26,0.26,0.26}
\definecolor{grey27}{rgb}{0.27,0.27,0.27}
\definecolor{grey28}{rgb}{0.28,0.28,0.28}
\definecolor{grey29}{rgb}{0.29,0.29,0.29}
\definecolor{grey2}{rgb}{0.02,0.02,0.02}
\definecolor{grey30}{rgb}{0.30,0.30,0.30}
\definecolor{grey31}{rgb}{0.31,0.31,0.31}
\definecolor{grey32}{rgb}{0.32,0.32,0.32}
\definecolor{grey33}{rgb}{0.33,0.33,0.33}
\definecolor{grey34}{rgb}{0.34,0.34,0.34}
\definecolor{grey35}{rgb}{0.35,0.35,0.35}
\definecolor{grey36}{rgb}{0.36,0.36,0.36}
\definecolor{grey37}{rgb}{0.37,0.37,0.37}
\definecolor{grey38}{rgb}{0.38,0.38,0.38}
\definecolor{grey39}{rgb}{0.39,0.39,0.39}
\definecolor{grey3}{rgb}{0.03,0.03,0.03}
\definecolor{grey40}{rgb}{0.40,0.40,0.40}
\definecolor{grey41}{rgb}{0.41,0.41,0.41}
\definecolor{grey42}{rgb}{0.42,0.42,0.42}
\definecolor{grey43}{rgb}{0.43,0.43,0.43}
\definecolor{grey44}{rgb}{0.44,0.44,0.44}
\definecolor{grey45}{rgb}{0.45,0.45,0.45}
\definecolor{grey46}{rgb}{0.46,0.46,0.46}
\definecolor{grey47}{rgb}{0.47,0.47,0.47}
\definecolor{grey48}{rgb}{0.48,0.48,0.48}
\definecolor{grey49}{rgb}{0.49,0.49,0.49}
\definecolor{grey4}{rgb}{0.04,0.04,0.04}
\definecolor{grey50}{rgb}{0.50,0.50,0.50}
\definecolor{grey51}{rgb}{0.51,0.51,0.51}
\definecolor{grey52}{rgb}{0.52,0.52,0.52}
\definecolor{grey53}{rgb}{0.53,0.53,0.53}
\definecolor{grey54}{rgb}{0.54,0.54,0.54}
\definecolor{grey55}{rgb}{0.55,0.55,0.55}
\definecolor{grey56}{rgb}{0.56,0.56,0.56}
\definecolor{grey57}{rgb}{0.57,0.57,0.57}
\definecolor{grey58}{rgb}{0.58,0.58,0.58}
\definecolor{grey59}{rgb}{0.59,0.59,0.59}
\definecolor{grey5}{rgb}{0.05,0.05,0.05}
\definecolor{grey60}{rgb}{0.60,0.60,0.60}
\definecolor{grey61}{rgb}{0.61,0.61,0.61}
\definecolor{grey62}{rgb}{0.62,0.62,0.62}
\definecolor{grey63}{rgb}{0.63,0.63,0.63}
\definecolor{grey64}{rgb}{0.64,0.64,0.64}
\definecolor{grey65}{rgb}{0.65,0.65,0.65}
\definecolor{grey66}{rgb}{0.66,0.66,0.66}
\definecolor{grey67}{rgb}{0.67,0.67,0.67}
\definecolor{grey68}{rgb}{0.68,0.68,0.68}
\definecolor{grey69}{rgb}{0.69,0.69,0.69}
\definecolor{grey6}{rgb}{0.06,0.06,0.06}
\definecolor{grey70}{rgb}{0.70,0.70,0.70}
\definecolor{grey71}{rgb}{0.71,0.71,0.71}
\definecolor{grey72}{rgb}{0.72,0.72,0.72}
\definecolor{grey73}{rgb}{0.73,0.73,0.73}
\definecolor{grey74}{rgb}{0.74,0.74,0.74}
\definecolor{grey75}{rgb}{0.75,0.75,0.75}
\definecolor{grey76}{rgb}{0.76,0.76,0.76}
\definecolor{grey77}{rgb}{0.77,0.77,0.77}
\definecolor{grey78}{rgb}{0.78,0.78,0.78}
\definecolor{grey79}{rgb}{0.79,0.79,0.79}
\definecolor{grey7}{rgb}{0.07,0.07,0.07}
\definecolor{grey80}{rgb}{0.80,0.80,0.80}
\definecolor{grey81}{rgb}{0.81,0.81,0.81}
\definecolor{grey82}{rgb}{0.82,0.82,0.82}
\definecolor{grey83}{rgb}{0.83,0.83,0.83}
\definecolor{grey84}{rgb}{0.84,0.84,0.84}
\definecolor{grey85}{rgb}{0.85,0.85,0.85}
\definecolor{grey86}{rgb}{0.86,0.86,0.86}
\definecolor{grey87}{rgb}{0.87,0.87,0.87}
\definecolor{grey88}{rgb}{0.88,0.88,0.88}
\definecolor{grey89}{rgb}{0.89,0.89,0.89}
\definecolor{grey8}{rgb}{0.08,0.08,0.08}
\definecolor{grey90}{rgb}{0.90,0.90,0.90}
\definecolor{grey91}{rgb}{0.91,0.91,0.91}
\definecolor{grey92}{rgb}{0.92,0.92,0.92}
\definecolor{grey93}{rgb}{0.93,0.93,0.93}
\definecolor{grey94}{rgb}{0.94,0.94,0.94}
\definecolor{grey95}{rgb}{0.95,0.95,0.95}
\definecolor{grey96}{rgb}{0.96,0.96,0.96}
\definecolor{grey97}{rgb}{0.97,0.97,0.97}
\definecolor{grey98}{rgb}{0.98,0.98,0.98}
\definecolor{grey99}{rgb}{0.99,0.99,0.99}
\definecolor{grey9}{rgb}{0.09,0.09,0.09}
\definecolor{grey}{rgb}{0.75,0.75,0.75}
\definecolor{honeydew1}{rgb}{0.94,1.00,0.94}
\definecolor{honeydew2}{rgb}{0.88,0.93,0.88}
\definecolor{honeydew3}{rgb}{0.76,0.80,0.76}
\definecolor{honeydew4}{rgb}{0.51,0.55,0.51}
\definecolor{honeydew}{rgb}{0.94,1.00,0.94}
\definecolor{hotpink}{rgb}{1.00,0.41,0.71}
\definecolor{indianred}{rgb}{0.80,0.36,0.36}
\definecolor{ivory1}{rgb}{1.00,1.00,0.94}
\definecolor{ivory2}{rgb}{0.93,0.93,0.88}
\definecolor{ivory3}{rgb}{0.80,0.80,0.76}
\definecolor{ivory4}{rgb}{0.55,0.55,0.51}
\definecolor{ivory}{rgb}{1.00,1.00,0.94}
\definecolor{khaki1}{rgb}{1.00,0.96,0.56}
\definecolor{khaki2}{rgb}{0.93,0.90,0.52}
\definecolor{khaki3}{rgb}{0.80,0.78,0.45}
\definecolor{khaki4}{rgb}{0.55,0.53,0.31}
\definecolor{khaki}{rgb}{0.94,0.90,0.55}
\definecolor{lavenderblush}{rgb}{1.00,0.94,0.96}
\definecolor{lavender}{rgb}{0.90,0.90,0.98}
\definecolor{lawngreen}{rgb}{0.49,0.99,0.00}
\definecolor{lemonchiffon}{rgb}{1.00,0.98,0.80}
\definecolor{lightblue}{rgb}{0.68,0.85,0.90}
\definecolor{lightcoral}{rgb}{0.94,0.50,0.50}
\definecolor{lightcyan}{rgb}{0.88,1.00,1.00}
\definecolor{lightgoldenrod}{rgb}{0.93,0.87,0.51}
\definecolor{lightgoldenrod}{rgb}{0.98,0.98,0.82}
\definecolor{lightgray}{rgb}{0.83,0.83,0.83}
\definecolor{lightgreen}{rgb}{0.56,0.93,0.56}
\definecolor{lightgrey}{rgb}{0.83,0.83,0.83}
\definecolor{lightpink}{rgb}{1.00,0.71,0.76}
\definecolor{lightsalmon}{rgb}{1.00,0.63,0.48}
\definecolor{lightsea}{rgb}{0.13,0.70,0.67}
\definecolor{lightsky}{rgb}{0.53,0.81,0.98}
\definecolor{lightslate}{rgb}{0.47,0.53,0.60}
\definecolor{lightslate}{rgb}{0.47,0.53,0.60}
\definecolor{lightslate}{rgb}{0.52,0.44,1.00}
\definecolor{lightsteel}{rgb}{0.69,0.77,0.87}
\definecolor{lightyellow}{rgb}{1.00,1.00,0.88}
\definecolor{limegreen}{rgb}{0.20,0.80,0.20}
\definecolor{linen}{rgb}{0.98,0.94,0.90}
\definecolor{magenta1}{rgb}{1.00,0.00,1.00}
\definecolor{magenta2}{rgb}{0.93,0.00,0.93}
\definecolor{magenta3}{rgb}{0.80,0.00,0.80}
\definecolor{magenta4}{rgb}{0.55,0.00,0.55}
\definecolor{magenta}{rgb}{1.00,0.00,1.00}
\definecolor{maroon1}{rgb}{1.00,0.20,0.70}
\definecolor{maroon2}{rgb}{0.93,0.19,0.65}
\definecolor{maroon3}{rgb}{0.80,0.16,0.56}
\definecolor{maroon4}{rgb}{0.55,0.11,0.38}
\definecolor{maroon}{rgb}{0.69,0.19,0.38}
\definecolor{mediumaquamarine}{rgb}{0.40,0.80,0.67}
\definecolor{mediumblue}{rgb}{0.00,0.00,0.80}
\definecolor{mediumorchid}{rgb}{0.73,0.33,0.83}
\definecolor{mediumpurple}{rgb}{0.58,0.44,0.86}
\definecolor{mediumsea}{rgb}{0.24,0.70,0.44}
\definecolor{mediumslate}{rgb}{0.48,0.41,0.93}
\definecolor{mediumspring}{rgb}{0.00,0.98,0.60}
\definecolor{mediumturquoise}{rgb}{0.28,0.82,0.80}
\definecolor{mediumviolet}{rgb}{0.78,0.08,0.52}
\definecolor{midnightblue}{rgb}{0.10,0.10,0.44}
\definecolor{mintcream}{rgb}{0.96,1.00,0.98}
\definecolor{mistyrose}{rgb}{1.00,0.89,0.88}
\definecolor{moccasin}{rgb}{1.00,0.89,0.71}
\definecolor{navajowhite}{rgb}{1.00,0.87,0.68}
\definecolor{navyblue}{rgb}{0.00,0.00,0.50}
\definecolor{navy}{rgb}{0.00,0.00,0.50}
\definecolor{oldlace}{rgb}{0.99,0.96,0.90}
\definecolor{olivedrab}{rgb}{0.42,0.56,0.14}
\definecolor{orange1}{rgb}{1.00,0.65,0.00}
\definecolor{orange2}{rgb}{0.93,0.60,0.00}
\definecolor{orange3}{rgb}{0.80,0.52,0.00}
\definecolor{orange4}{rgb}{0.55,0.35,0.00}
\definecolor{orangered}{rgb}{1.00,0.27,0.00}
\definecolor{orange}{rgb}{1.00,0.65,0.00}
\definecolor{orchid1}{rgb}{1.00,0.51,0.98}
\definecolor{orchid2}{rgb}{0.93,0.48,0.91}
\definecolor{orchid3}{rgb}{0.80,0.41,0.79}
\definecolor{orchid4}{rgb}{0.55,0.28,0.54}
\definecolor{orchid}{rgb}{0.85,0.44,0.84}
\definecolor{palegoldenrod}{rgb}{0.93,0.91,0.67}
\definecolor{palegreen}{rgb}{0.60,0.98,0.60}
\definecolor{paleturquoise}{rgb}{0.69,0.93,0.93}
\definecolor{paleviolet}{rgb}{0.86,0.44,0.58}
\definecolor{papayawhip}{rgb}{1.00,0.94,0.84}
\definecolor{peachpuff}{rgb}{1.00,0.85,0.73}
\definecolor{peru}{rgb}{0.80,0.52,0.25}
\definecolor{pink1}{rgb}{1.00,0.71,0.77}
\definecolor{pink2}{rgb}{0.93,0.66,0.72}
\definecolor{pink3}{rgb}{0.80,0.57,0.62}
\definecolor{pink4}{rgb}{0.55,0.39,0.42}
\definecolor{pink}{rgb}{1.00,0.75,0.80}
\definecolor{plum1}{rgb}{1.00,0.73,1.00}
\definecolor{plum2}{rgb}{0.93,0.68,0.93}
\definecolor{plum3}{rgb}{0.80,0.59,0.80}
\definecolor{plum4}{rgb}{0.55,0.40,0.55}
\definecolor{plum}{rgb}{0.87,0.63,0.87}
\definecolor{powderblue}{rgb}{0.69,0.88,0.90}
\definecolor{purple1}{rgb}{0.61,0.19,1.00}
\definecolor{purple2}{rgb}{0.57,0.17,0.93}
\definecolor{purple3}{rgb}{0.49,0.15,0.80}
\definecolor{purple4}{rgb}{0.33,0.10,0.55}
\definecolor{purple}{rgb}{0.63,0.13,0.94}
\definecolor{red1}{rgb}{1.00,0.00,0.00}
\definecolor{red2}{rgb}{0.93,0.00,0.00}
\definecolor{red3}{rgb}{0.80,0.00,0.00}
\definecolor{red4}{rgb}{0.55,0.00,0.00}
\definecolor{red}{rgb}{1.00,0.00,0.00}
\definecolor{rosybrown}{rgb}{0.74,0.56,0.56}
\definecolor{royalblue}{rgb}{0.25,0.41,0.88}
\definecolor{saddlebrown}{rgb}{0.55,0.27,0.07}
\definecolor{salmon1}{rgb}{1.00,0.55,0.41}
\definecolor{salmon2}{rgb}{0.93,0.51,0.38}
\definecolor{salmon3}{rgb}{0.80,0.44,0.33}
\definecolor{salmon4}{rgb}{0.55,0.30,0.22}
\definecolor{salmon}{rgb}{0.98,0.50,0.45}
\definecolor{sandybrown}{rgb}{0.96,0.64,0.38}
\definecolor{seagreen}{rgb}{0.18,0.55,0.34}
\definecolor{seashell1}{rgb}{1.00,0.96,0.93}
\definecolor{seashell2}{rgb}{0.93,0.90,0.87}
\definecolor{seashell3}{rgb}{0.80,0.77,0.75}
\definecolor{seashell4}{rgb}{0.55,0.53,0.51}
\definecolor{seashell}{rgb}{1.00,0.96,0.93}
\definecolor{sienna1}{rgb}{1.00,0.51,0.28}
\definecolor{sienna2}{rgb}{0.93,0.47,0.26}
\definecolor{sienna3}{rgb}{0.80,0.41,0.22}
\definecolor{sienna4}{rgb}{0.55,0.28,0.15}
\definecolor{sienna}{rgb}{0.63,0.32,0.18}
\definecolor{skyblue}{rgb}{0.53,0.81,0.92}
\definecolor{slateblue}{rgb}{0.42,0.35,0.80}
\definecolor{slategray}{rgb}{0.44,0.50,0.56}
\definecolor{slategrey}{rgb}{0.44,0.50,0.56}
\definecolor{snow1}{rgb}{1.00,0.98,0.98}
\definecolor{snow2}{rgb}{0.93,0.91,0.91}
\definecolor{snow3}{rgb}{0.80,0.79,0.79}
\definecolor{snow4}{rgb}{0.55,0.54,0.54}
\definecolor{snow}{rgb}{1.00,0.98,0.98}
\definecolor{springgreen}{rgb}{0.00,1.00,0.50}
\definecolor{steelblue}{rgb}{0.27,0.51,0.71}
\definecolor{tan1}{rgb}{1.00,0.65,0.31}
\definecolor{tan2}{rgb}{0.93,0.60,0.29}
\definecolor{tan3}{rgb}{0.80,0.52,0.25}
\definecolor{tan4}{rgb}{0.55,0.35,0.17}
\definecolor{tan}{rgb}{0.82,0.71,0.55}
\definecolor{thistle1}{rgb}{1.00,0.88,1.00}
\definecolor{thistle2}{rgb}{0.93,0.82,0.93}
\definecolor{thistle3}{rgb}{0.80,0.71,0.80}
\definecolor{thistle4}{rgb}{0.55,0.48,0.55}
\definecolor{thistle}{rgb}{0.85,0.75,0.85}
\definecolor{tomato1}{rgb}{1.00,0.39,0.28}
\definecolor{tomato2}{rgb}{0.93,0.36,0.26}
\definecolor{tomato3}{rgb}{0.80,0.31,0.22}
\definecolor{tomato4}{rgb}{0.55,0.21,0.15}
\definecolor{tomato}{rgb}{1.00,0.39,0.28}
\definecolor{turquoise1}{rgb}{0.00,0.96,1.00}
\definecolor{turquoise2}{rgb}{0.00,0.90,0.93}
\definecolor{turquoise3}{rgb}{0.00,0.77,0.80}
\definecolor{turquoise4}{rgb}{0.00,0.53,0.55}
\definecolor{turquoise}{rgb}{0.25,0.88,0.82}
\definecolor{violetred}{rgb}{0.82,0.13,0.56}
\definecolor{violet}{rgb}{0.93,0.51,0.93}
\definecolor{wheat1}{rgb}{1.00,0.91,0.73}
\definecolor{wheat2}{rgb}{0.93,0.85,0.68}
\definecolor{wheat3}{rgb}{0.80,0.73,0.59}
\definecolor{wheat4}{rgb}{0.55,0.49,0.40}
\definecolor{wheat}{rgb}{0.96,0.87,0.70}
\definecolor{whitesmoke}{rgb}{0.96,0.96,0.96}
\definecolor{white}{rgb}{1.00,1.00,1.00}
\definecolor{yellow1}{rgb}{1.00,1.00,0.00}
\definecolor{yellow2}{rgb}{0.93,0.93,0.00}
\definecolor{yellow3}{rgb}{0.80,0.80,0.00}
\definecolor{yellow4}{rgb}{0.55,0.55,0.00}
\definecolor{yellowgreen}{rgb}{0.60,0.80,0.20}
\definecolor{yellow}{rgb}{1.00,1.00,0.00}
\usepackage{amsfonts}
\usepackage{bm}
\usepackage{longtable}
\usepackage{multirow}
\usepackage{pifont}
\usepackage{gensymb}
\usepackage{amsmath}
\usepackage[]{txfonts}

\usepackage{booktabs} 
 
\newcommand{\gskfont}{
  \bfseries 
  \color{red}
}
\newcommand{\jcafont}{
  \bfseries 
  \color{green}
}
\newcommand{\afkfont}{
  \bfseries 
  \color{orange}
}
\newcommand{\romfont}{
   \bfseries 
  \color{violet}
}
\newcommand{\tkfont}{
  \bfseries 
  \color{blue}
}
\newcommand{\nzpfont}{
  \color{blue}
}
\newcommand{\jwbfont}{
  \color{grey}
}

\DeclareTextFontCommand{\gsk}{\gskfont}
\DeclareTextFontCommand{\jca}{\jcafont}
\DeclareTextFontCommand{\tk}{\tkfont}
\DeclareTextFontCommand{\afk}{\afkfont}
\DeclareTextFontCommand{\rom}{\romfont}
\DeclareTextFontCommand{\nzp}{\nzpfont}
\DeclareTextFontCommand{\jwb}{\jwbfont}

\newcommand{\radyn}{\texttt{RADYN}}
\newcommand{\radynfp}{\texttt{RADYN+FP}}
\newcommand{\fpc}{\texttt{FP}}
\newcommand{\rhpar}{\texttt{RH15D}}
\newcommand{\rh}{\texttt{RH}}
\newcommand{\ozpy}{\texttt{OrrallZirkerPy}}
\newcommand{\lya}{Ly~$\alpha$}
\newcommand{\lyb}{Ly~$\beta$}

\shorttitle{Revisiting the Orrall-Zirker Effect}
\shortauthors{Kerr et al}

\begin{document}


	\title{Prospects of Detecting Non-thermal Protons in Solar Flares via Lyman Line Spectroscopy: Revisiting the Orrall-Zirker Effect}
	\author[0000-0001-5316-914X]{Graham~S. Kerr}
	\email{graham.s.kerr@nasa.gov}
	\email{kerrg@cua.edu}
	\affil{NASA Goddard Space Flight Center, Heliophysics Science Division, Code 671, 8800 Greenbelt Rd., Greenbelt, MD 20771, USA}
 	\affil{Department of Physics, Catholic University of America, 620 Michigan Avenue, Northeast, Washington, DC 20064, USA}
	
	\author[0000-0003-4227-6809]{Joel~C. Allred}
	\affil{NASA Goddard Space Flight Center, Heliophysics Science Division, Code 671, 8800 Greenbelt Rd., Greenbelt, MD 20771, USA}
       
        \author[0000-0001-7458-1176]{Adam~F. Kowalski}
	\affil{National Solar Observatory, University of Colorado Boulder, 3665 Discovery Drive, Boulder CO 80303, USA}
        \affil{Department of Astrophysical and Planetary Sciences, University of Colorado, Boulder 2000 Colorado Ave, CO 80305, USA}
 
        \author[0000-0001-5031-1892]{Ryan~O. Milligan}
	\affil{Queen's University Belfast, University Rd, Belfast BT7 1NN, Northern Ireland}
	\affil{Department of Physics, Catholic University of America, 620 Michigan Avenue, Northeast, Washington, DC 20064, USA}  
	
      	 \author[0000-0001-5685-1283]{Hugh~S. Hudson}
	 \affil{SUPA School of Physics \& Astronomy, University of Glasgow, Glasgow, G12 8QQ, UK}
	 \affil{Space Sciences Laboratory, UC Berkeley, CA 94720, USA}	
 
	 \author[0000-0001-6395-7115]{Natalia Zambrana Prado}
	 \affil{NASA Goddard Space Flight Center, Heliophysics Science Division, Code 671, 8800 Greenbelt Rd., Greenbelt, MD 20771, USA}
 	 \affil{Department of Physics, Catholic University of America, 620 Michigan Avenue, Northeast, Washington, DC 20064, USA}
	
	 \author[0000-0001-9632-447X]{Therese~A. Kucera}
	 \affil{NASA Goddard Space Flight Center, Heliophysics Science Division, Code 671, 8800 Greenbelt Rd., Greenbelt, MD 20771, USA}
	 
      	\author{Jeffrey~W. Brosius}
	\affil{NASA Goddard Space Flight Center, Heliophysics Science Division, Code 671, 8800 Greenbelt Rd., Greenbelt, MD 20771, USA}
        	\affil{Department of Physics, Catholic University of America, 620 Michigan Avenue, Northeast, Washington, DC 20064, USA}
	  
	\date{Received / Accepted}
	
	
	\begin{abstract}	
	Solar flares are efficient particle accelerators, with a substantial fraction of the energy released manifesting as non-thermal particles. While the role that non-thermal electrons play in transporting flare energy is well studied, the properties and importance of non-thermal protons is rather less well understood. This is in large part due to the paucity of diagnostics, particularly at the lower-energy (deka-keV) range of non-thermal proton distributions in flares. One means to identify the presence of deka-keV protons is by an effect originally described by \cite{1976ApJ...208..618O}. In the Orrall-Zirker effect, non-thermal protons interact with ambient neutral hydrogen, and via charge exchange produce a population of energetic neutral atoms (ENAs) in the chromosphere. These ENAs subsequently produce an extremely redshifted photon in the red wings of hydrogen spectral lines. We revisit predictions of the strength of this effect using modern interaction cross-sections, and numerical models capable of self-consistently simulating the flaring non-equilibrium ionization stratification, and the non-thermal proton distribution (and, crucially, their feedback on each other). We synthesize both the thermal and non-thermal emission from \lya\ and \lyb, the most promising lines that may exhibit a detectable signal. These new predictions are are weaker and more transient than prior estimates, but the effects should be detectable in fortuitous circumstances. We degrade the \lyb\ emission to the resolution of the Spectral Imaging of the Coronal Environment (SPICE) instrument on board Solar Orbiter, demonstrating that though likely difficult, it should be possible to detect the presence of non-thermal protons in flares observed by SPICE.
	\end{abstract}


\section{Introduction}\label{sec:intro}

Solar flares and other forms of solar magnetic activity often generate high-energy particles, with energies far in excess of the highest mean
energies ($kT \approx 1$~keV) for any collisionally relaxed solar plasma population \citep[e.g.,][]{2012RSPTA.370.3241V}. 
The bremsstrahlung continuum allows us to observe electrons above $E\sim10$~keV by the techniques of X-ray astronomy, and to characterize them spatially, spectrally, and temporally \citep[reviews in the solar flare context include][]{2011SSRv..159..107H,2011SSRv..159..301K}. This knowledge underpins the development of ``thick target'' beam models 
\citep[e.g.][]{1971SoPh...18..489B,1972SoPh...24..414H,1978ApJ...224..241E}, which have had an extensive theoretical development.  The particularly close spatial and temporal association between deka-keV hard X-ray sources and flare ribbons and footpoints observed in the UV/optical/infrared \citep[e.g.][]{2011SSRv..159...19F} supports the idea that accelerated electrons dominate the  transport and dissipation of flare energy during the impulsive phase \citep{1970ApJ...162.1003K} of an event.

In some flares we also observe strong emissions from high-energy ions -- protons and, to a lesser extent, $\alpha$ particles and other heavier ions -- \citep[e.g.][]{2009ApJ...698L.152S}. In such cases they may well carry energy comparable to non-thermal electrons \citep[][]{2012ApJ...759...71E}. By excluding non-thermal ions in flare models, we are potentially ignoring up to half of the flare energy transported through the Sun’s atmosphere, or at least an energetically important constituent.

Unfortunately, characterising their properties (or even their presence in chromospheric flare footpoints), is more challenging as non-thermal ions do not have such convenient signatures as non-thermal electrons do. \cite{1981ApJ...244L.171C} initially discovered solar $\gamma$-ray line emission in the form of the 2.223~MeV line produced by deuterium formation ($n + p \rightarrow {^2}H$).
This implied the existence of free neutrons as a part of the flare process, and these could only have come from nuclear reactions: hence, the acceleration of high-energy ions (as already well known from ``solar cosmic rays,'' now known as ``solar energetic particles'', SEPs). These first observations also showed the presence of the positron-annihilation line at 511~keV and of inelastic-scattering (spallation) lines, such as the 4.43~MeV line of ${^{12}}C$, as well. These processes had all been anticipated theoretically, based on the prior knowledge of SEPs, but the $\gamma$-radiation gave us hints about the interactions of such particles in the lower solar atmosphere and of their possible role in flare dynamics. Limitations of detector sensitivity generally make it almost impossible to use $\gamma$~rays to detect ions from thin-target processes in the corona. The positron-annihilation and deuterium-formation lines result from secondary processes; the initial spallation reactions produce unstable nuclei that emit positrons and neutrons, respectively, and these must slow down collisionally before they interact. The interpretation of these delayed emission lines therefore involves additional model assumptions \citep[e.g.][]{1975SSRv...18..341R}.

The $\gamma$-ray observations (for ions) have severe limitations in comparison with the hard X-ray bremsstrahlung (for electrons), however. In addition to the major technical problems of astronomy in this photon energy range,  the inelastic-scattering lines alone provide almost no spectral information. Thus our remote-sensing knowledge of non-thermal ions actually in the solar flare volume
basically only really reflects the proton population at about 10~MeV, the typical energy threshold for these reactions. Given coronal values of $kT_e \approx 100$~eV, or at most a few keV in flares, this leaves a huge gap in our knowledge. Extrapolating down to 1~MeV/nucleon, the energy of the flare ion population already rivals that of the non-thermal electrons in some cases at least \citep[e.g.,][]{1995ApJ...455L.193R}, suggesting orders-of-magnitude uncertainty in total particle energy if one were to consider (still highly non-thermal) particles at 10-100~keV/nucleon.

In spite of the challenges of working in the few-MeV range of photon energies, the Reuven Ramaty High-Energy Spectroscopic Imager \citep[RHESSI;][]{2002SoPh..210....3L} succeeded in obtaining useful images of solar flares in a small number of events \citep{2003ApJ...595L..77H,2006ApJ...644L..93H}. These results localized high-energy ions, with their sources consistent with the flare loops \citep{2011SSRv..159..421L}, confirming the conclusion drawn from the rapid time variability previously observed. The 2.223~MeV line proved to come from flare footpoint regions, as expected theoretically, though with centroids curiously offset from the 200-300~keV X-ray (i.e. electron) sources by some $[14-20]^{\prime\prime}\pm5^{\prime\prime}$ in the three events studied. The RHESSI data also strongly suggested that the acceleration mechanism for the flare ions had a close relationship with that observed via the relatively well-observed X-ray bremsstrahlung above about 300~keV \citep{2009ApJ...698L.152S}. This key observation, identifying high-energy particles within closed magnetic field regions (the flare loops), makes the study of ion beams interesting, many decades after this was proposed on the basis of early observations of SEPs \citep{1970SoPh...13..471S,1970SoPh...15..176N,1995SSRv...73..387S}.

\begin{figure}
	\centering 
	{\includegraphics[width = 0.5\textwidth, clip = true, trim = 0cm 0.cm 0cm 0.cm]{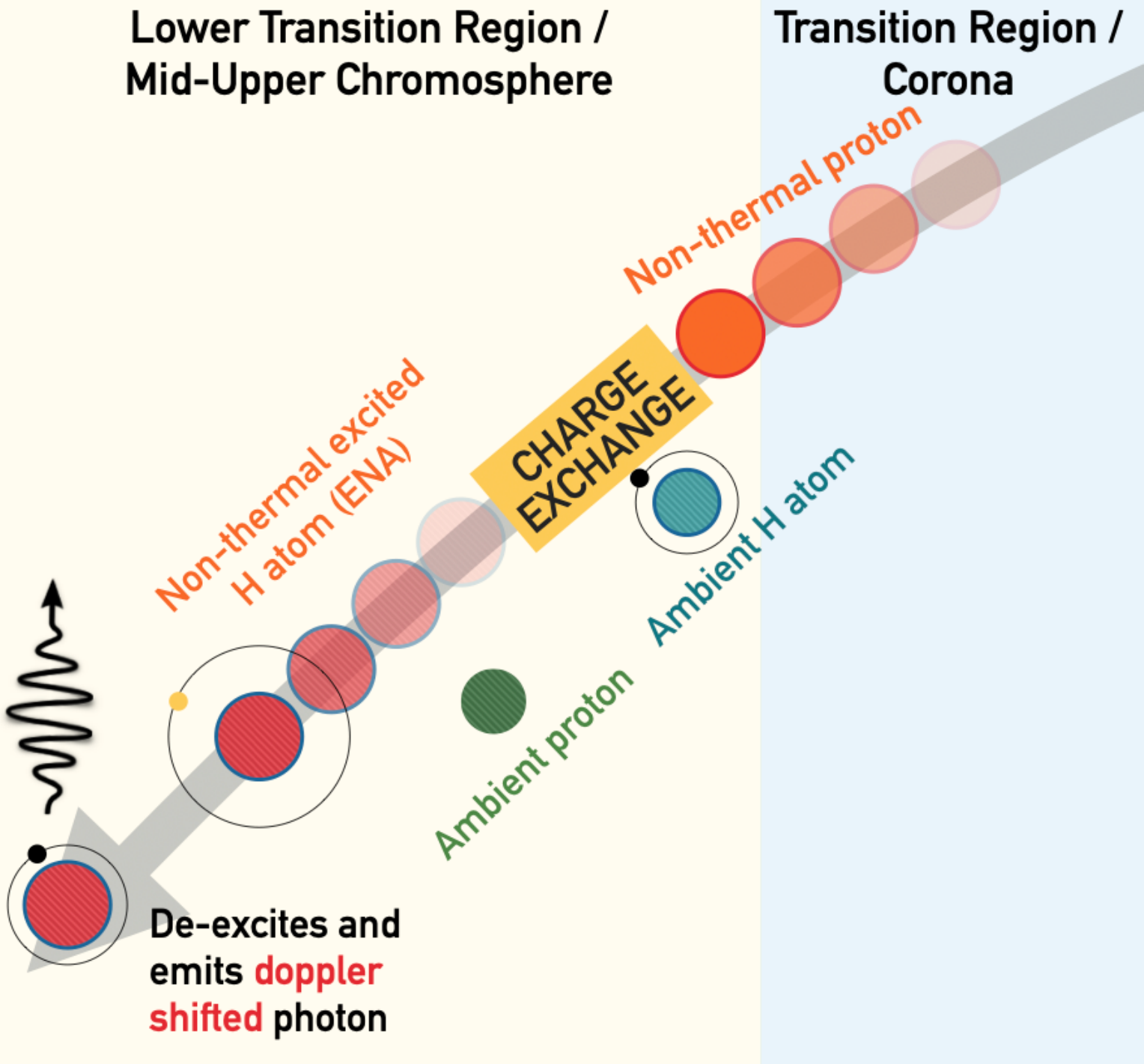}}	
	\caption{\textsl{A schematic of the Orrall-Zirker (OZ) Effect, in which a distribution of non-thermal protons streams through the solar atmosphere until it reaches a population of neutral hydrogen atoms. Due to charge exchange with this ambient population, and subsequent collisions with ambient particles (hydrogen, protons, electrons) some fraction of the precipitating non-thermal protons is transformed into a non-thermal hydrogen atom (an energetic neutral atom, ENA) in an excited state. Spontaneous radiative de-excitation subsequently produces a highly Doppler shifted photon (e.g. of \lya, \lyb, H~$\alpha$).}}
	\label{fig:OZSchematic}
\end{figure}

The likely existence of ion beams in flares provides a crucial motivation for the work presented in this paper. Here we study the Orrall-Zirker effect (hereinafter OZ), as proposed by \cite{1976ApJ...208..618O}. Briefly, successful observations of OZ, via the \lya\ line wings, would permit remote sensing of the potentially dominant low-energy ions in the energy range $\sim0.01-1$~MeV for the first time. Figure~\ref{fig:OZSchematic} sketches the OZ process: precipitating ions enter the partially ionized lower atmosphere, pick up electrons from neutral H atoms by atomic charge exchange reactions, and then spontaneously radiate a bound-bound photon in flight. This photon would be Doppler shifted by some amount depending on the original energy (velocity) of the proton. In the simplest picture this would enhance the red wing of \lya\ or \lyb\ because of the essentially downward motion of the beam particle. At exact beam alignment (the observer on the beam axis) 10-100 keV protons would appear as displacements of 5.6--17.7~\AA\ in the red wing of \lya, well beyond likely flaring line widths of a few \AA; we do not yet have enough direct observation of this line to be more precise about this. We note that stellar \lya\ spectra obtained by the Hubble Space Telescope \citep{2005ApJS..159..118W} show flare-star \lya\ line profiles no more than a few \AA\ wide in quiescent conditions.

Follow on studies by \cite{1985ApJ...295..275C}, \cite{1995A&A...297..854F}, and \cite{1999ApJ...514..430B} confirmed and expanded upon the initial theoretical work of \cite{1976ApJ...208..618O}. \cite{1995A&A...297..854F} noted that the predictions of the strength of non-thermal emission from models that assumed a homogeneous, and low, ionization fraction, were overstimated. However, even though they used a more realistic ionization stratification, they used static semi-empirical flare atmospheres and did not model self-consistently the feedback between the non-thermal protons and the atmospheric stratification. \cite{1985ApJ...295..275C} and \cite{1999ApJ...514..430B}, investigated the effect that different proton beam distributions had on the appearance of the non-thermal feature, indicating diagnostic potential. \cite{1999ApJ...514..430B} looked at the temporal evolution of the non-thermal emission, resulting from the atmospheric evolution, using a more sophisticated treatment of the non-thermal proton thermalisation than employed by the other studies. They focused on weaker energy fluxes than \cite{1985ApJ...295..275C}, and noted an initially strong source, that quickly diminished but then increased somewhat. However, they did not calculate the ambient emission\footnote{When we refer to `ambient' emission throughout, we mean the intensity in the absence of non-thermal emission following charge exchange; that is the Lyman line enhancements following flare heating in the original \radynfp\ simulations.} in their flare simulations, to compare against the strength of the non-thermal emission.  

The only tantalizing signature of the OZ effect from \textsl{bona-fide} red-shifted feature of the OZ effect was reported by \cite{1992ApJ...397L..95W}, who used data from the Goddard High Resolution spectrograph (GHRS) on the Hubble Space Telescope (HST). Those observations of \lya\ from a moderately strong stellar flare that occurred on the dMe star AU Mic revealed a short ($t\sim3$~s), but clear, enhancement to redward of line core relative to the blueward side. Unfortunately, follow up studies of flares on AU Mic had null detections of non-thermal \lya\ emission \citep{1993ApJ...414..872R,2001ApJ...554..368R,2022AJ....164..110F}. As regards solar flares, we have lacked routine high-quality observations of the Lyman lines in flares, and those that we do possess have not revealed the OZ effect. Searching for the equivalent feature from charge exchange between a non-thermal $\alpha$-particle beam producing non-thermal \ion{He}{2}~304~\AA\ emission \citep{1990ApJ...351..317P}, \cite{2001ApJ...555..435B} found no evidence using SOHO/CDS data (which observed this line in second order). Similarly, \cite{2012ApJ...752...84H} was unable to detect non-thermal \ion{He}{2}~304~\AA\ emission in large $\gamma$-ray events using SDO/EVE observations. Outwith the solar flare context, a similar process results in proton aurorae on Mars, where charge exchange between the solar wind and the Martian exosphere results in enhancements to \lya\ \citep[see the recent observations and discussions of this effect by][]{2019JGRA..12410533H}.

While there is not a track record of successful solar flare detections, it is important that we revisit the Orrall-Zirker effect now that we have more sophisticated numerical tools with which to attack the problem. Performing rigorous modern studies, of which this manuscript is the first, could lead to constraints on low-energy cutoffs of the non-thermal proton distribution, and thus total energetics and acceleration mechanisms. If our modern calculations confirm prior numerical studies that suggest that the OZ effect is, in principle, eminently observable then we must address the null detections. The solar physics community also has a renewed interest in solar Lyman line observations with several new and upcoming missions set to provide \lya\ or \lyb\ data (including Solar Orbiter, Solar-C/EUVST, ASO-S, and the SNIFS sounding rocket). In this study we revisit the predictions of the strength of the non-thermal Lyman line emission over time in flare simulations, and importantly we compare this to the predicted ambient flare emission in the same wavelength range. 

In Section~\ref{sec:radynfp} we describe our proton beam-driven flare models, and in Section~\ref{sec:ozeffect} discuss in detail the processes relevant to the Orrall-Zirker effect. The results of applying our Orrall-Zirker numerical code are shown in Section~\ref{sec:nthmemission}, where we synthesize \lya\ and \lyb\ emission. Our results confirm the earlier findings that non-thermal H$\alpha$ emission is far too weak to be detected \citep{1985ApJ...295..275C,1999ApJ...514..430B}, so we do not discuss that line in detail here. Finally, in Section~\ref{sec:spice} we degrade our predictions to the resolution of the SPICE instrument on board Solar Orbiter.

\section{Numerical Flare Experiments}\label{sec:radynfp}

\begin{figure*}
	\centering 
	{\includegraphics[width = 0.8\textwidth, clip = true, trim = 0.cm 0.cm 0.cm 0.cm]{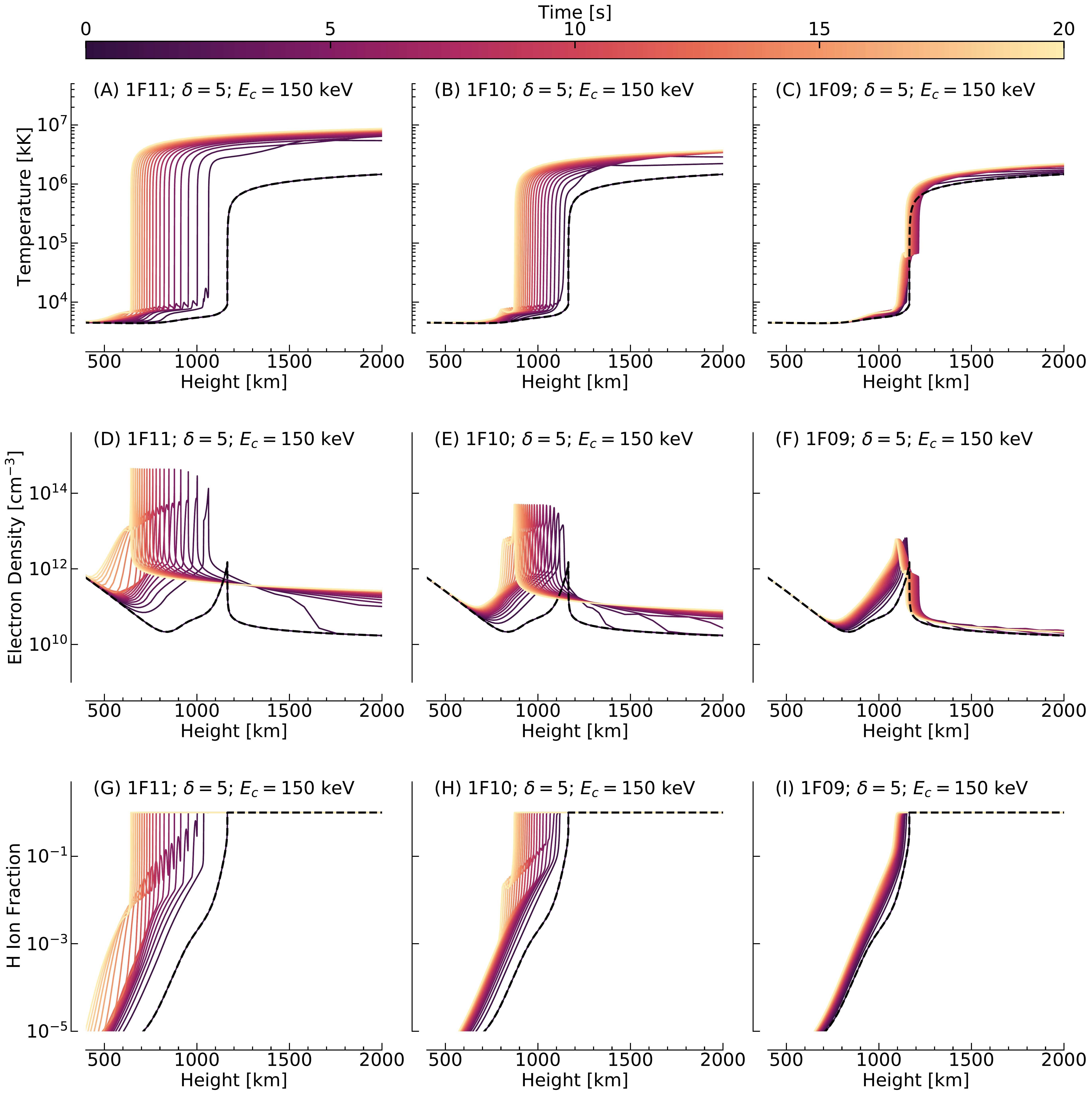}}
	\caption{\textsl{The atmospheric stratification for each simulation in our study. The top row (A-C) shows temperature, middle row (D-F) shows electron density, and bottom row (G-I) shows ionization fraction. Color represents time, where we show lines at $t=1$~s cadence during the heating phase of the flares. The first column shows the $F = 1\times10^{11}$~erg~s$^{-1}$~cm$^{-2}$ simulation, the middle column shows the $F = 1\times10^{10}$~erg~s$^{-1}$~cm$^{-2}$  simulation, and the final column shows the $F = 1\times10^{9}$~erg~s$^{-1}$~cm$^{-2}$  simulation.}}
	\label{fig:protonflares}
\end{figure*}

The field-aligned radiation hydrodynamics (RHD) code \radyn\ \citep{1992ApJ...397L..59C,1995ApJ...440L..29C,1997ApJ...481..500C,1999ApJ...521..906A,2005ApJ...630..573A,2015ApJ...809..104A} is a well established resource that models the solar atmosphere's response to energy injection. The coupled, non-linear equations of hydrodynamics, charge conservation, radiation transfer and non-equilibrium atomic level populations are solved on an adaptive grid (capable of resolving the steep gradients and shocks that form during flares) spanning the sub-photosphere through corona. In the pre-flare atmosphere the $z=0$ point is defined as where the standard optical depth is unity, $\tau_{5000} = 1$. Crucially, a NLTE, non-equilibrium chromosphere is modelled, with feedback between radiative heating and cooling, the hydrodynamics, and on the energy transport mechanisms.  \radyn\ has been used extensively to study many aspects of flares, from details of energy transport, driving of mass flows, and the evolution of the flaring plasma, to the formation of radiation and sources of energy losses \citep[e.g.][]{2016ApJ...827..101K,2019ApJ...885..119K,2019ApJ...883...57K,2020ApJ...900...18K,2021ApJ...912..153K,2022ApJ...931...60A,2018ApJ...856..178P,2019ApJ...879L..17P,2018ApJ...862...59B,2020ApJ...895....6G,2015SoPh..290.3487K,2017ApJ...836...12K,2022ApJ...928..190K,2015ApJ...813..125K,2017A&A...605A.125S}. For a complete description of the base level of the code see \cite{2015ApJ...809..104A}. \radyn\ has been modified somewhat since \cite{2015ApJ...809..104A}, for example to include suppression of thermal conduction \citep{2022ApJ...931...60A}, more accurate Stark broadening of Balmer lines \citep{2022ApJ...928..190K}, and to model the transport of energetic particles via the \fpc\ (Fokker-Planck) code \citep[][]{2020ApJ...902...16A}. Of those we employ the latter in this work.  
For a comprehensive review of flare loop modelling, including \radyn, in the context of observations from the Interaface Region Imaging Spectrograph \citep[IRIS;][]{2014SoPh..289.2733D} see \cite{2022FrASS...960856K,2023FrASS...960862K}.  

When simulating solar or stellar flares driven by non-thermal particles \radyn\ now uses \fpc\ to model the propagation, evolution, and thermalisation of that particle population. \fpc\ is a standalone, open source, code that has been merged with \radyn, that while being similar to the approach used in \cite{2015ApJ...809..104A} offers an upgraded treatment of the solution of the Fokker-Planck kinetic theory, in particular more accurate warm target physics, the inclusion of return currents, and the ability to model particles of arbitrary mass and charge. A non-thermal particle distribution is injected at loop apex, which propagates through the loop, with energy lost via Coulomb collisions transferred to the plasma. This distribution is a power-law defined by the total energy flux $F$ [erg~s$^{-1}$~cm$^{-2}$] above some low-energy cutoff $E_{c}$ [keV], with a spectral index $\delta$. \fpc\ (and hence \radynfp) now has the ability to model a multi-species beam, but here we restrict ourselves to a proton-only distribution to simplify this initial study. Also, the only flare footpoints imaged at MeV $\gamma$-ray energies by RHESSI (from which we infer the presence of non-thermal protons) have an offset relative to the centroid positions of hard X-rays (i.e. the electron population). With the caveat that these are only a few events, such a separation may suggest that different species were accelerated along different flux tubes. In addition to energy losses, non-thermal particle distributions can cause collisional ionization or excitation of ambient species. For non-thermal collisional rates from the ground state of neutral hydrogen impacted by the proton beams we follow the treatment of \cite{1993A&A...274..923H}.

\radynfp\ provides us with the time-dependent NLTE ionization stratification alongside the full spectral distribution of non-thermal protons, and self-consistently models the feedback between them. Armed with these we can make predictions of the production of energetic neutral atoms in a more accurate manner than previous codes. In this paper we focus on studying how detectable the non-thermal Lyman line emission might be for different injected energy fluxes. As mentioned in the Introduction, we do not have a firm grasp of the total energy flux carried by non-thermal protons and so, for the time being, elect to study the impact on the various physical processes of three regimes of flare heating where the energy flux varies over three orders of magnitude, $F = [10^{9}, 10^{10}, 10^{11}]$~erg~s$^{-1}$~cm$^{-2}$. The smallest energy flux could represent the initial stages of energy injection into a flare footpoint before some ramp up to a higher energy flux, or it could represent the total flux injected in some small scale heating event. 

\subsection{Proton Beam-Driven Flares}
Three proton beam-driven flares were simulated using \radynfp, varying the injected flux $F = [10^{9}, 10^{10}, 10^{11}]$~erg~s$^{-1}$~cm$^{-2}$ with fixed $E_{c} = 150$~keV and $\delta = 5$. In this initial study we explore the potential of detection based on energy flux, and a fuller exploration of the parameters defining the proton distributions can be performed at a later date. Hereafter we refer to these experiments as 1F9, 1F10, 1F11, for the $F = [10^{9}, 10^{10}, 10^{11}]$~erg~s$^{-1}$~cm$^{-2}$ models, respectively. Energy was injected at a constant rate for $t_{inj} = 20$~s, with each loop allowed to cool over a further $t_{decay}=80$~s. Return current effects were not included in these models. A return current would appear as a downward propagating electron beam in order to neutralize the current introduced by the non-thermal proton beam. As discussed in \cite{2020ApJ...902...16A}, this could result in energy losses of the proton beam itself, or in Joule heating (operating primarily in the corona or upper transition region). Since return current losses are typically on the order of $10$~keV \citep{2020ApJ...902...16A}, which is a small fraction of the energy of the protons, it is safe to neglect return-current-induced energy losses of the non-thermal protons in our experiments. Any Joule heating of the corona or upper transition region would also likely be small and not impact our conclusions regarding the chromospheric footpoints.

The evolution of the stratification of temperature, electron density, and H ionization fraction is shown in Figure~\ref{fig:protonflares} for each experiment, where color represents time. It is clear that increasing the magnitude of the energy flux carried by the non-thermal protons has more dramatic effect on the atmospheres.

In the 1F10 and 1F11 flares the chromosphere is strongly compressed, and the transition region is pushed deeper in altitude, with a steeper transition region in the 1F11 case. The electron density increases by orders of magnitude, with a narrow region exceeding $n_{e} = [2.5\times10^{13},2.5\times10^{14}]$~cm$^{-3}$ for F10 and 1F11, respectively, with the location of the peak being pushed ever deeper as the chromospheres compress. There is a more extended region of more moderately elevated electron density towards greater depths, of a few $\times10^{12-13}$ cm$^{-3}$. Due to both the increased temperatures and the presence of non-thermal collisional ionisations the ionization fraction rapidly increases, particularly so for the 1F11 simulation. Even by $t=1$~s there is a `wall' of ionization present in both of those simulations, where the ionization climbs from $\chi_{H} \sim 0.5$ to $\chi_{H} \sim 1$ in only a few meters, which occurs $z\sim100$~km deeper for the 1F11 simulation compared to the 1F10. As time goes on the location of this ionization front pushes increasingly deeper, such that it is located at $0.71$~Mm and $0.96$~Mm in the 1F11 and 1F10 simulations, respectively, by the end of the heating phase. The gradient from $\chi_{H} \sim 0.1$ to $\chi_{H} \sim 1$ is steeper for the 1F11 simulation, so that for $t \gtrsim10$~s seconds this change in ionization fraction occurs over a span of only $\Delta Z \sim [10-30]$~km, compared to $\Delta z\sim[50-130]$~km for the 1F10 simulation. 

Compared to those simulations, the dynamics in the 1F9 simulation is much more modest. The flare transition region does not really change location substantially, eventually residing $z\sim100$~km higher in altitude than it did before the flare. However, it is more extended, with $40$~kK$<T<1$~MK over a span of $\Delta z \sim130$~km. Given the modest temperature rise and smaller rate of non-thermal collisions, electron densities were commensurately smaller than the other simulations, peaking at only $n_{e} \sim 4.5\times10^{12}$~cm$^{-3}$, and with a tail of only a few $\times10^{11}$~cm$^{-3}$ through the lower chromosphere. Consequently, the ionization stratification is less sharp, with the `wall' feature being softened to occur over a span of $\Delta z\sim10-20$~km, with a similarly more gentle gradient down to $\chi_{H} \sim 0.1$.

Since we inject non-thermal protons with energies $\sim150$~keV, their speed is intermediate between the thermal electron and ion speeds for plasma with temperature greater than $\sim1$ MK. This is the warm-target regime described in \citet{1986ApJ...309..409T}. Warm-target collisions are much less effective at slowing non-thermal particles than cold target collisions \citep{1986ApJ...309..409T, 2020ApJ...902...16A}. Therefore, the protons are able to transport through the corona relatively unimpeded until they reach the transition region where they encounter cooler plasma and are stopped, thereby heating the transition region and upper chromosphere, evaporating more plasma into the corona and forcing the transition region increasingly deeper. 

We briefly note that the momentum of a non-thermal proton is larger than that of a non-thermal electron, which should be accounted for. With \fpc\ we include the particle momentum self-consistently \citep{1986ApJ...309..409T, 2020ApJ...902...16A}. Comparing the acceleration induced to that of mass flows resulting from plasma heating indicates that plasma heating dominates, and that it is plasma heating that ultimately results in pushing the transition region to greater depths.

\subsection{Ambient Lyman Emission in Flares}
Since \radyn\ truncates the Lyman line profiles at 10 Doppler widths in order to mimic the effects of partial frequency redistribution (PRD) for the purposes of energetics (to avoid overestimating radiative losses in the wings), we use \rhpar\ \citep{2001ApJ...557..389U,2015A&A...574A...3P} with our \radyn\ flare atmospheres as input to synthesize the Lyman lines and the region around them. 

\rhpar\ solves the radiation transport equation and atomic level populations for a desired species, given an atmosphere (the depth scale, temperature, electron density, gas velocity and hydrogen populations). We have modified \rhpar\ to keep the hydrogen and \ion{Ca}{2} populations from \radyn\ fixed so that non-equilibrium effects and non-thermal collisions are accounted for, while still solving the statistical equilibrium atomic level populations of other species. Full NLTE radiation transport was solved for the following species \citep[with PRD for certain transitions as appropriate, including \lya\ and \lyb, using the hybrid scheme of][]{2012A&A...543A.109L}: \ion{H}{1}, \ion{C}{1}+\ion{C}{2}, \ion{O}{1}, \ion{Si}{1}+\ion{Si}{2}, \ion{Ca}{2}, \ion{He}{1}+\ion{He}{2} \& \ion{Mg}{2}. Other species were included as sources of background opacity. The opacity `fudge' factors of \cite{1992A&A...265..237B}, that are typically included in \rh\ and \rhpar\ by default to mimic the UV line haze, were not used as these are rather specific to a VAL-C type atmosphere \citep{1981ApJS...45..635V} and likely not appropriate for more complex flare-like atmospheres. Instead we include opacity from the myriad of bound-bound lines not solved in detail via the Kurucz linelists\footnote{\url{http://kurucz.harvard.edu/linelists.html}}, where hundreds of thousands of lines were included in the range $\lambda = [20-8000]$~\AA. The intensities of those lines surrounding the \lya\ and \lyb\ lines were estimated in LTE (though with two-level scattering) with a wavelength sampling of at least 0.1~\AA\ (finer sampling was present where there was overlap with the model atoms included in the solution). Microturbulent broadening of $v_{\mathrm{nthm}} = 8$~km~s$^{-1}$ was included, and lines were Doppler shifted as appropriate. 

In addition to the \rhpar\ solution we synthesize the \ion{O}{6} doublet at $\lambda = 1031.9$~\AA\ and $\lambda = 1037.6$~\AA, which are two strong optically thin lines that appear in the vicinity of the peak of the non-thermal emission in the red wing of \lyb. The contribution functions, $G(n_{e}, T)$, which encapsulate various atomic processes that populate the relevant levels, were computed for each line from the CHIANTI atomic database \citep[V8.07;][]{1997A&AS..125..149D,2015A&A...582A..56D} for various values of temperature and electron density. These were tabulated with grid spacings of $\delta \log T$ [K]$ = 0.05$ and $\delta \log n_{e} $ [cm$^{-3}$] $=0.5$, and at each grid cell in the \radyn\ atmospheres we interpolate $G(n_{e}, T)$ to the local values of temperature and electron density. The intensity emitted within each cell, $I_{\lambda,z}$ is then:

\begin{equation}
    I_{\lambda,z} = A_{O}G(n_{e},T)n_{e(z)}n_{H}(z)\delta z,
\end{equation}

\noindent where $A_{O}$ is the elemental abundance relative to hydrogen\footnote{We used the abundance value from \cite{2012ApJ...755...33S}, $A_{O,log}=8.61$, defined on the usual logarithmic scale, where $A_{H,log}=12$ (the abundance is then $A_{O} = 10^{A_{O,log} - A_{H,log}} = 4.1\times10^{-4}$). There is much debate over whether to use the coronal or photospheric abundance values during flares, with the intensity varying by some factor based on the choice. Since we are synthesising these lines for the purpose of determining if there is a region between them that permits identification of non-thermal emission, it will not affect our conclusions if their intensity is over- or under-estimated by some relatively small factor.}, $\delta z$ is the length of the cell, and $n_{H}$ is the hydrogen density. The spectral lines were thermally broadened according to the local temperature, and Doppler-shifted as appropriate. The total emergent intensity was then $I_{\lambda} = \Sigma_z I_{\lambda,z}$. Note that we only include these lines to help assess whether the non-thermal emission can be identified at wavelengths surrounding them, and that a more rigorous sythesis would include resonant scattering of chromospheric radiation.

From \radynfp+\rhpar\ we obtained the ambient Lyman line plus surrounding emission as a function of time in our flares, on top of which the non-thermal OZ emission would appear.

\section{Orrall-Zirker Effect}\label{sec:ozeffect}
\subsection{Overview of the Problem}
As discussed in Section~\ref{sec:intro} and sketched in Figure~\ref{fig:OZSchematic}, the non-thermal protons streaming through the solar atmosphere may undergo charge exchange with ambient hydrogen once they reach the chromosphere. These charge exchange interactions may occur both to the ground or excited states. Subsequent collisions can also excite a non-thermal hydrogen atom into an excited state, or cause transitions between bound excited states. 

Following the discussions laid out in \cite{1976ApJ...208..618O} and \cite{1985ApJ...295..275C}, we assume that the non-thermal hydrogen atoms created through charge exchange share the same velocity (i.e. energy) as their parent non-thermal proton, and that the de-excitation of a non-thermal hydrogen atom occurs over a very short distance. This latter point means that the creation and destruction of non-thermal hydrogen is an inherently local process, such that advection terms can be ignored along with time-dependent terms  \citep[see the demonstration that de-excitation occurs much faster than deceleration in][]{1976ApJ...208..618O}. These two assumptions allow us to write the population equation of non-thermal hydrogen levels in the form of a statistical equilibrium equation, with the total number of non-thermal particles (protons plus hydrogen atoms) within some energy range $[E,E+\delta E]$,  located within some height range $[z, z+\delta z]$, conserved. This also means that we can assume any emitted photon is red-shifted by an amount equal to the velocity of the original non-thermal proton that underwent charge exchange. 

The number density of a non-thermal hydrogen atom, with energy $E$ at height $z$, is $\eta_{j}(E,z)$ [cm$^{-3}$], where $j = (1, ..., m)$ is the atomic level. Similarly, the non-thermal proton density is $\eta_{p}(E,z)$ [cm$^{-3}$]. Here we follow the convention in \cite{1985ApJ...295..275C} that $\eta$ symbols refer to number densities in $(E,z)$ space as opposed to the thermal population with number densities of particles defined solely on $(z)$ space, for example the ambient electron density $n_{e}(z)$ or ambient hydrogen densities $n_{j}(z)$.  

Assuming statistical equilibrium for a total creation term $C$ (e.g. charge exchange, collisional excitation, etc.,) and total destruction term $D$ (e.g. spontaneous radiative de-excitation, collisional de-excitation, etc.,):

\begin{equation}
\sum_{j \ne i}^{m} \eta_{j} C_{ji} - \eta_{i}D_{i} = 0
\end{equation}

Collating creation and destruction terms (detailed below) to a total rates matrix $P$, and including $j = (1, ..., m)$ then:

\begin{equation}\label{eq:nonthermpops}
P\cdot N = X,
\end{equation}
 
\begin{equation*}
\begin{pmatrix} P_{11} & P_{12} & P_{13} & P_{1p} \\ P_{21} & P_{22} & P_{23} & P_{2p} \\ P_{31} & P_{31}  & P_{32} & P_{33} \\ P_{p1} & P_{p2} & P_{p3} & P_{pp}\end{pmatrix} 
\begin{pmatrix} \eta_{1} \\ \eta_{2} \\ \eta_{3} \\ \eta_{p} \end{pmatrix} 
 = 
 \begin{pmatrix} 0 \\ 0 \\ 0 \\ 0 \end{pmatrix} 
\end{equation*}

\noindent where we dropped the $(E,z)$ dependence for clarity.  To solve this system we replace one of the equations with the particle conservation equation so that the total number of non-thermal particles at $(E,z)$ is conserved.  We choose the final equation so that $P_{p1} = P_{p2} = P_{p3} = P_{pp} = 1$ and $X_{p} = \eta_{p,FP}$, for $\eta_{p,FP}$ the non-thermal proton distribution number density from \radynfp. 

Once we know the non-thermal populations $\eta$, the photon emissivity, $\Phi(\Delta \lambda, z)$, of transition $j\rightarrow i$, at some $\Delta \lambda$~\AA\ from line center, at height $z$ in the atmosphere, in photons~s$^{-1}$~cm$^{-3}$~sr$^{-1}$~\AA$^{-1}$ is given by 

\begin{equation}
\Phi(\Delta \lambda, z) = \frac{\sqrt{2m_{p}c^{2}}}{4\pi \lambda_{ji}}~E^{1/2}~\eta_{j}(E,z)A_{ji},
\end{equation}

\noindent where $A_{ji}$ is the Einstein coefficient for spontaneous emission and $m_{p}$ is the proton mass (or particle mass in the more general sense). In the above we have used the fact that $\mathrm{d} (\Delta \lambda) = \lambda_{ji} \mathrm{d}\nu/c =  \lambda_{ji} \mathrm{d}E/(c\sqrt{2m_{p}E})$.  This becomes an emissivity in terms of energy, $\psi_{ji}(\Delta \lambda, z)$ erg~s$^{-1}$~cm$^{-3}$~sr$^{-1}$~\AA$^{-1}$, via the conversion,

\begin{equation}
	\psi_{ji}(\Delta \lambda, z) = \frac{hc}{\lambda_{ji}} \Phi_{ji}(\Delta \lambda, z).
\end{equation}

\noindent the emergent intensity is then simply,

\begin{equation}
	I_{\mathrm{nthm},ji}(\Delta \lambda) = \int_{z_1}^{z_2}\psi_{ji}(\Delta \lambda, z)\mathrm{d}z, 
\end{equation}

\noindent where $z_{1}$ an $z_{2}$ are two pre-selected height ranges (e.g. $z_{1} = 0$ and $z_{2} = \infty$ if we sum over the full atmosphere). If we assume the chromosphere is optically thin a few angstroms from the Lyman line cores then this non-thermal intensity can be added as straight sum to the emergent intensity from the ambient flaring emission. Note also the assumption here that the protons are streaming almost vertically near disk center. In the event that the flare is located with increasing heliocentric angle towards the solar limb then the $\Delta \lambda$ would be somewhat different due to the dependence on the non-thermal particle velocity, $v$. Despite the ideal nature of the setup we are presently investigating, including angular effects would not be expected to markedly reduce the peak intensity of the emission.

\subsection{Atomic Processes and Cross-Sections}
\subsubsection{Populating non-thermal Hydrogen Excited Levels}
Here we lay out the relevant atomic processes that populate a three level non-thermal hydrogen atom, allowing us to then synthesize \lya\ and \lyb\ emission. These processes are: charge exchange (both to ground and excited levels); collisional excitation of bound levels by ambient particles; collisional ionization by ambient particles; collisional de-excitation by ambient particles; spontaneous radiative de-excitation of an excited level; stimulated radiative de-excitation of an excited level; radiative recombination of $\eta_{p} \rightarrow \eta_{j}$; three-body recombination of $\eta_{p} \rightarrow \eta_{j}$. 

\cite{1985ApJ...295..275C} looked in detail at these processes, noting that charge exchange predominantly occurred to the ground levels, and that collisional excitation of $\eta_{1}\rightarrow \eta_{2},\eta_{3}$ was an important pathway to creating excited non-thermal hydrogen. Charge exchange was orders of magnitude more efficient than both three-body and spontaneous radiative recombination so we ignore those processes going forward. Spontaneous radiative de-excitation from the excited levels was far more rapid than any other destruction process so we additionally ignore collisional de-excitation. Radiative excitation of $\eta_{1}$ is also safely neglected. So, we are left with charge exchange, collisional excitation and ionisation, and spontaneous radiative de-excitation as the important atomic processes to include in our model.

In the expressions that follow, terms such as $C_{ij}$ refer to rates of collisional processes from level $i \rightarrow j$, in s$^{-1}$. For our three level model hydrogen atom $j = (1, 2, 3)$, and  $p$ refers to a non-thermal proton. Charge exchange rates in s$^{-1}$ are labelled $CX_{pj}$, where the electron is captured to the $j$-th level. Radiative terms are listed in terms of the Einstein coefficients for spontaneous emission rates $A_{ji}$ in s$^{-1}$. Collisional rates are individually comprised of interactions between the beam particle, with velocity $v$ in cm~s$^{-1}$, and a target of ambient protons (density $n_{p}$), hydrogen atoms (density $n_{H}$) and electrons (density $n_{e}$). Atomic cross sections are defined as $Q_{ij}^{P,E,H,CX}$ cm$^{2}$ where the index refers to a collision with a proton, electron, or neutral hydrogen respectively, or that a charge exchange interaction takes place.

The ground state of non-thermal hydrogen $\eta_{1}$ is populated by charge exchange, and de-populated by collisional excitation or ionisation;

\begin{equation}
\eta_{1} \rightarrow \eta_{p}CX_{p1} - \eta_{1}(C_{12} + C_{13} + C_{1p}) = 0,
\end{equation}

\noindent where in the above 
\begin{equation}
\begin{split}
CX_{p1} = n_{H}Q_{p1}^{CX}v\\
C_{12} = (n_{p}Q_{12}^{P} + n_{H}Q_{12}^{H} + n_{e}Q_{12}^{E})v\\
C_{13} = (n_{p}Q_{13}^{P} + n_{H}Q_{13}^{H} + n_{e}Q_{13}^{E})v\\
C_{1p} = (n_{p}Q_{1p}^{P} + n_{H}Q_{1p}^{H} + n_{e}Q_{1p}^{E})v
\end{split}
\end{equation}

The first excited state of non-thermal hydrogen $\eta_{2}$ is populated by direct charge exchange and by collisional excitation from $\eta_{1}$, and de-populated by collisional excitation or ionization plus spontaneous emission;

\begin{equation}
\eta_{2} \rightarrow \eta_{p}CX_{p2} + \eta_{1}C_{12} - \eta_{2}(C_{23} + C_{2p} + A_{21}) = 0,
\end{equation}

\noindent where in the above 
\begin{equation}
\begin{split}
CX_{p2} = n_{H}Q_{p2}^{CX}v\\
C_{23} = (n_{p}Q_{23}^{P} + n_{H}Q_{23}^{H} + n_{e}Q_{23}^{E})v\\
C_{2p} = (n_{p}Q_{2p}^{P} + n_{H}Q_{2p}^{H} + n_{e}Q_{2p}^{E})v
\end{split}
\end{equation}

The second excited state of non-thermal hydrogen $\eta_{3}$ is populated by direct charge exchange and by collisional excitation from $\eta_{1}$ and $\eta_{2}$, and de-populated by collisional ionization plus spontaneous emission;

\begin{equation}
\eta_{3} \rightarrow \eta_{p}CX_{p3} + \eta_{1}C_{13} + \eta_{2}C_{23} - \eta_{3}(C_{3p} + A_{31} + A_{32}) = 0,
\end{equation}

\noindent where in the above 
\begin{equation}
\begin{split}
CX_{p3} = n_{H}Q_{p3}^{CX}v\\
C_{3p} = (n_{p}Q_{3p}^{P} + n_{H}Q_{3p}^{H} + n_{e}Q_{3p}^{E})v
\end{split}
\end{equation}

\subsubsection{Cross-Sections Used in This Study}\label{sec:csec}
\begin{figure}
	\centering 
	\vbox{
	\subfloat{\includegraphics[width = .5\textwidth, clip = true, trim = 0.cm 0.cm 0.cm 0.cm]{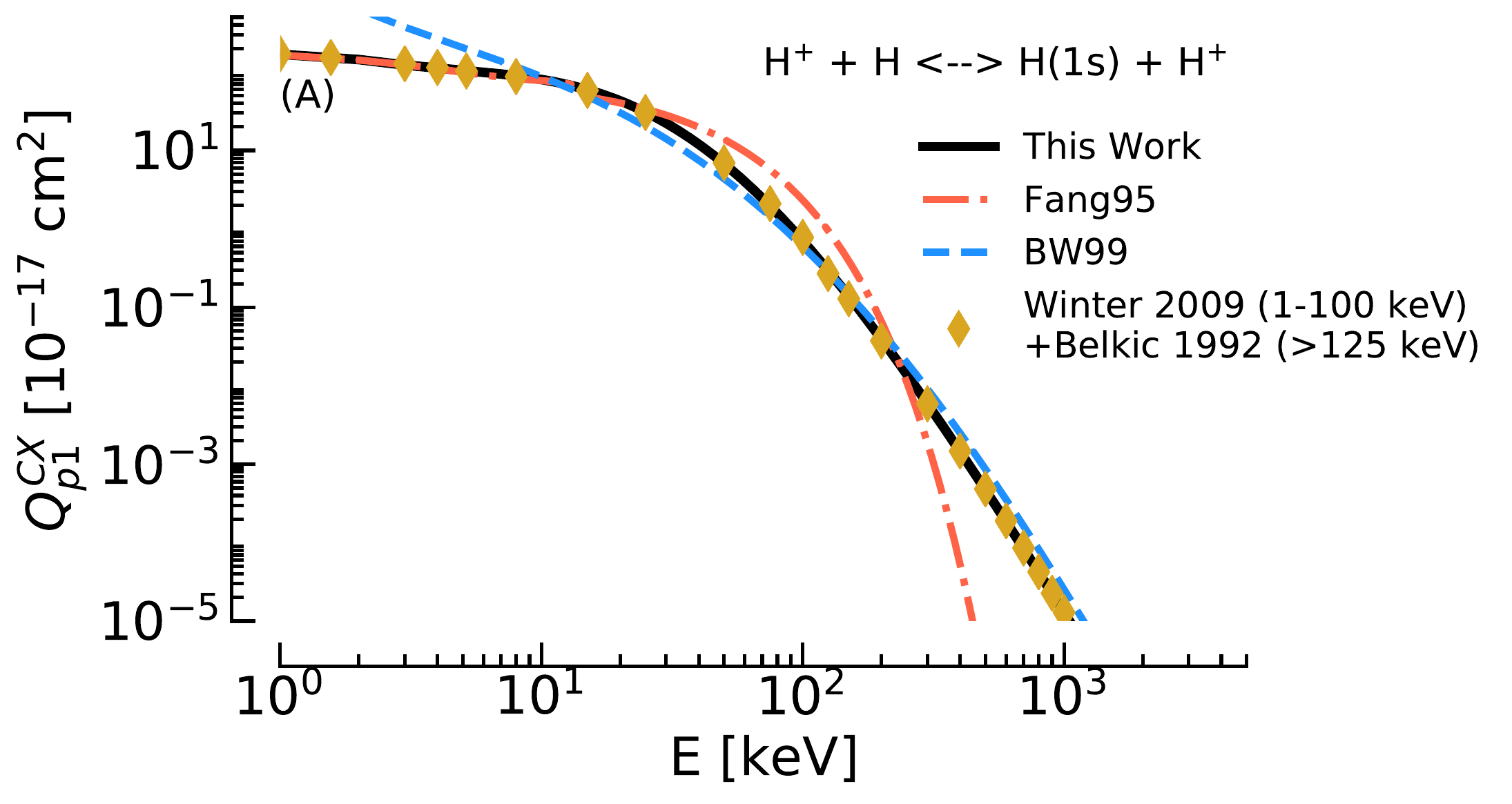}}	
	}
	\vbox{
	\subfloat{\includegraphics[width = .5\textwidth, clip = true, trim = 0.cm 0.cm 0.cm 0.cm]{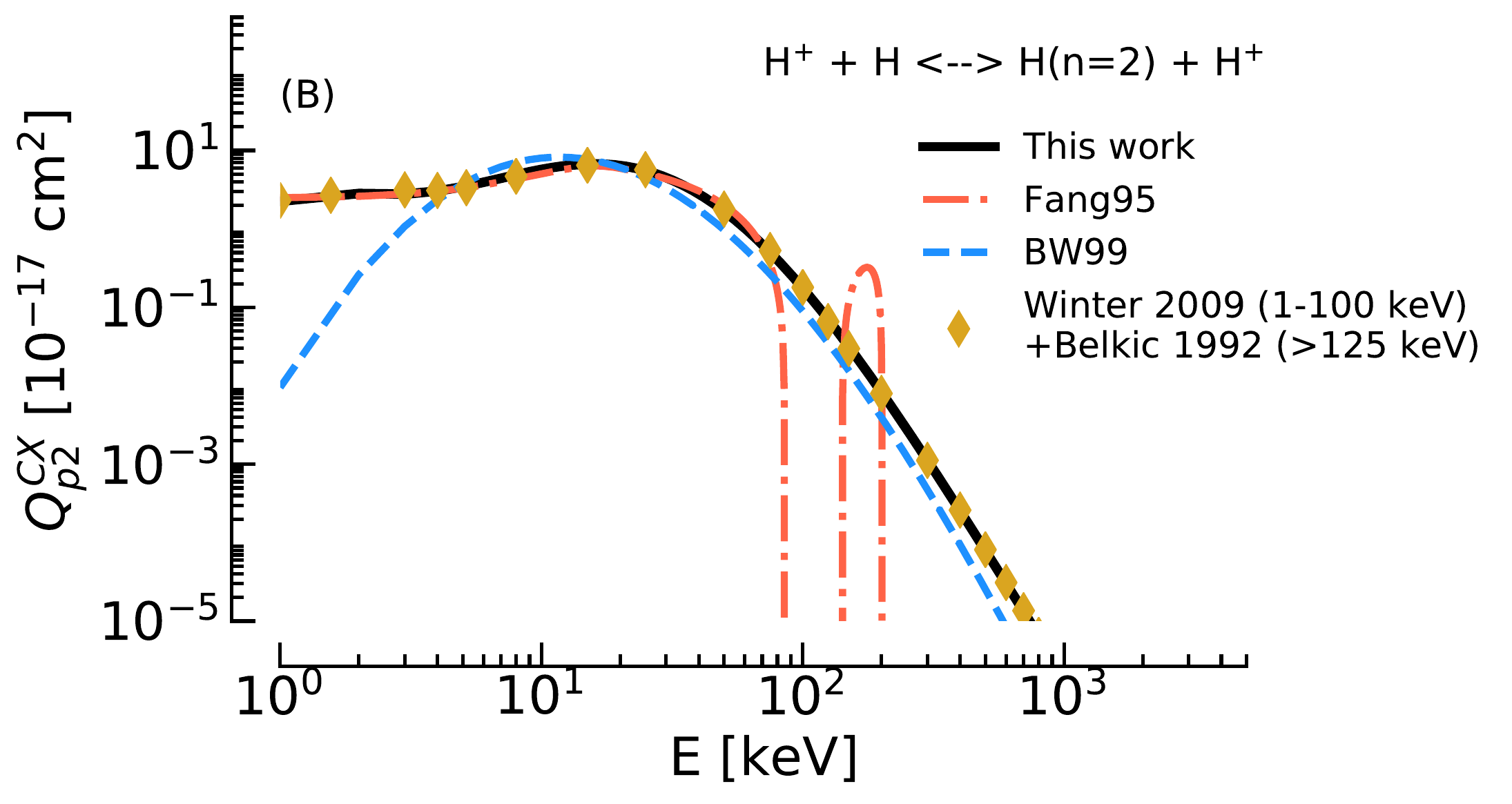}}	
	}
	\vbox{
	\subfloat{\includegraphics[width = .5\textwidth, clip = true, trim = 0.cm 0.cm 0.cm 0.cm]{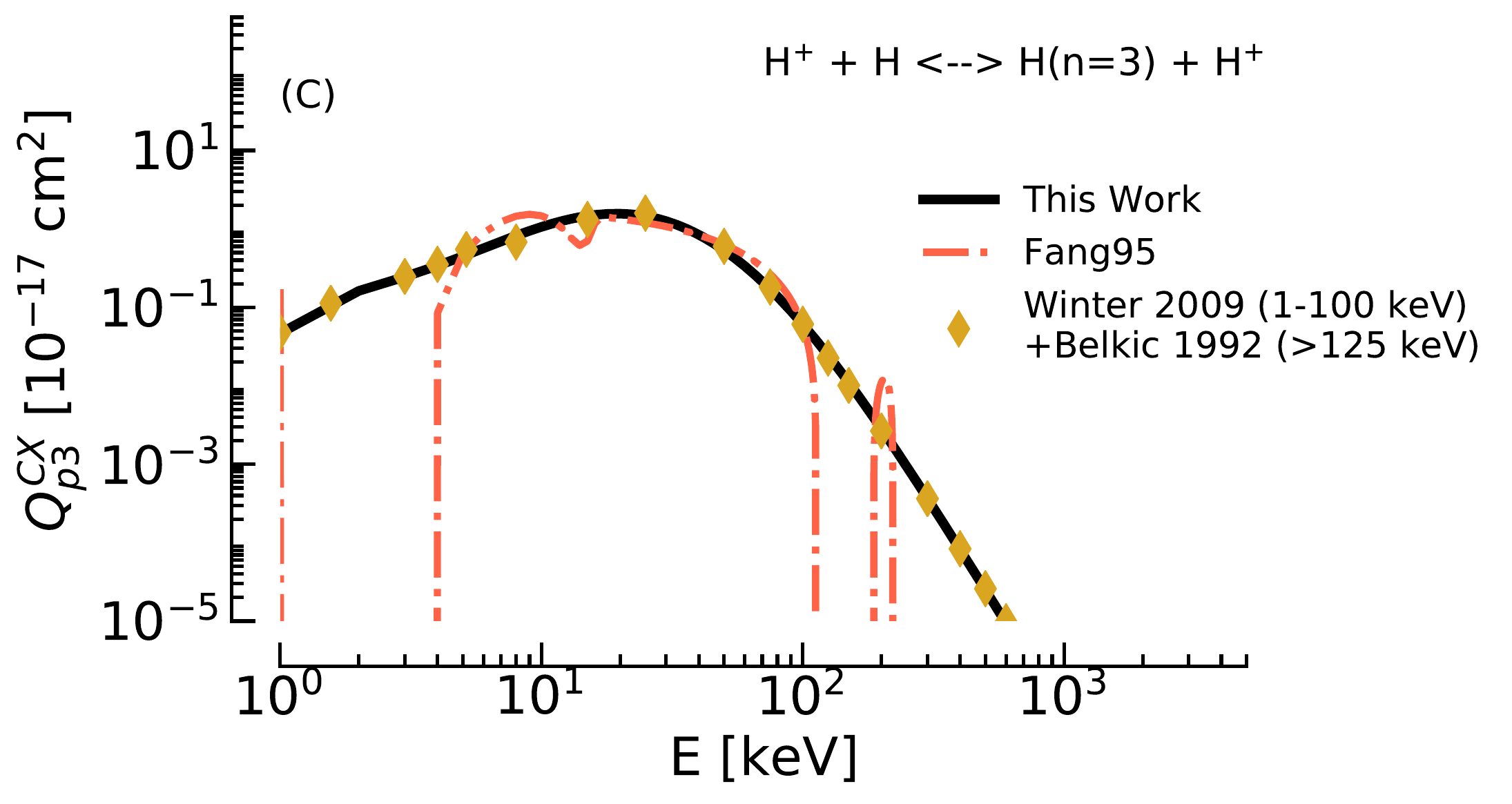}}	
	}
	\caption{\textsl{Charge exchange cross-sections used in this study, where the black line shows the fit to data (diamonds). The functions used by \cite{1995A&A...297..854F} are shown as dot-dashed red lines, and by \cite{1999ApJ...514..430B} as blue dashed lines. Panel (A) is charge exchange to the ground state, panel (B) is to the first excited state (n=2), and panel (C) is to the second excited state (n=3).}}
	\label{fig:csec_cx}
\end{figure}

\begin{figure*}
	\centering 
	\vbox{
	\hbox{
	\subfloat{\includegraphics[width = .5\textwidth, clip = true, trim = 0.cm 0.cm 0.cm 0.cm]{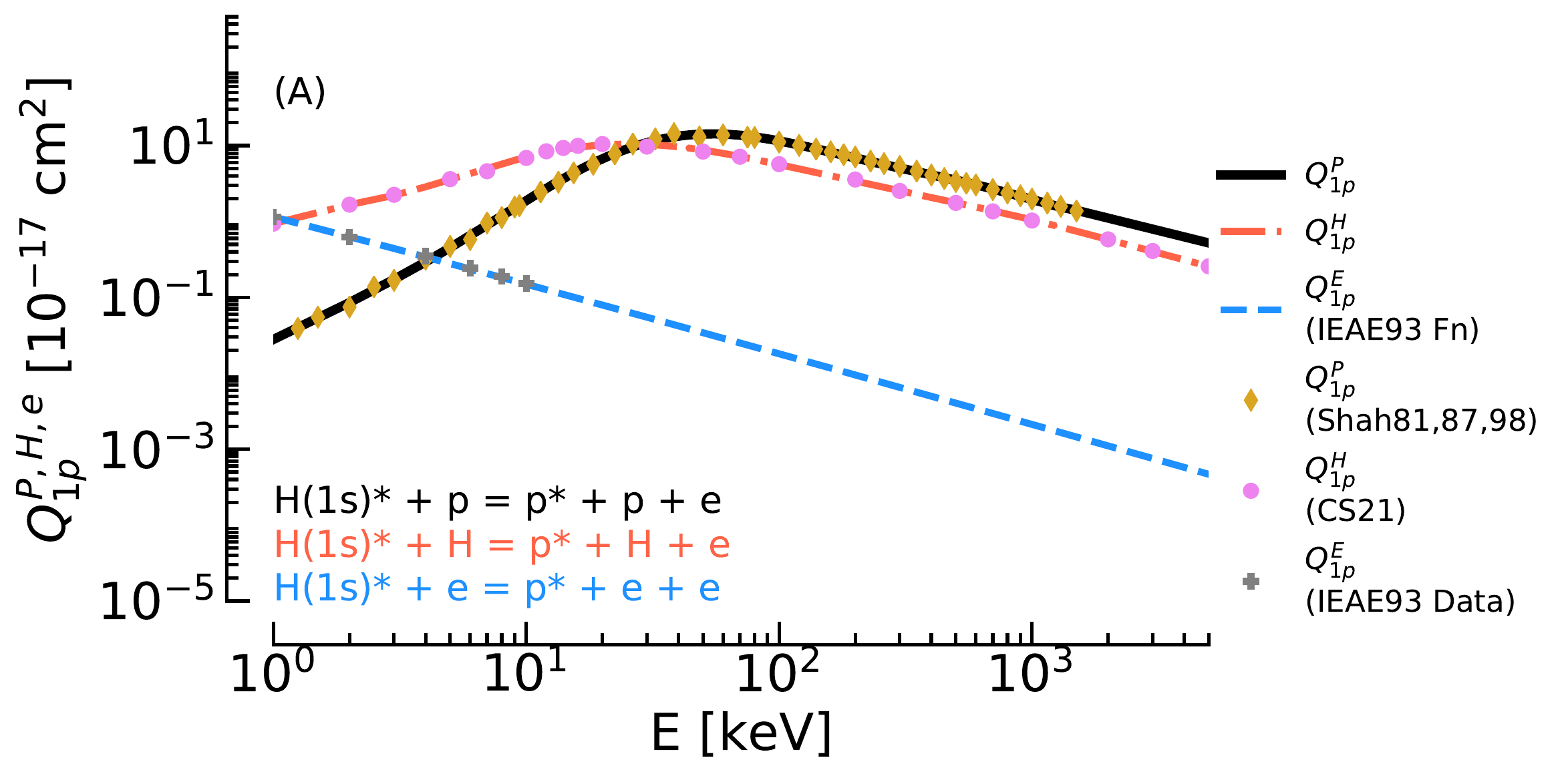}}	
	\subfloat{\includegraphics[width = .5\textwidth, clip = true, trim = 0.cm 0.cm 0.cm 0.cm]{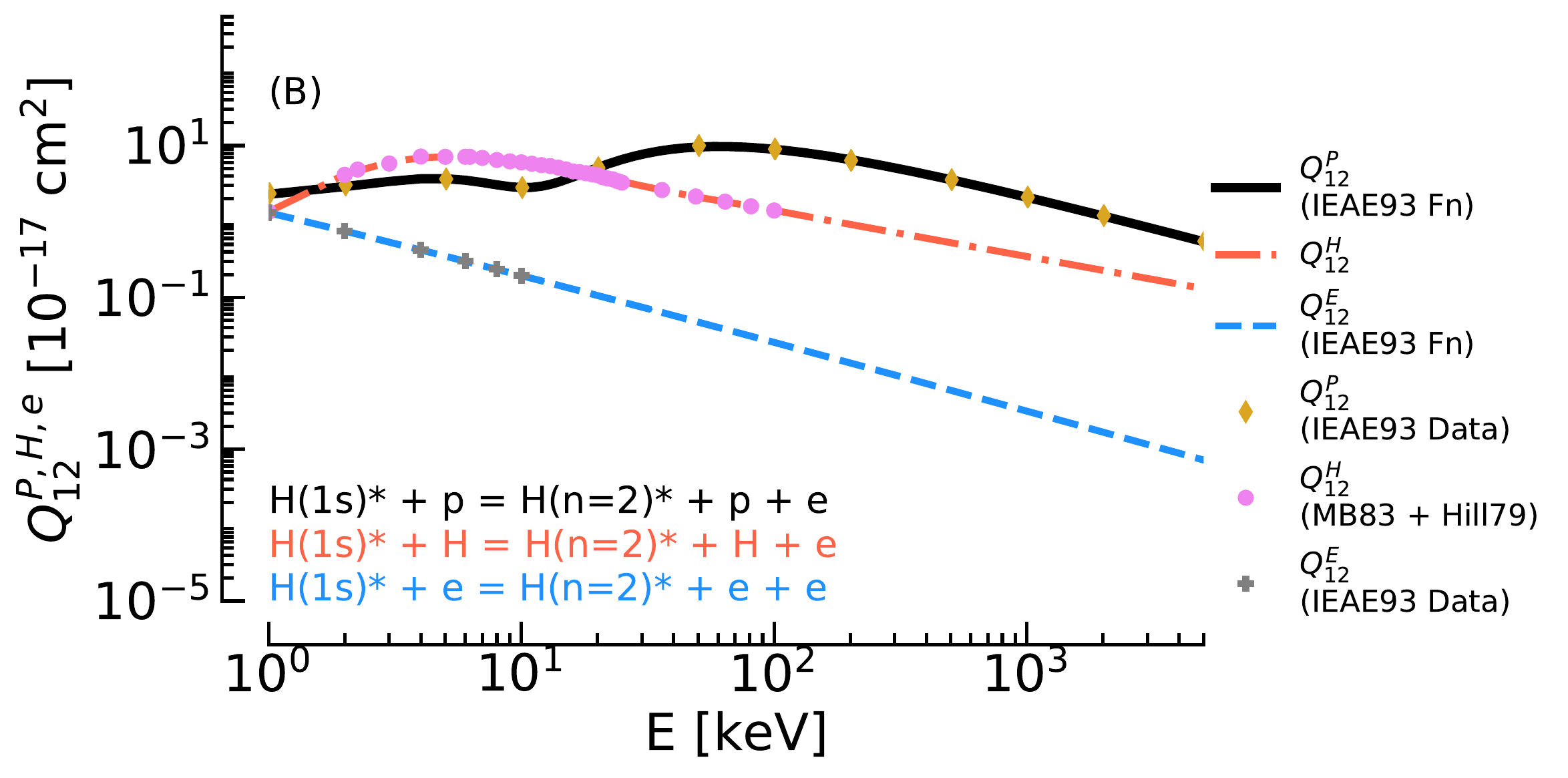}}	
	}
	}
	\vbox{
	\hbox{
	\subfloat{\includegraphics[width = .5\textwidth, clip = true, trim = 0.cm 0.cm 0.cm 0.cm]{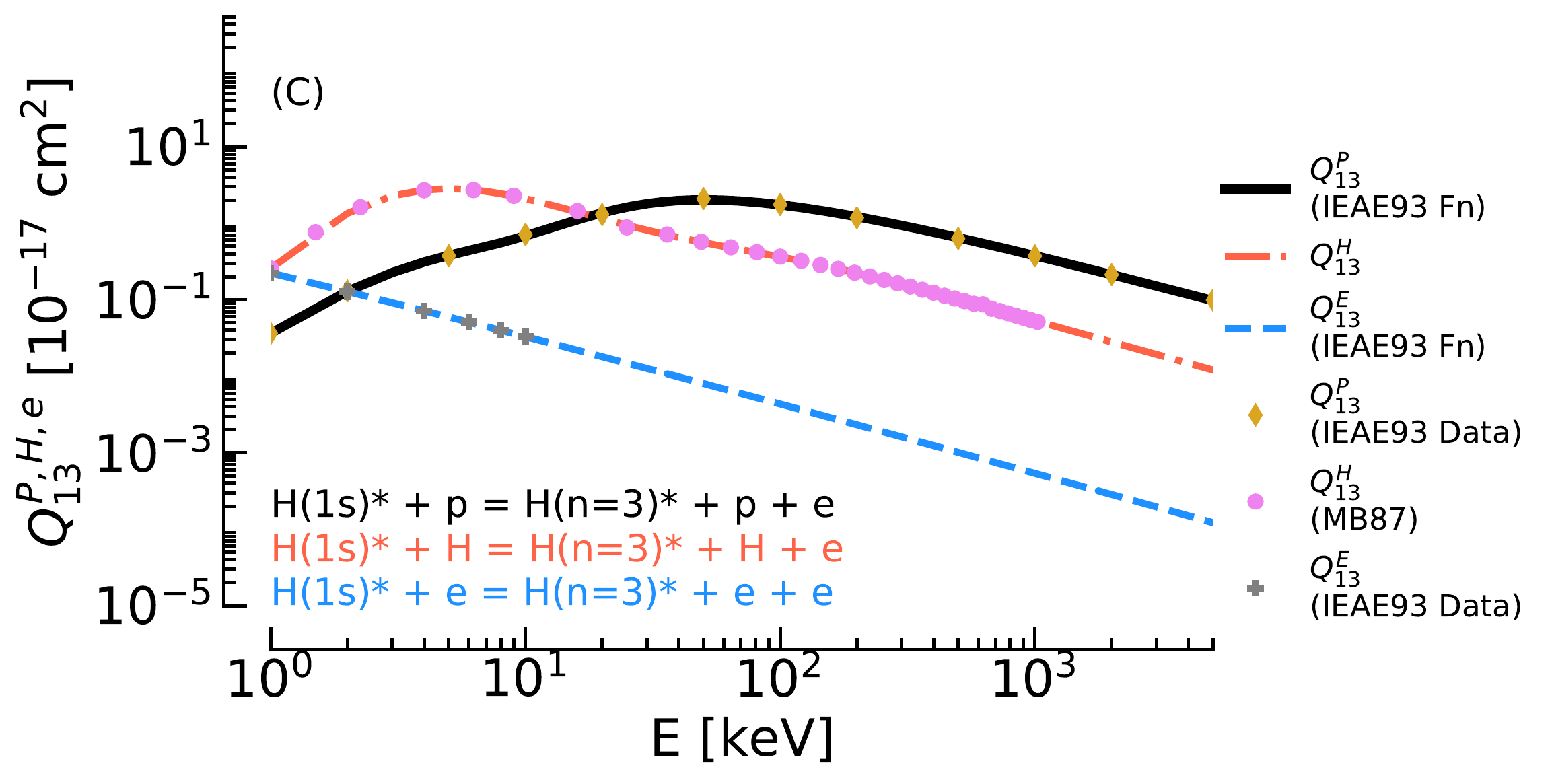}}	
	\subfloat{\includegraphics[width = .5\textwidth, clip = true, trim = 0.cm 0.cm 0.cm 0.cm]{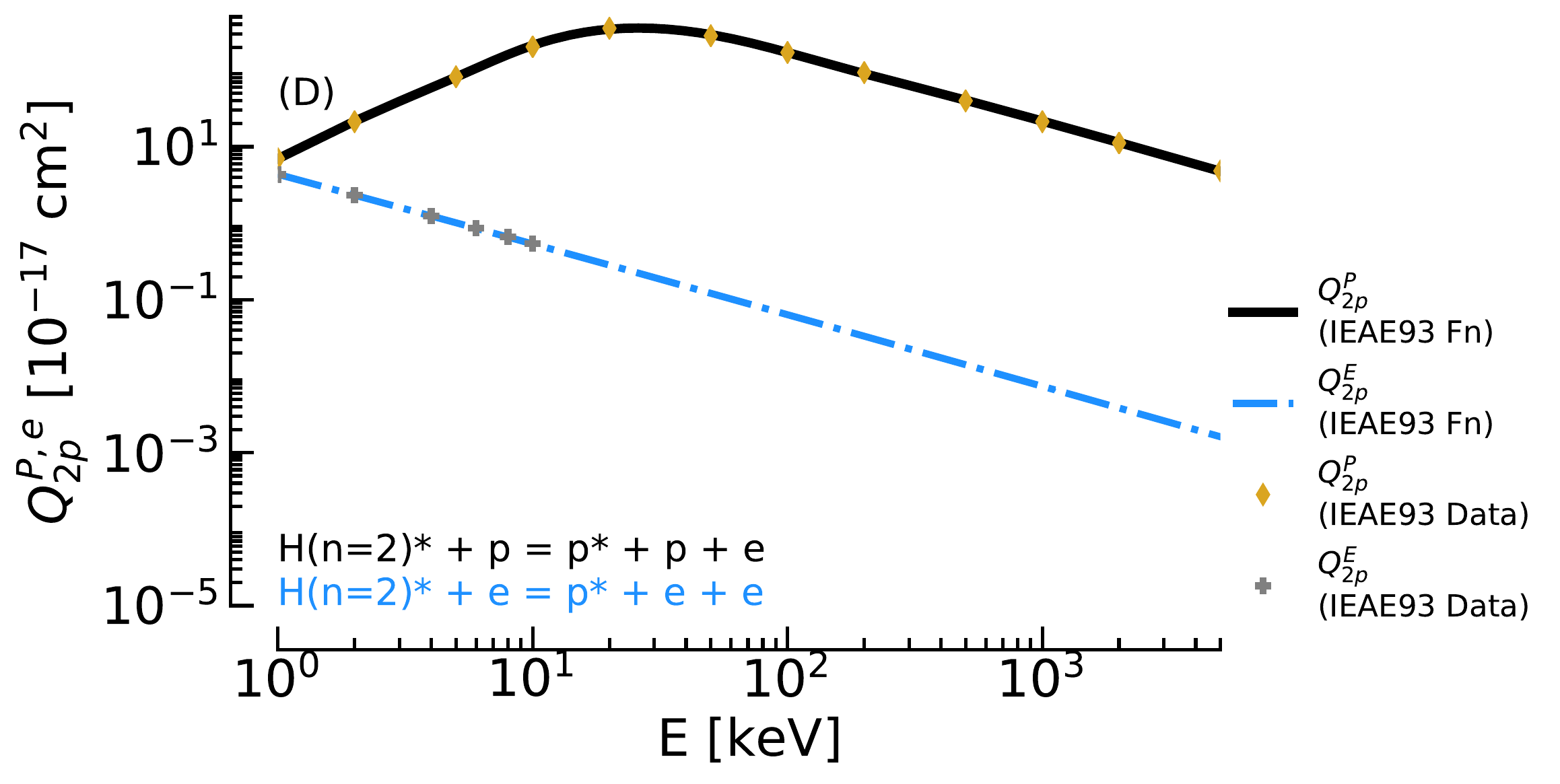}}	
	}
	}
	\vbox{
	\hbox{
	\subfloat{\includegraphics[width = .5\textwidth, clip = true, trim = 0.cm 0.cm 0.cm 0.cm]{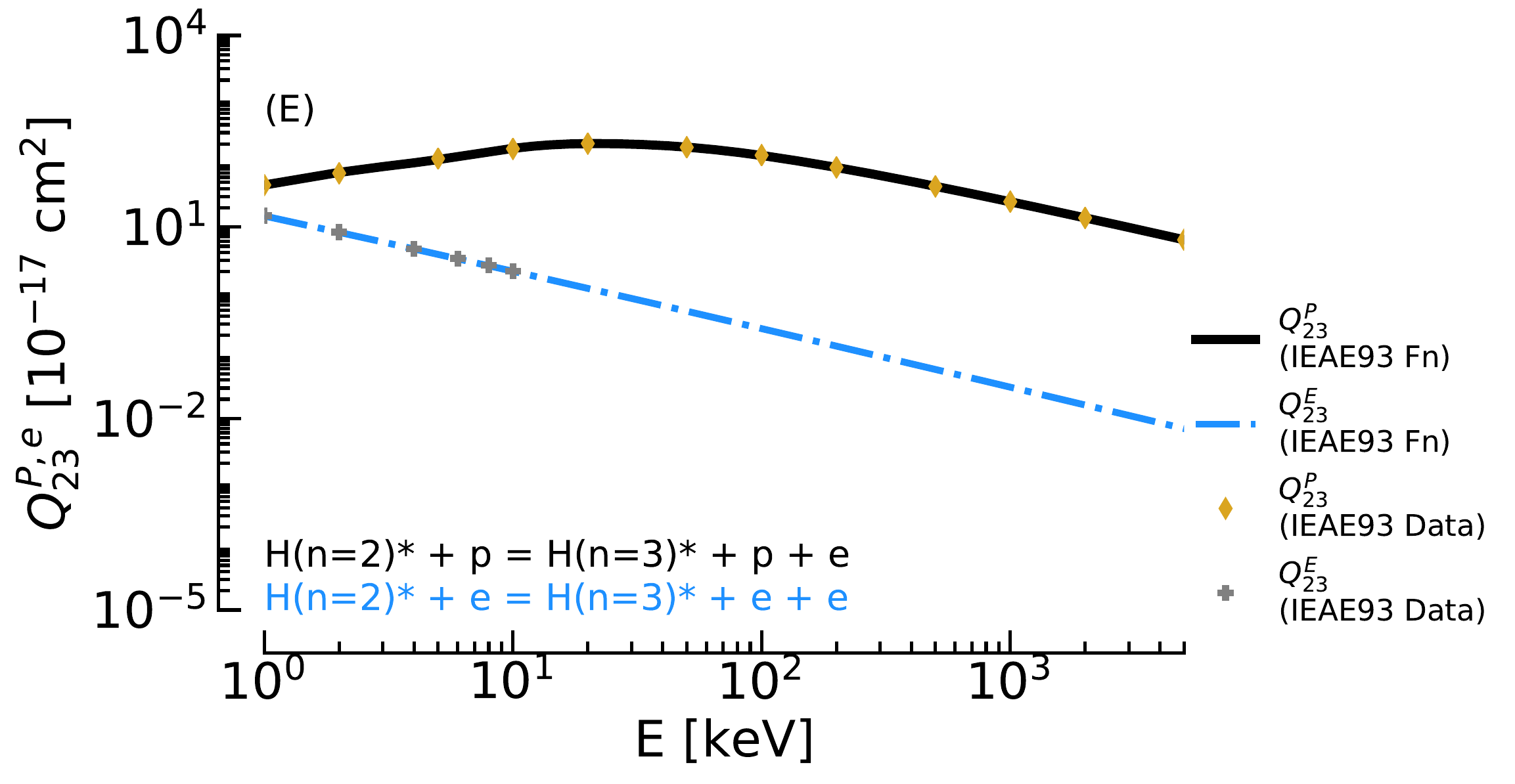}}	
	\subfloat{\includegraphics[width = .5\textwidth, clip = true, trim = 0.cm 0.cm 0.cm 0.cm]{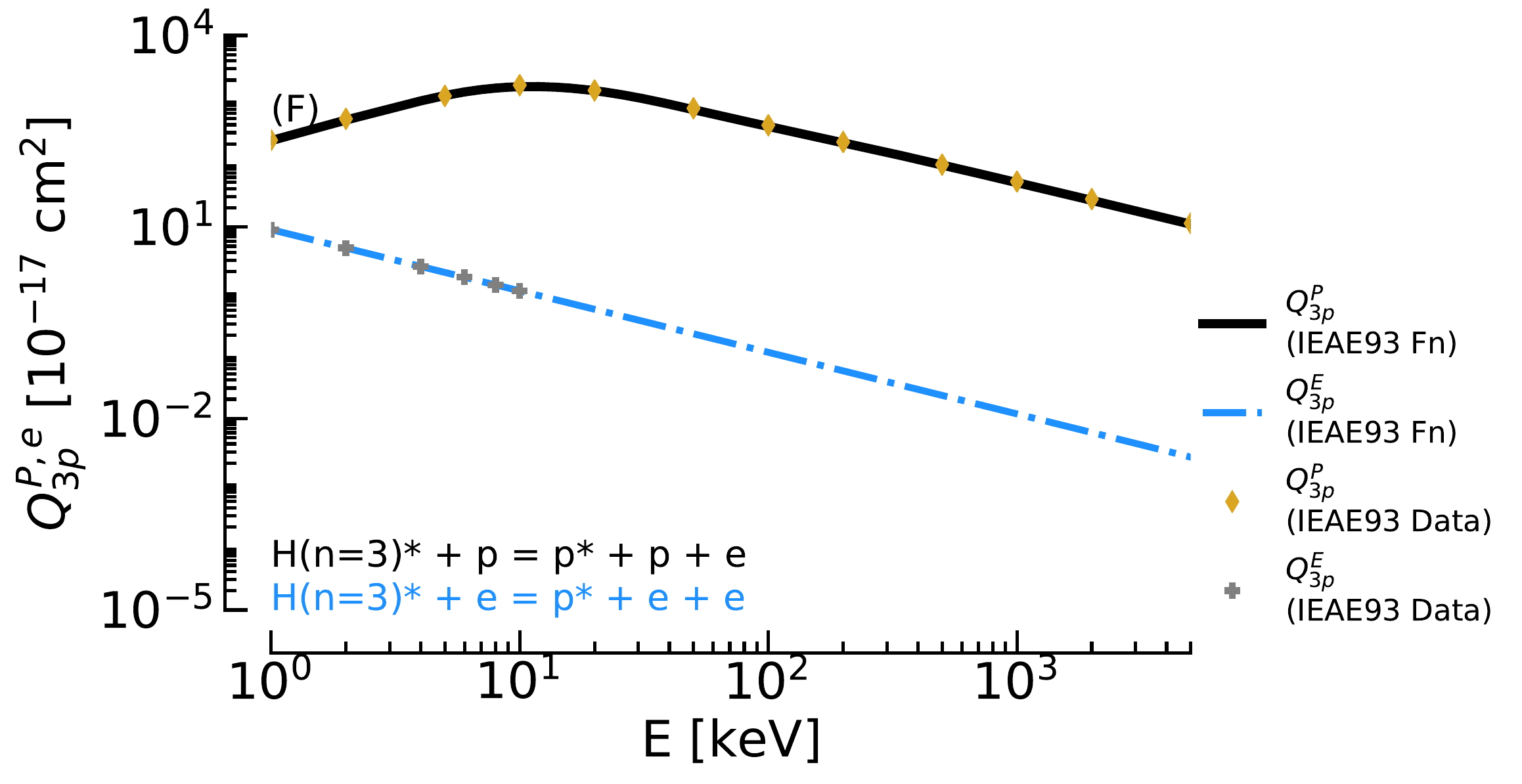}}	
	}
	}
	\caption{\textsl{Cross-sections for collisional excitation and ionization with ambient electrons, protons, and hydrogen ($E, P, H$ superscripts, respectively) used in this study. Solid lines are our own polynomial fits, or are the functions from the IAEA \citep[IAEA93 on each panel;][]{international1993iaea}, as indicated. The underlying data are shown as symbols, and are as described in the text. Shah81,87,98 refers to \cite{1981JPhB...14.2361S}, \cite{1987JPhB...20.2481S}, \cite{1998JPhB...31L.757S}. MB83 refers to \cite{McLaughlin_1983}. Hill79 refers to \cite{Hill_1979}. MB87 refers to \cite{McLaughlin_1987}. An asterisk in the transition descriptions indicate a non-thermal particle.}}
	\label{fig:csec_other}
\end{figure*}

The cross-sections are an important element in the prediction of the non-thermal hydrogen level populations, so we compiled more up-to-date values than used previously, where available, particularly for the charge exchange cross sections. As seen in Figures~\ref{fig:csec_cx} and \ref{fig:csec_other}, both charge exchange and collisional excitation/ionization peak in the deka-keV to 100~keV range, so that the relevant wavelengths for the Lyman lines are roughly $\Delta \lambda \approx [5-50]$~\AA. For certain cross-sections we fit polynomial functions to data (sometimes from multiple sources), allowing us to subsequently evaluate the cross-sections at any energy within the bounds of the data, for example at the non-thermal energies used by the \radynfp\ grid. Those polynomial fits were performed in $\log Q$ and $\log E$ space, for $Q$ in $10^{-17}$~cm$^{2}$ and $E$ in keV, and have the form:

\begin{equation}
	\log Q = a_{0} + \sum_{i} a_{i}~(\log E)^{i}.
\end{equation}

\noindent For other cross-sections we used functions from the International Atomic Energy Agency's (IAEA) supplement series \cite{international1993iaea}, who provide functions fit to data that they compiled from multiple sources for each process (see their references). As discussed below, where relevant we extrapolated beyond the energy bounds quoted by various authors but fit the decay so that extrapolations are a straight line in log-log space, which seems justified by the $Q(E)$ values at high energies. We did not fit beyond the IAEA functions but did check that extrapolating beyond the energy bounds that they quote provided sensible results. This is only really required when considering collisions with ambient electrons, and for other processes we were able to locate $Q(E)$ values in the energy range appropriate for the problem at hand. We were unable to locate cross-sections for $Q_{2p}^{H}$, $Q_{23}^{H}$, or $Q_{3p}^{H}$, at suitably high energies, and so omit those in the rates matrix. This should not have a major impact as we do consider the same collisional processes where the targets are protons or electrons. Those missing processes were not included by previous models. Below we show the cross-sections, and comment on their sources. For those that we fit polynomial functions to, their coefficients are listed in Table~\ref{tab:csec_tab}. These data, either fit by us or by the IAEA, come from both experiments and theoretical calculations.

\begin{table*}
\begin{rotatetable*}
\caption{Parameters of polynomial fits of certain cross-sections used in this study. Numbers in parentheses are the indices of the powers of ten to which the values are raised. For example $2.19952(-1) = 2.19952\times10^{-1}$. The valid energy ranges for each fit are indicated in the final column. Fits were performed in $\log Q$ \& $\log E$ space, for $Q$ in $\times10^{-17}$~cm$^{2}$ and $E$ in keV.}
\begin{center}
\begin{tabular}{{ l | l  | l | l | l | l | l | l | l | l | l }}
\toprule
 & {\multirow{2}{*}{$a_0$}} & {\multirow{2}{*}{$a_1$}} & {\multirow{2}{*}{$a_2$}} & {\multirow{2}{*}{$a_3$}} & {\multirow{2}{*}{$a_4$}} & {\multirow{2}{*}{$a_5$}} & {\multirow{2}{*}{$a_6$}} & {\multirow{2}{*}{$a_7$}} & {\multirow{2}{*}{$a_8$}}& E [keV]  \\
 &  & & & & & & &  & &   \\
     \toprule
     \toprule
$\mathbf{Q_{p1}^{CX}}$ &  $2.22694$ & $2.19952(-1)$ & $-2.61594$ & $4.88315$ & $-4.06255$ & $1.49092$ & $-2.50756(-1)$ & $1.43542(-1)$ & $3.20664(-4)$ & $1-8000$ \\
$\mathbf{Q_{p2}^{CX}}$ &  $ 3.52822(-1)$ & $1.64356$ & $-7.98958$ & $1.55848(1)$ & $-1.33785(1)$ & $5.69352$ & $-1.28934$ & $1.48437(-1)$ & $-6.77411(-3)$ & $1-8000$ \\
$\mathbf{Q_{p3}^{CX}}$ &  $ -1.33340$ & $1.81849$ & $-3.20438$ & $8.73242$ & $-8.72014$ & $3.83963$ & $-8.43943(-1)$ & $8.84133(-2)$ & $-3.29639(-3)$ & $1-8000$ \\
\midrule
$\mathbf{Q_{1p}^{P}}$ &  $-1.55347$ & $1.36699$ & $1.85672$ & $-5.69725$ & $9.64059$ & $-8.12511$ & $3.44612$ & $-7.16542(-1)$ & $5.84099(-2)$ & $1-1500$ \\
 & $2.76008$ & $-8.22412(-1)$ & & & & & & & &  $>1500$\\
 $\mathbf{Q_{1p}^{H}}$ &  $-2.64545(-2)$ & $1.52462$ & $-4.62208$ & $1.03086(1)$ & $-1.04104(1)$ & $5.36558$ & $-1.49664$ & $2.15950(-1)$ & $-1.26636(-2)$ & $1-10^{4}$ \\
\midrule
$\mathbf{Q_{12}^{H}}$ &  $1.29154(-1)$ & $1.59680$ & $2.19750$ & $-1.06375(1)$ & $1.50168(1)$ & $-1.13679(1)$ & $4.78532$ & $-1.01709$ & $8.03237(-2)$ & $2-100$ \\
 & $1.35631$ & $-6.05696(-1)$ & & & & & & & &  $>100$\\
\midrule
$\mathbf{Q_{13}^{H}}$ &  $-5.91530(-1)$ & $3.09401$ & $-2.40878$ & $7.40410(-1)$ & $-2.21064$ & $2.89666$ & $-1.53333$ & $3.67348(-1)$ & $-3.33341(-2)$ & $1-1024$ \\
 & $1.49527$ & $-9.23204(-1)$ & & & & & & & &  $>1024$\\
\midrule
\bottomrule
\end{tabular}\label{tab:csec_tab}
\end{center}
\end{rotatetable*}
\end{table*}

\noindent \textbf{Cross-Sections for Charge Exchange}: Charge exchange cross-section data were taken from \cite{2009PhRvA..80c2701W} for $E = [1-100]$~keV, and \cite{1992ADNDT..51...59B} for $E = [125-8000]$~keV. Those energy ranges were combined and an 8-degree polynomial fit from $E=[1-8000]$~keV performed, the results of which are shown in Figure~\ref{fig:csec_cx}. Prior studies of the OZ effect used data from various sources including \cite{1953PPSA...66..972B}, \cite{1974eiip.book.....M}, \cite{1956PhRv..103..896S}, \cite{1970JPhB....3..813C}, \cite{1978PhRvA..18.1930S}, and \cite{1982JPhB...15.2703L} For comparison, on each panel of Figure~\ref{fig:csec_cx} we also show the polynomial fits by \cite{1995A&A...297..854F} and by \cite{1999ApJ...514..430B}, the latter of which are comparable to those used by \cite{1976ApJ...208..618O} and \cite{1985ApJ...295..275C}, to illustrate the differences with the more up-to-date cross-sections (which can be large at times). Though the functions from \cite{1995A&A...297..854F} are not ideal at certain energies, the most important energies were mostly adequately fit.\\

\noindent \textbf{Cross-Sections for Collisional ionization of $\eta_{1}~ (\mathrm{Q_{1p}})$}: Cross-sections for collisional ionization of $\eta_{1}$ by ambient protons ($Q_{1p}^{P}$) were combined from \cite{1981JPhB...14.2361S}, \cite{1987JPhB...20.2481S}, \cite{1998JPhB...31L.757S}, with $E = [38-1500]$~keV, $E = [9-75]$~keV, $E = [1.25-9]$~keV, respectively. An 8-degree polynomial fit from $E = [1-1500]$~keV was performed. Since in $\log Q - \log E$ space the range $E = [500-1500]$~keV is a straight line, a 1-degree polynomial was fit to that range, so that an extrapolation to $E > 1500$~keV can be performed if required (though in practice we really only care about $E< 1$~MeV).  $Q_{1p}^{P}$ is shown in Figure~\ref{fig:csec_other}(A), alongside the underlying data.

Cross-sections for collisional ionization of $\eta_{1}$ by ambient hydrogen ($Q_{1p}^{H}$) are the recommended values from \cite{2021ApJS..252....7C}, in the energy range $E = [0.0362 - 10000]$~keV. An 8-degree polynomial fit was performed over that range, the result of which is shown in Figure~\ref{fig:csec_other}(A), alongside the underlying data.

Cross-sections for collisional ionization of $\eta_{1}$ by ambient electrons ($Q_{1p}^{E}$) are the values from the IAEA. We used their functional form, which is shown in Figure~\ref{fig:csec_other}(A), and can be found on page 28 of \cite{international1993iaea}.\\

\noindent \textbf{Cross-Sections for Collisional Excitation of $\eta_{1} \rightarrow \eta_{2}~ (\mathrm{Q_{12}})$}: Cross-sections for collisional excitation of $\eta_{1}\rightarrow \eta_{2}$ by ambient protons ($Q_{12}^{P}$) are the values from the IAEA. We used their functional form, which is shown in Figure~\ref{fig:csec_other}(B), and can be found on page 46 of \cite{international1993iaea}.

Cross-sections for collisional excitation of $\eta_{1}\rightarrow \eta_{2}$ by ambient hydrogen atoms ($Q_{12}^{H}$) are taken from \cite{Hill_1979} and \cite{McLaughlin_1983}. For the $2s$ level the \cite{Hill_1979} experimental results are used for $E = [2-25]$~keV with \cite{McLaughlin_1983}'s calculations used for $E = [36-100]$~keV. For the $2p$ level \cite{McLaughlin_1983} is used for the full range $2-100$~keV. The $2s$ and $2p$ cross-sections are summed to obtain the total $n=2$ values. These data were fit with an 8-degree polynomial within the range $E = [2-100]$~keV. Since there is a linear decay (in log space) at higher energies, a 1-degree polynomial was fit to $E = [36-100]$~keV allowing extrapolation to $E>100$~keV. The fit results are shown in Figure~\ref{fig:csec_other}(B) alongside the underlying data.

Cross-sections for collisional excitation of $\eta_{1}\rightarrow \eta_{2}$ by ambient electrons ($Q_{12}^{E}$) are the values from the IAEA. We used their functional form, which is shown in Figure~\ref{fig:csec_other}(B), and can be found on page 6 of \cite{international1993iaea}.\\

\noindent \textbf{Cross-Sections for Collisional Excitation of $\eta_{1} \rightarrow \eta_{3}~ (\mathrm{Q_{13}})$}: Cross-sections for collisional excitation of $\eta_{1}\rightarrow \eta_{3}$ by ambient protons ($Q_{13}^{P}$) are the values from the IAEA. We used their functional form, which is shown in Figure~\ref{fig:csec_other}(C), and can be found on page 48 of \cite{international1993iaea}.

Cross-sections for collisional excitation of $\eta_{1}\rightarrow \eta_{2}$ by ambient hydrogen atoms ($Q_{12}^{H}$) are taken from \cite{McLaughlin_1987}, in the energy range $E = [1-1024]$~keV. Data from the $3s, 3p, 3d$ levels were summed to obtain the cross-sections for $n=3$. These data were fit with an 8-degree polynomial. Since there is a linear decay (in log space) at higher energies, a 1-degree polynomial was fit to $E = [324-1024]$~keV allowing extrapolation to $E>1024$~keV.The fit results are shown in Figure~\ref{fig:csec_other}(C) alongside the underlying data.

Cross-sections for collisional excitation of $\eta_{1}\rightarrow \eta_{3}$ by ambient electrons ($Q_{13}^{E}$) are the values from the IAEA. We used their functional form, which is shown in Figure~\ref{fig:csec_other}(C), and can be found on page 14 of \cite{international1993iaea}.\\

\noindent \textbf{Cross-Sections for Collisional ionization of $\eta_{2}~ (\mathrm{Q_{2p}})$}: Cross-sections for collisional ionization of $\eta_{2}$ by ambient protons ($Q_{2p}^{P}$) are the values from the IAEA. We used their functional form, which is shown in Figure~\ref{fig:csec_other}(D), and can be found on page 70 of \cite{international1993iaea}.

Cross-sections for collisional ionization of $\eta_{2}$ by ambient electrons ($Q_{2p}^{E}$) are the values from the IAEA. We used their functional form, which is shown in Figure~\ref{fig:csec_other}(D), and can be found on page 34 of \cite{international1993iaea}.\\

\noindent \textbf{Cross-Sections for Collisional Excitation of $\eta_{2}\rightarrow \eta_{3}~ (\mathrm{Q_{23}})$}: Cross-sections for collisional excitation of $\eta_{2}\rightarrow \eta_{3}$ by ambient protons ($Q_{23}^{P}$) are the values from the IAEA. We used their functional form, which is shown in Figure~\ref{fig:csec_other}(E), and can be found on page 56 of \cite{international1993iaea}.

Cross-sections for collisional excitation of $\eta_{2}\rightarrow \eta_{3}$ by ambient electrons ($Q_{23}^{E}$) are the values from the IAEA. We used their functional form, which is shown in Figure~\ref{fig:csec_other}(E), and can be found on page 22 of \cite{international1993iaea}.\\

\noindent \textbf{Cross-Sections for Collisional ionization of $\eta_{3}~ (\mathrm{Q_{3p}})$}: Cross-sections for collisional ionization of $\eta_{3}$ by ambient protons ($Q_{3p}^{P}$) are the values from the IAEA. We used their functional form, which is shown in Figure~\ref{fig:csec_other}(F), and can be found on page 72 of \cite{international1993iaea}.

Cross-sections for collisional ionization of $\eta_{3}$ by ambient electrons ($Q_{3p}^{E}$) are the values from the IAEA. We used their functional form, which is shown in Figure~\ref{fig:csec_other}(F), and can be found on page 36 of \cite{international1993iaea}.\\

\subsection{Comments on \ozpy}
The model described above was implemented via a python package that we call \ozpy, which can be used to post-process flare simulations such as \radynfp\ to obtain the non-thermal \ion{H}{1} emission resulting from bombardment of the chromosphere by non-thermal protons. This code takes as input a model atmosphere, from any source not just from \radynfp, that is the ambient hydrogen, electron, and proton number densities in cm$^{-3}$. Those can be defined either as a point, or on a depth scale. If input as a depth scale, the height of each grid cell must be provided in also (in cm). A single snapshot, or a series of snapshots can be input, where the time of each snapshot should be provided (in seconds). Additionally, the distribution of precipitating non-thermal particles must be input, (in units of particles~cm$^{-3}$~sr$^{-1}$~keV$^{-1}$), along with the energy grid on which that distribution is defined (in keV), and the pitch angle grid if the particle distribution's pitch angle is resolved. The non-thermal particle distribution should be defined on the same height and time grids as the ambient atmosphere. 

A default set of cross-sections is included (those described in Section~\ref{sec:csec}), defined for either a two- or three-level hydrogen atom alongside other necessary atomic information (particle mass, wavelengths, Einstein coefficients etc.,). Those defaults can be straightforwardly substituted to any cross-sections contained within the \texttt{CrossSections} module, to which the user can append any additional cross-sections. That module also contains methods to fit the data from the various sources of cross-sections. 

From the input atmosphere and non-thermal particle distribution, and defined cross-sections, \ozpy\ solves the set of Equations~\ref{eq:nonthermpops} to obtain the non-thermal hydrogen atom level populations, where by default the final equation is replaced by the non-thermal particle conservation equation though this can be defined by the user. Using those populations the emissivity is calculated as a function of energy, and converted to a Doppler shift (in \AA\ and km~s$^{-1}$). Currently \ozpy\ can model the emissivity of \lya, \lyb, and H~$\alpha$. There are further methods to return either the intensity within each grid cell, or to integrate through the entire loop to obtain the total emergent intensity. Of course, the user can integrate the emissivity over any desired height range outside of \ozpy. 

There are plans to extend this to model non-thermal \ion{He}{2}~304~\AA\ emission, created in the same manner as non-thermal \ion{H}{1} emission, but by a precipitating $\alpha$-particle beam rather than protons. \ozpy\ is available from \url{https://github.com/grahamkerr/OrrallZirkerPy/releases/tag/v1.0.1}.  

\section{Non-thermal Lyman Emission}\label{sec:nthmemission}
Outputs from each of the \radynfp\ flare simulations were used as input to \ozpy, producing the non-thermal \lya\ \& \lyb\ emission as functions of height and time, that is a non-thermal emissivity produced by the proton beam. The non-thermal intensity produced by the beam is then the integral of the emissivity through the full extent of the chromosphere. Here the general characteristics and formation properties of the non-thermal emission are presented, along with a comparison to the ambient thermal Lyman line emission.

\subsection{Characteristics of Non-Thermal Ly Emission}
\begin{figure*}
	\centering 
	{\includegraphics[width = \textwidth, clip = true, trim = 0.cm 0.cm 0.cm 0.cm]{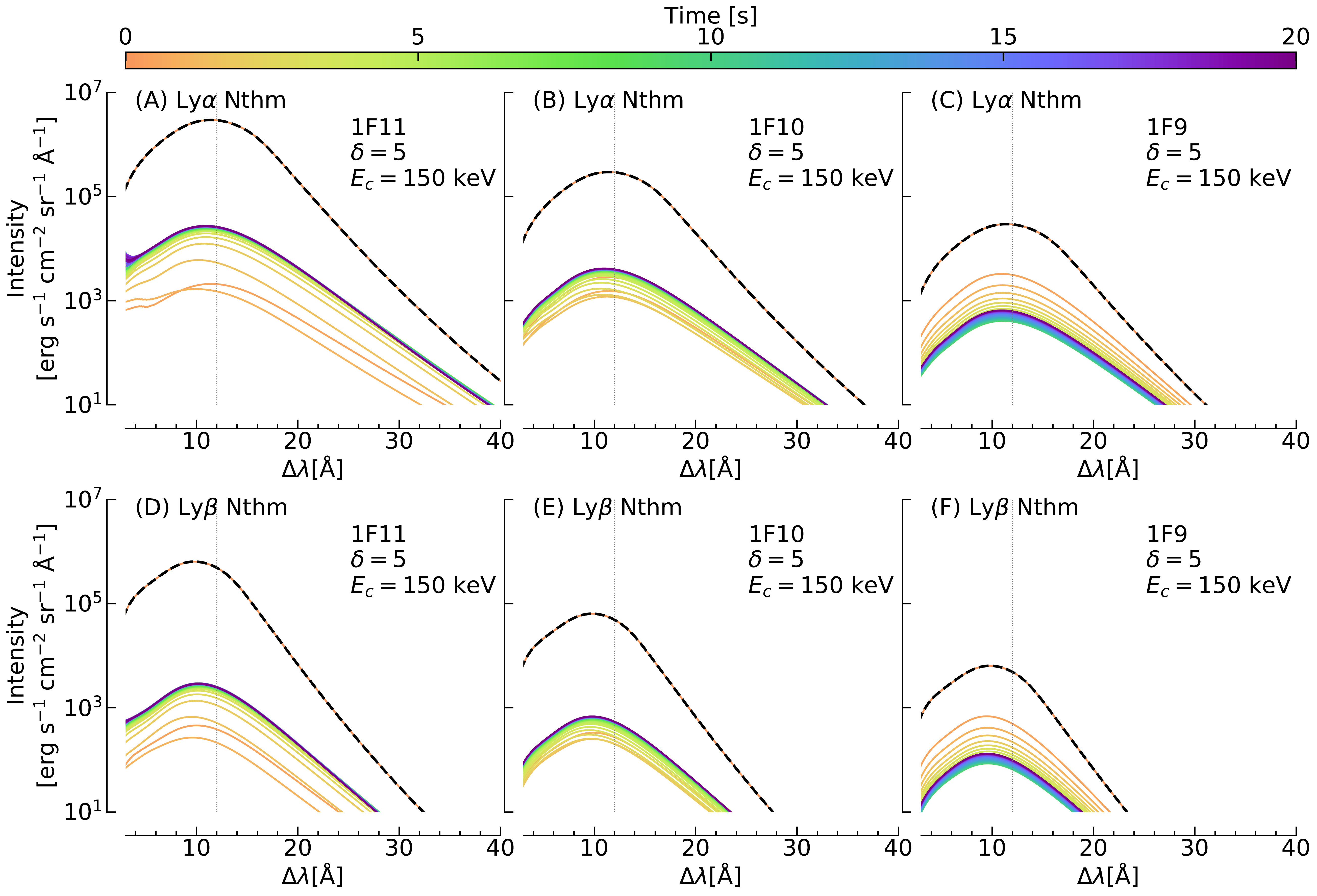}}	
	\caption{\textsl{Non-thermal spectra of \lya\ (top row, panels A-C) and \lyb\ (bottom, panels D-F) as a function of time where color represents time at $0.5$~s cadence. The black dashed line is the emission at $t=0$~s which is the moment that the proton beam first impacts the chromosphere. Each column represents a different injected energy flux (i.e. flare strength). Panels (A,D) are the 1F11 simulations, panels (B,E) are 1F10, and panels (C,F) are 1F9. The x-axis is the wavelength from the core of each line. The dotted vertical line on each panel shows $\Delta\lambda = 12$~\AA, which is the wavelength used as reference when showing various formation properties in other figures.}}
	\label{fig:OZSpectra_nthmonly}
\end{figure*}

\begin{figure}
	\centering 
		{\includegraphics[width = 0.5\textwidth, clip = true, trim = 0.cm 0.cm 0.cm 0.cm]{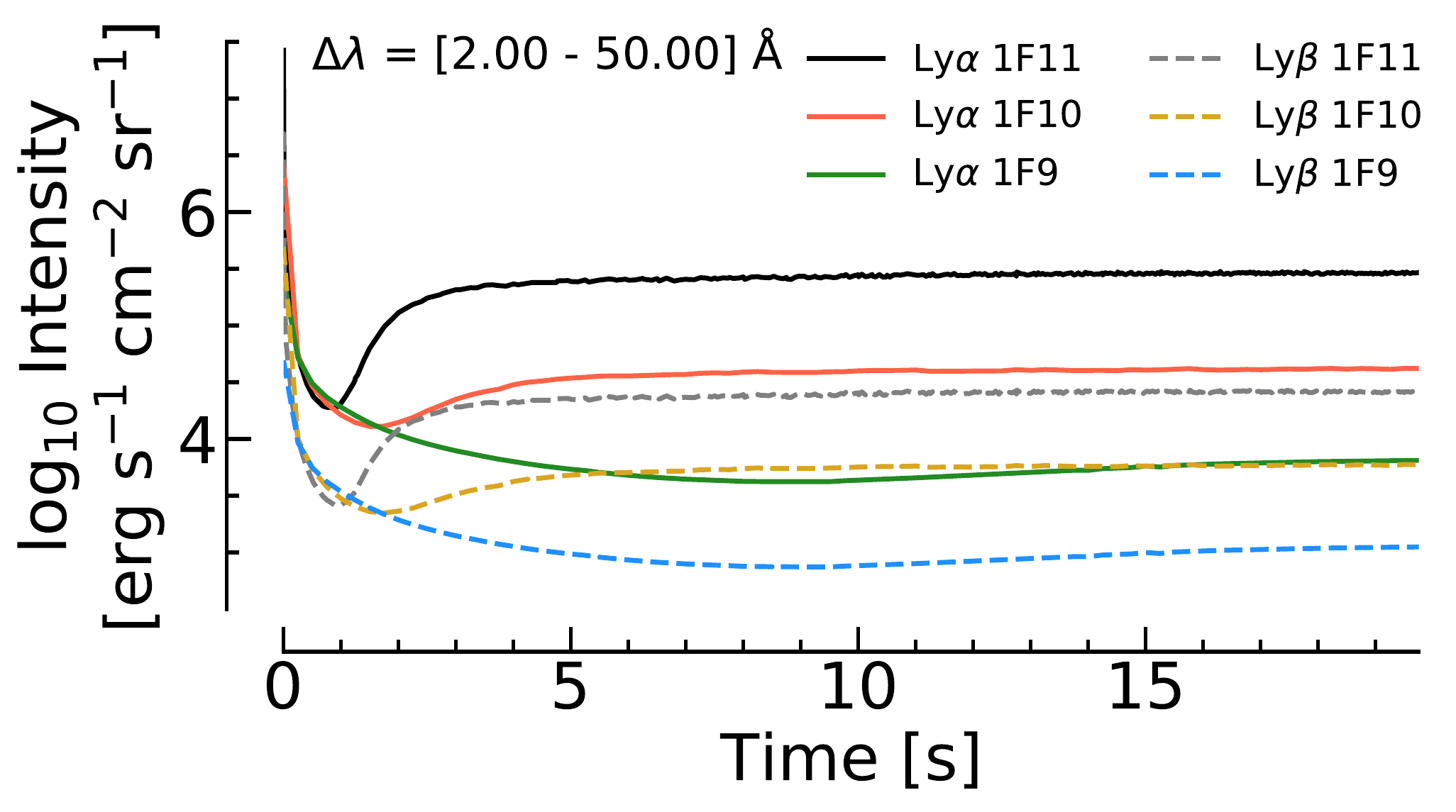}}	
	\caption{\textsl{Lightcurves of non-thermal \lya\ (solid lines) and \lyb\ (dashed lines) integrated over the range $\Delta\lambda = [2-50]$~\AA. Black and grey are the 1F11 simulations, red and gold are the 1F10 simulations and green and blue are the 1F9 simulations.}}
	\label{fig:OZ_lightcurves}
\end{figure}

\begin{figure*}
	\centering 
	{\includegraphics[width = 0.85\textwidth, clip = true, trim = 0.cm 0.cm 0.cm 0.cm]{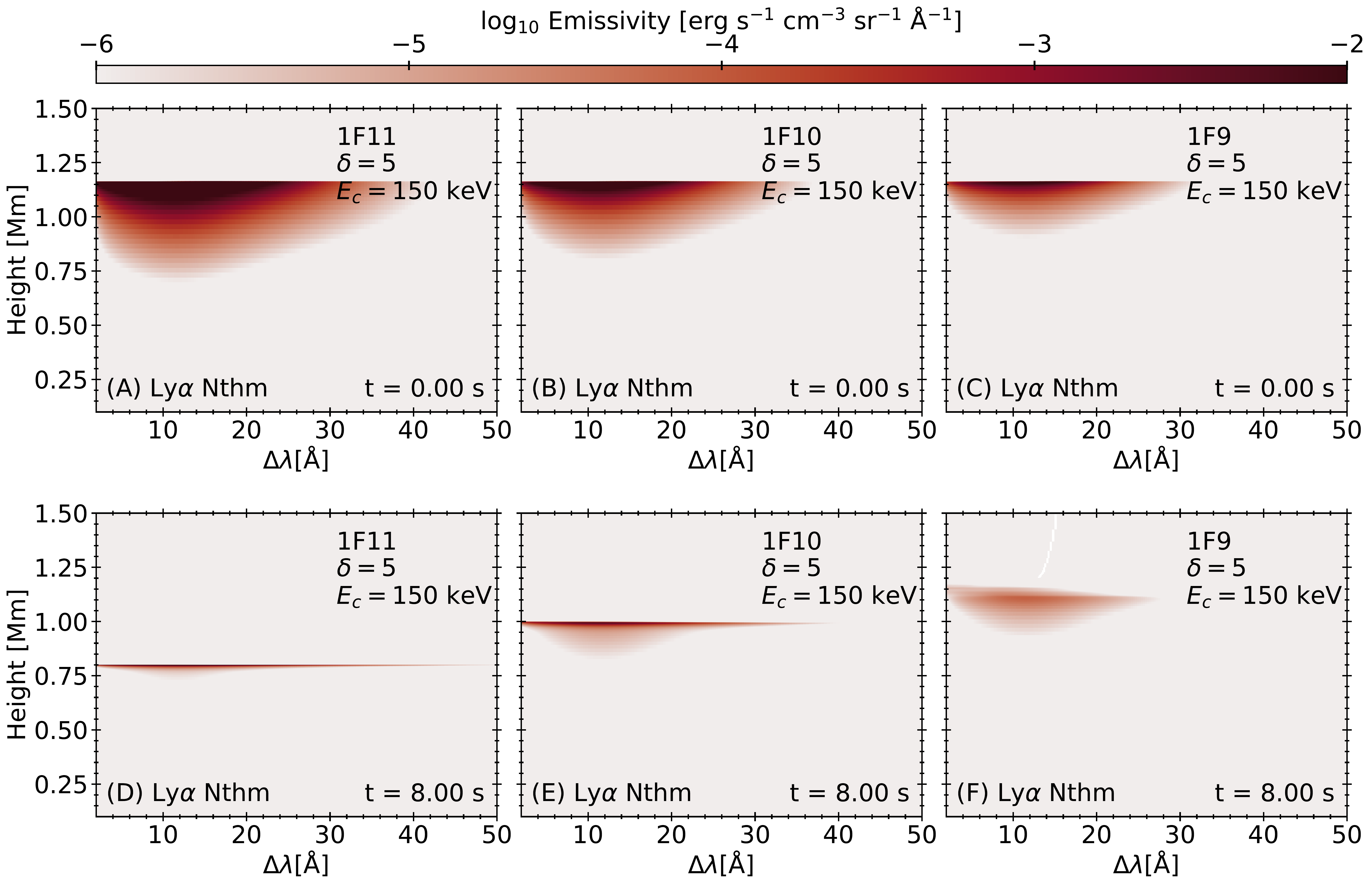}}	
	\caption{\textsl{The emissivity on a $\log_{10}$ scale (represented by color) of non-thermal \lya\ emission at $t = 0$~s (top row, A-C) and $t = 8$~s (bottom row, D-F) as functions of both wavelength from line center and height in the atmosphere. Integrating through height yields the total emergent non-thermal Lyman line intensity shown in Figure~\ref{fig:OZSpectra_nthmonly}. Each column is a different flare simulation, (A,D) are the 1F11, (B,E) are the 1F10, and (C,F) are the 1F9 flares. The $t=0$~s panels show the first instant of non-thermal emission caused by the proton beams. The sharp upper boundary represents the location of the flare transition region, above which there are insufficient neutrals to allow significant charge exchange interactions to take place (i.e. there is little to no non-thermal emission). The $t=8$~s shows a time during the heating phase where atmospheric dynamics have greatly changed the interaction of the non-thermal distribution and ambient plasma.}}
	\label{fig:OZSpectra_contrib1}
\end{figure*}

Non-thermal spectra are shown in Figure~\ref{fig:OZSpectra_nthmonly}, as functions of $\Delta\lambda$ from line center of \lya\ (A-C; top row) and \lyb\ (D-F; bottom row). Color represents time. Though there is a slight drift over time, the wavelength of peak intensity is relatively stable. For \lya\ the peak is initially $\Delta\lambda = 11.4$~\AA. In the 1F11 simulation this initially rises to $\Delta\lambda = 11.6$~\AA, then shortens to $\Delta\lambda = 9.6$~\AA\ for the first few seconds, before settling to $\Delta\lambda \sim11$~\AA. The weaker simulations show less scatter, with the 1F10 simulation initially exhibiting a shortening to $\Delta\lambda = 10.6$~\AA, before settling to $\Delta\lambda = 11$~\AA\ by $t = 5.1$~s. The 1F9 shows even less scatter, though settles somewhat to a longer wavelength at $\Delta\lambda = 11.15$~\AA.

Since the line core of \lyb\ is around $\sim190$~\AA\ shorter than \lya\, the non-thermal \lyb\ emission peaks with a somewhat shorter displaceent, initially at  $\Delta\lambda = 9.8$~\AA. As before, in the 1F11 flare this initially lengthens (within the first $t = 1$~s) to $\Delta\lambda = 10$~\AA, before shortening to a minimum value of $\Delta\lambda = 9.3$~\AA, finally settling at around $\Delta\lambda = 10.25$~\AA\ by $t = 5$~s. The 1F10 simulation initially has a decreasing peak to $\Delta\lambda = 9.5$~\AA, but by $t=5$~s has settled back to $\Delta\lambda = 9.8$. Finally, the 1F9 simulation settles quickly to $\Delta\lambda = 9.5$.

While the variations in the Doppler shift of the peak emission are small, there is a significant change in the intensity over time, with the strongest emission occurring immediately at flare onset. At that time the non-thermal protons have reached the chromosphere and some portion of the beam is neutralized before the ambient flaring Lyman emission has brightened and before the ionization fraction increases. Non-thermal \lya\ emission peaks at $I_{nthm,Ly\alpha} = [3\times10^{6}, 3\times10^{5},3\times10^{4}]$~erg~s$^{-1}$~cm$^{-2}$~sr$^{-1}$~\AA$^{-1}$, for the 1F11, 1F10, 1F9 flares respectively. Just 0.25~s later this emission has dropped significantly so that each simulation actually has a comparable peak intensity of $I_{nthm,Ly\alpha} \approx 5\times10^{3}$~erg~s$^{-1}$~cm$^{-2}$~sr$^{-1}$~\AA$^{-1}$. Over the next few seconds the 1F10 and 1F11 flares continue to drop in intensity, but then both increase to peak values of $I_{nthm,Ly\alpha} = [4\times10^{3}, 2.6\times10^{4}]$~erg~s$^{-1}$~cm$^{-2}$~sr$^{-1}$~\AA$^{-1}$, respectively. The 1F9 simulation instead continues to decrease, such that by $t=10$~s, $I_{nthm,Ly\alpha} = 4.25\times10^{2}$~erg~s$^{-1}$~cm$^{-2}$~sr$^{-1}$~\AA$^{-1}$. This temporal evolution can be more clearly seen in Figure~\ref{fig:OZ_lightcurves}, which shows lightcurves integrated over the range $\Delta\lambda = [2-50]$~\AA.

Non-thermal \lyb\ emission behaves similarly, with the following peak intensities: at $t = 0$~s for 1F11, 1F10, and 1F9, $I_{nthm,Ly\beta} = [6.4\times10^{5}, 6.4\times10^{4},6.4\times10^{3}]$~erg~s$^{-1}$~cm$^{-2}$~sr$^{-1}$~\AA$^{-1}$; at $t = 0.25$~s all flares have a peak near $I_{nthm,Ly\beta} \approx 1.16\times10^{3}$~erg~s$^{-1}$~cm$^{-2}$~sr$^{-1}$~\AA$^{-1}$; at $t = 10$~s for 1F11, 1F10, and 1F9, the peaks intensities are $I_{nthm,Ly\beta} = [2.9\times10^{3}, 6.5\times10^{2},8.8\times10^{1}]$~erg~s$^{-1}$~cm$^{-2}$~sr$^{-1}$~\AA$^{-1}$. Lightcurves of \lyb\ are shown as dashed lines on Figure~\ref{fig:OZ_lightcurves}, indicating that non-thermal \lyb\ emission follows the temporal pattern of \lya, but is weaker. 

\begin{figure*}
	\centering 
	{\includegraphics[width = 0.85\textwidth, clip = true, trim = 0.cm 0.cm 0.cm 0.cm]{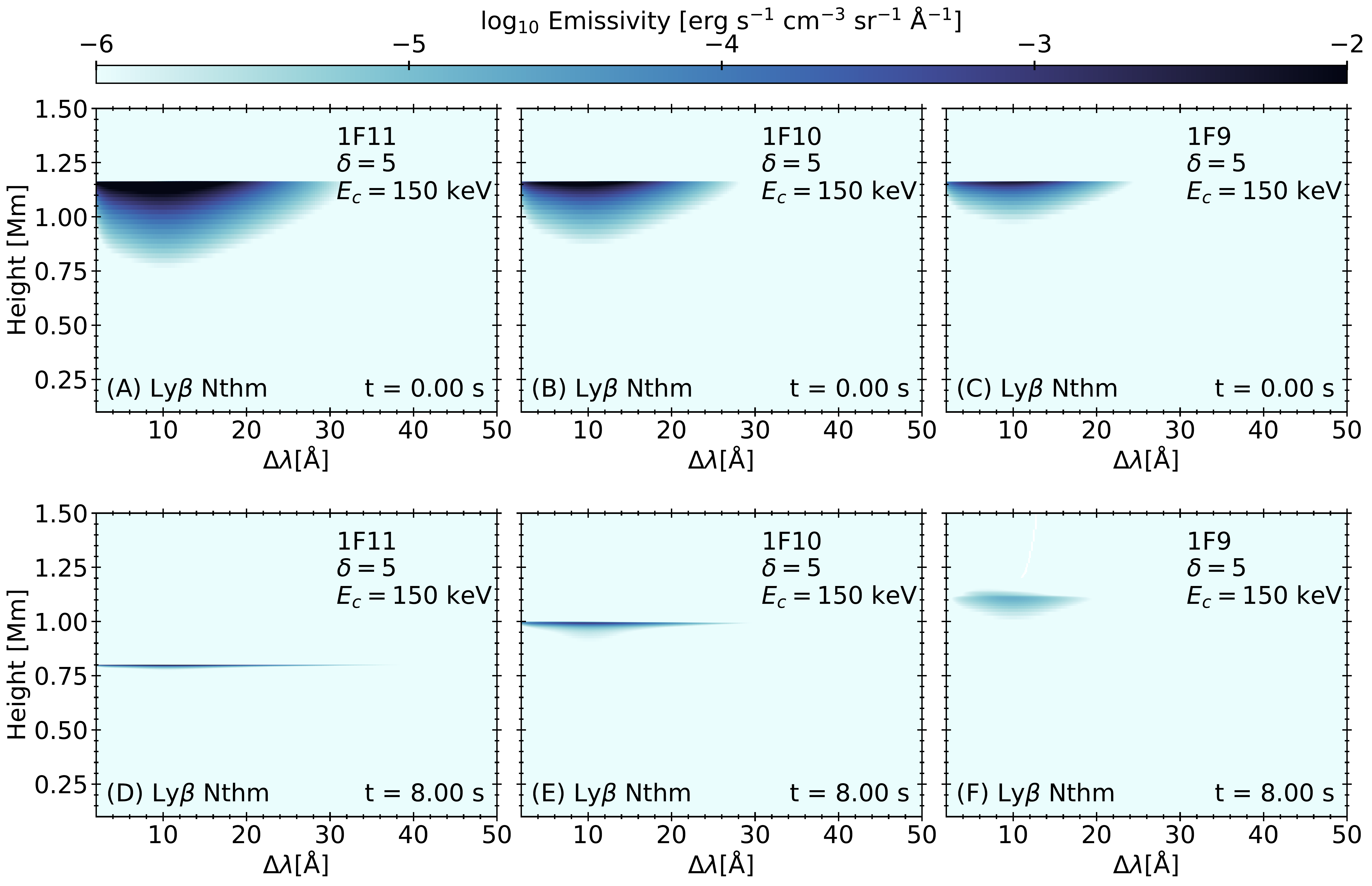}}	
	\caption{\textsl{Same as Figure~\ref{fig:OZSpectra_contrib1} but showing non-thermal \lyb\ emission.}}
	\label{fig:OZSpectra_contrib3}
\end{figure*}

\subsection{Formation Properties of Non-Thermal Ly Emission}

To better understand the non-thermal emission characteristics we look at where in the atmosphere this emission forms, and what the plasma properties are at those times. Figures~\ref{fig:OZSpectra_contrib1} and \ref{fig:OZSpectra_contrib3} show the emissivity of non-thermal \lya\ and \lyb, respectively, for each simulation (1F11, 1F10, 1F9 for each column left-to-right), as functions of height and wavelength, at $t = 0$~s (top rows), and $t = 8$~s (bottom rows). Integrating through height for each wavelength yields the total emergent intensity. Initially, there is a broad emitting region around the peak wavelengths, spanning a few tens of km (note the log scale on the color bar of these images). Very quickly the intensity drops, and the altitude of peak emissivity shifts deeper, more so for the stronger flare. By the latter stages of the simulation, emission from the stronger flares forms over a vanishingly narrow extent, and form progressively deeper, whereas the 1F9 simulation has a broader emitting region located at higher altitude. 

\begin{figure}
	\centering 
	\vbox{
	\hbox{
	\subfloat{\includegraphics[width = .5\textwidth, clip = true, trim = 0.cm 0.cm 0.cm 0.cm]{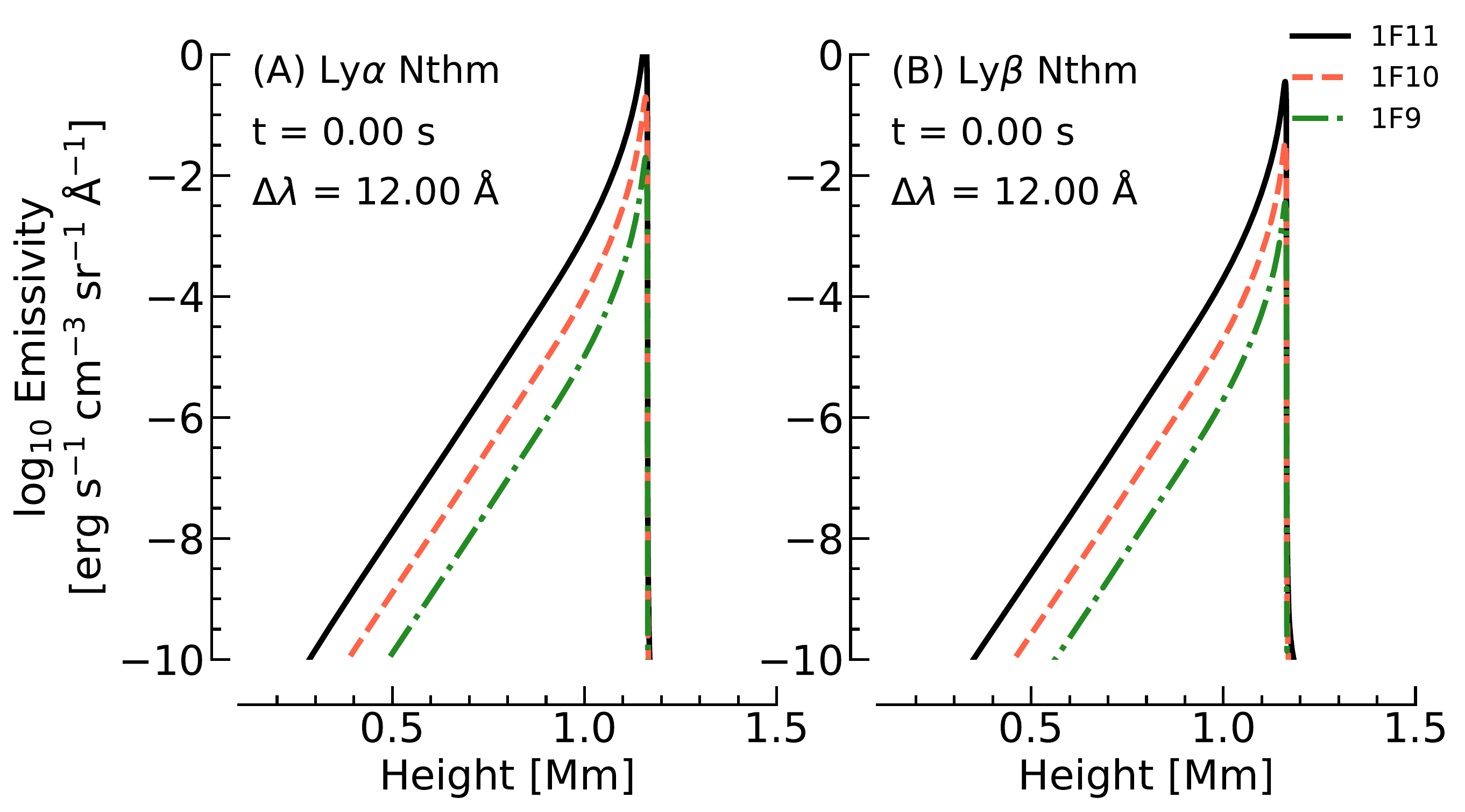}}	
	}
	}
	\vbox{
	\hbox{
	\subfloat{\includegraphics[width = .5\textwidth, clip = true, trim = 0.cm 0.cm 0.cm 0.cm]{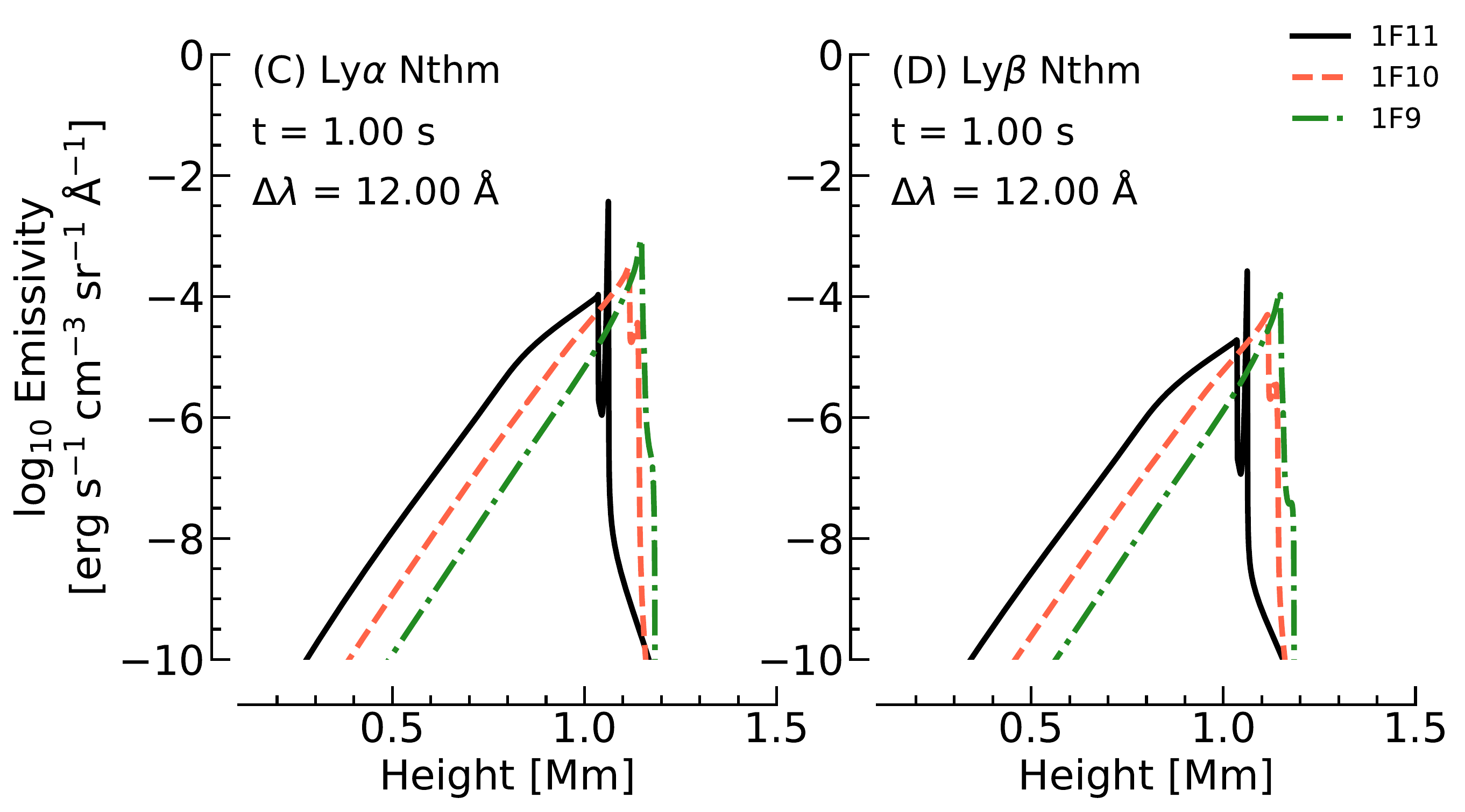}}	
	}
	}
	\vbox{
	\hbox{
	\subfloat{\includegraphics[width = .5\textwidth, clip = true, trim = 0.cm 0.cm 0.cm 0.cm]{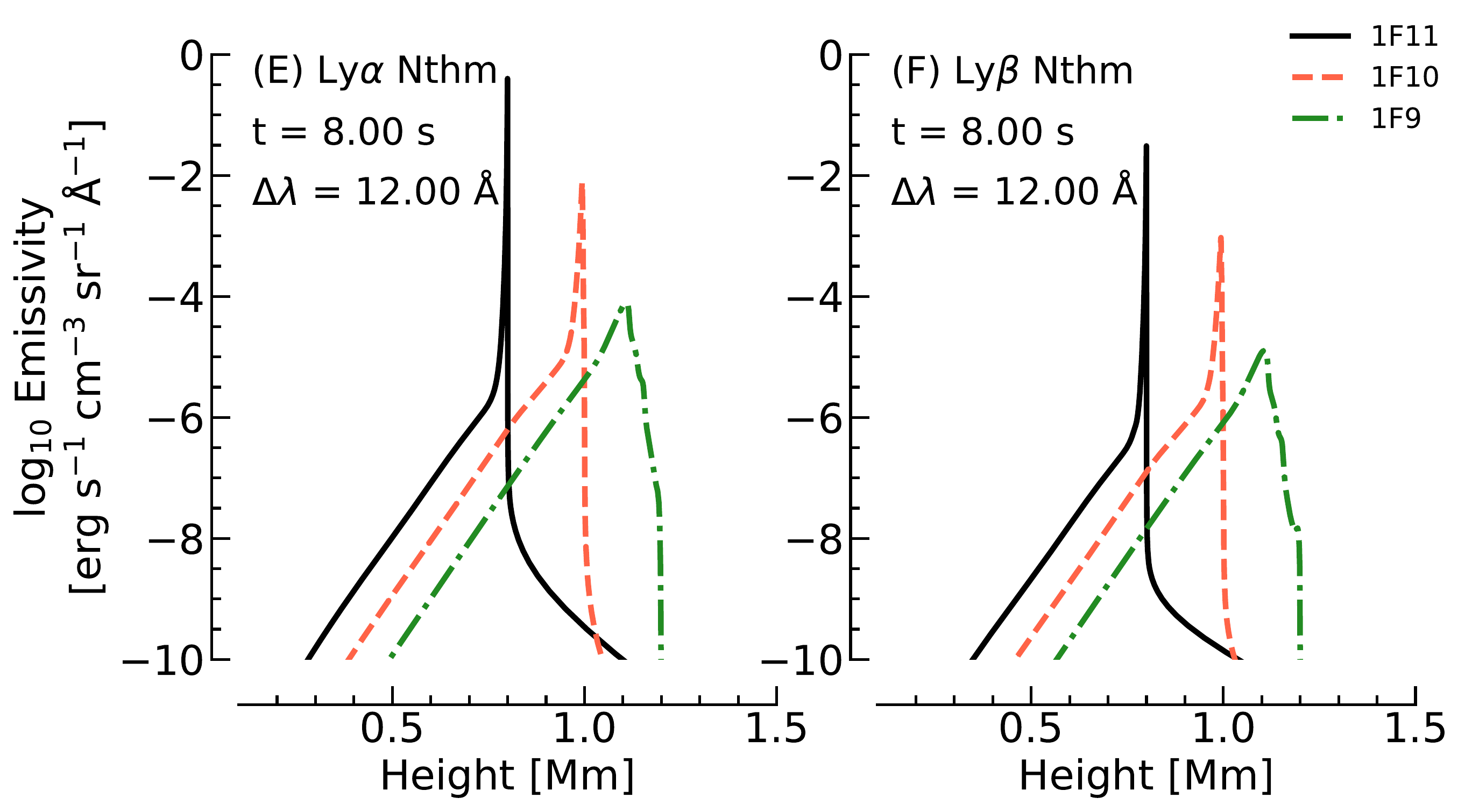}}	
	}
	}
	\caption{\textsl{The emissivity as a function of height for a wavelength $\Delta\lambda = 12$~\AA\ from line cores of \lya\ (first column, panels: A,C,E) and \lyb\ (second column, panels: B,D,F) for three different times. Color represents the flare strength, with the black solid line being the 1F11 simulation, the red dashed line being the 1F10 simulation, and the green dot-dashed line being the 1F9 simulation.}}
	\label{fig:OZSpectra_contrib_wref}
\end{figure}

Selecting a reference wavelength of $\Delta\lambda = 12$~\AA, Figure~\ref{fig:OZSpectra_contrib_wref} shows a more detailed view of the stratification of emissivity. Here we see that the emissivity initially drops through the mid-upper chromosphere, but the lower atmosphere does not vary greatly. In the stronger simulations the peak emissivity rebounds, regaining some strength while the peak height occurs deeper and deeper. Once it has rebounded, though, the peak region is vanishingly narrow compared to the broader emitting region of the 1F9 simulation. 

\begin{figure}
	\centering 
	\vbox{
	\hbox{
	\subfloat{\includegraphics[width = .5\textwidth, clip = true, trim = 0.cm 0.cm 0.cm 0.cm]{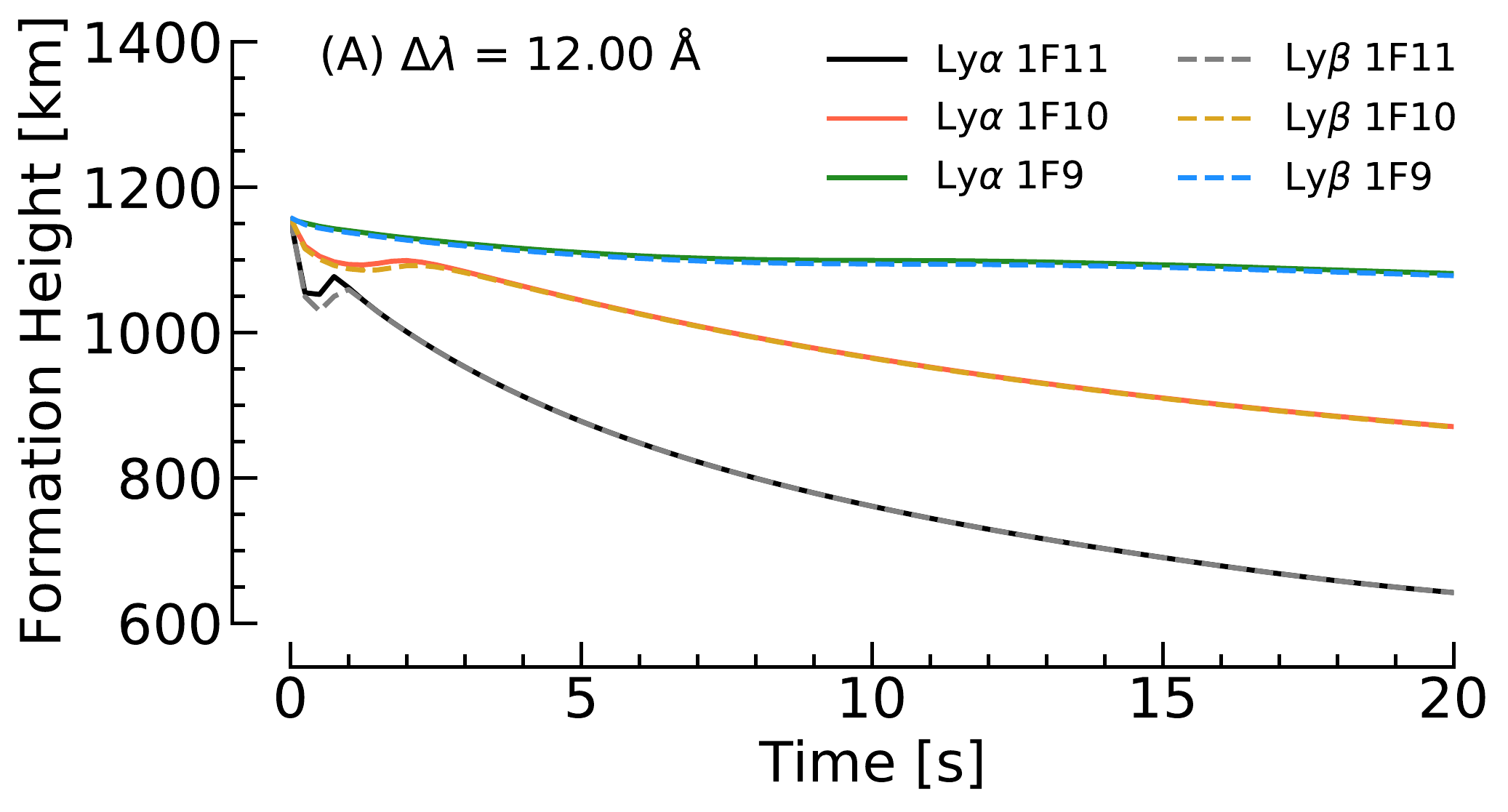}}	
	}
	}
	\vbox{
	\hbox{
	\subfloat{\includegraphics[width = .5\textwidth, clip = true, trim = 0.cm 0.cm 0.cm 0.cm]{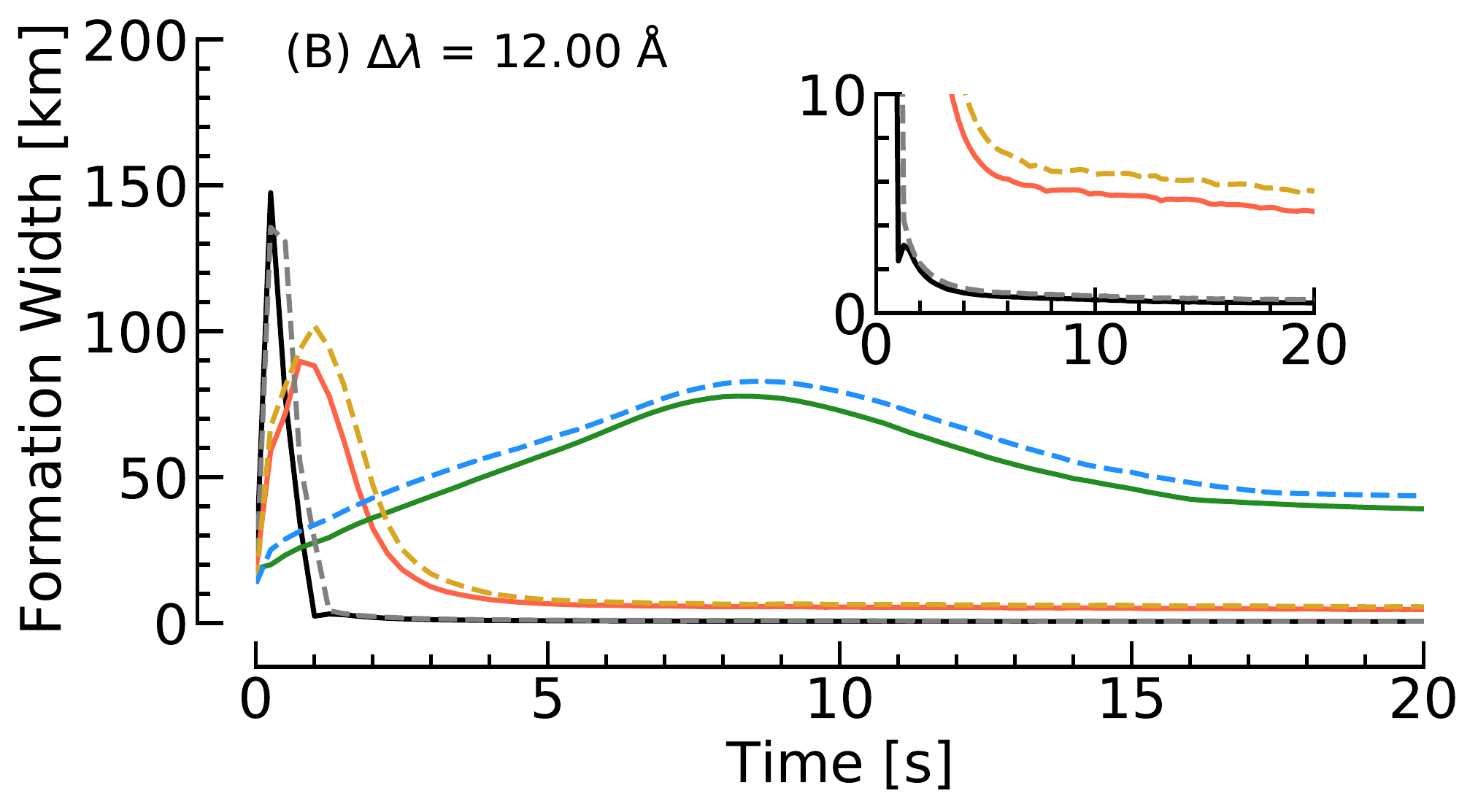}}	
	}
	}
	\caption{\textsl{Panel (A) shows the mean formation height of $\Delta\lambda = 12$~\AA\ from line cores of \lya\ (solid lines) and \lyb\ (dashed lines) as a function of time. The black and grey lines show the 1F11 simulation, the red and gold lines show the 1F10 simulation, and the green and blue lines show the 1F9 simulation. Panel (B) shows the widths of the formation region, where colors and lines are the same as panel (A).}}
	\label{fig:OZSpectra_formheights}
\end{figure}

Following \cite{2017ApJ...836...12K} we construct the normalized cumulative distribution function of the emissivity through height $NCDF_{j}(z)$, and define the emitting region as the region bounded by $NCDF_{j}(z)=0.05$ and $NCDF_{j}(z)=0.95$; that is, the region where the bulk of the emergent intensity originates. Labelling these heights as $z_{upp}$ and $z_{low}$, the mean formation height can be found by 

 \begin{equation}
<z> = \frac{\int_{z_{low}}^{z_{upp}}j(z)~z~\mathrm{d}z}{\int_{z_{low}}^{z_{upp}}j(z)\mathrm{d}z},
\label{eq:avz}
\end{equation}

\noindent where we weight by the emissivity. This is shown for $\Delta\lambda = 12$~\AA\ in Figure~\ref{fig:OZSpectra_formheights}, where we see that non-thermal emission for the stronger flares forms significantly deeper in the atmosphere compared to the 1F9 simulation, which barely changes in altitude. Both Lyman lines form essentially at the same altitude. If we then take the width of the formation region to be $\Delta z = z_{upp} - z_{low}$, shown in panel (B), we see that for the two stronger flares when the non-thermal Lyman emission intensity drops in strength, the formation region widens up to $100-150$~km. Over time, the widths of the formation region drop to only a few km, or down to metres, at which times the magnitude of the non-thermal intensity has risen somewhat but is not as strong as at the early phase of the flare. The weaker flare, however, exhibits a gradually increasing $\Delta z$ such that the non-thermal \lyb\ emission originates from a somewhat broader region.

Performing a similar calculation to obtain the mean ionization fraction in the non-thermal emission forming region we can understand the patterns we see above. The mean ionization fraction is:

 \begin{equation}
<\chi_{H}> = \frac{\int_{z_{low}}^{z_{upp}}j(z)~\chi_{H}~\mathrm{d}z}{\int_{z_{low}}^{z_{upp}}j(z)\mathrm{d}z}.
\label{eq:avion}
\end{equation}

\begin{figure}
	\centering 
	{\includegraphics[width = 0.5\textwidth, clip = true, trim = 0.cm 0.cm 0.cm 0.cm]{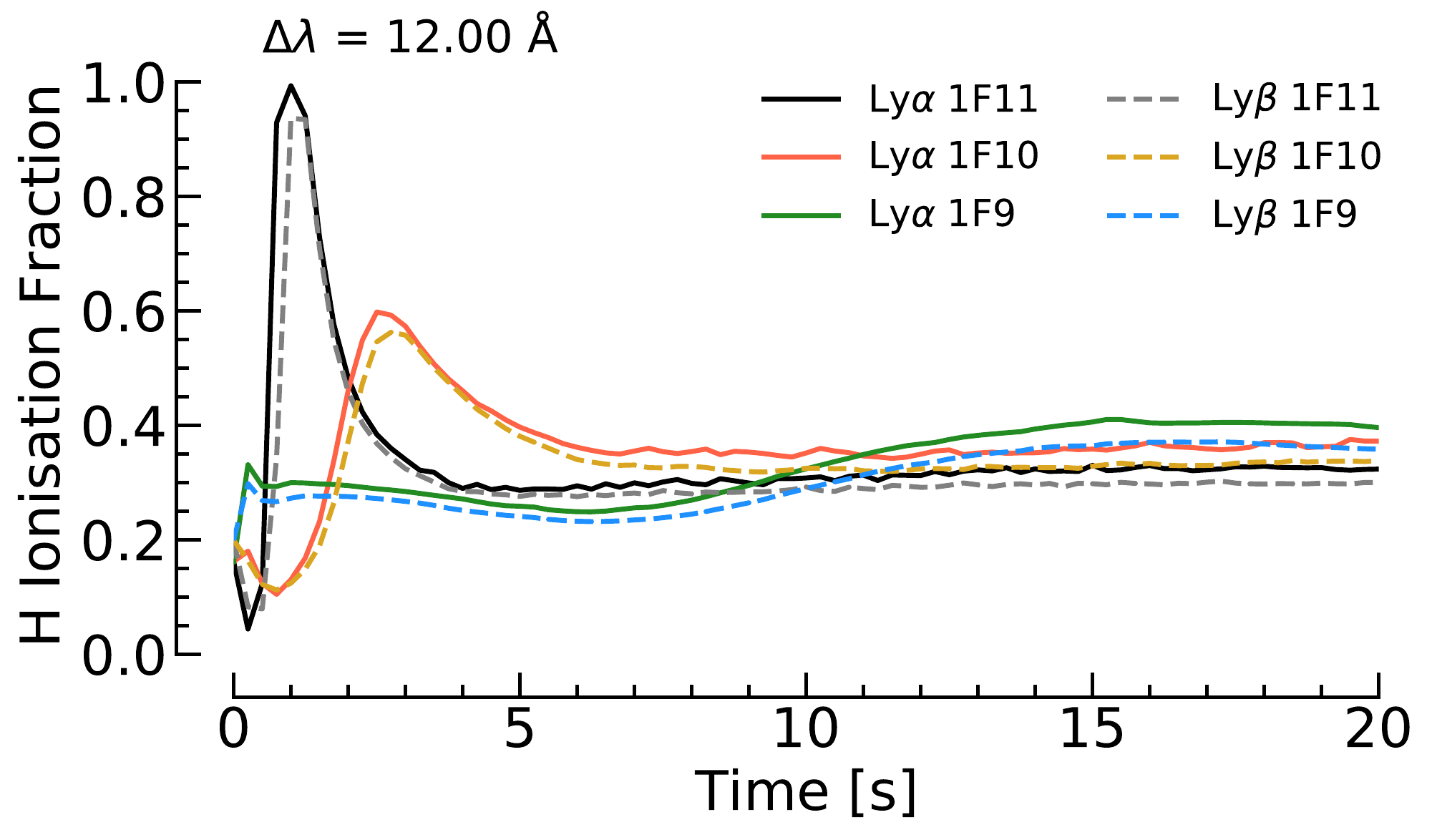}}	
	\caption{\textsl{The mean hydrogen ionization fraction in the formation region of $\Delta\lambda = 12$~\AA\ from line cores of \lya\ (solid lines) and \lyb\ (dashed lines) as a function of time. The black and grey lines show the 1F11 simulation, the red and gold lines show the 1F10 simulation, and the green and blue lines show the 1F9 simulation.}}
	\label{fig:OZSpectra_ionfrac}
\end{figure}

\noindent These results are shown in Figure~\ref{fig:OZSpectra_ionfrac}. For the 1F11 simulation, the mean ionization fraction initially drops before peaking close to unity. Recall, though, that this is the mean over the full formation region, and that the formation region of the 1F11 non-thermal emission rapidly broadened, such that a larger fraction of the lower atmosphere influences the mean. The ionization fraction in the upper chromosphere has, however, increased. Consequently there are fewer charge exchange interactions, which require a supply of neutrals, and the non-thermal emissivity drops. As the temperature in the mid-upper chromosphere climbs, the ionization fraction climbs and emissivity continues to fall. However, as we noted before there is a rebounding of non-thermal emission in the 1F11 simulation. This occurs due to the severe compression of the chromosphere. The proton beam encounters a fresh supply of neutrals once the transition region is pushed deeper, which picks up pace after a few seconds. At these times  $<\chi_{H}>$ is only somewhat larger than at the time of peak emission, but crucially, the formation region width $\Delta z$ is significantly reduced (itself due to where protons are predominantly thermalized by the increase in atmospheric density there). Consequently, integrating the emissivity through height produces only a modest total intensity. The 1F10 simulation behaves similarly and with the same end result, though in less dramatic fashion, only achieving a maximum ionization fraction of $<\chi_{H}> \approx 0.6$. 

Since the atmospheric dynamics are less extreme in the 1F9 flare, the reduction in emissivity is smaller compared to the drop seen in the strong flares, owing to the smaller increase in ionization fraction. While the peak emissivity continues to drop due to the increasing ionization fraction, the formation region widens due to lack of compression in the chromosphere, which combats the decreasing peak. A gradually declining emergent intensity results. 

\begin{figure*}
	\centering 
	{\includegraphics[width = \textwidth, clip = true, trim = 0.cm 0.cm 0.cm 0.cm]{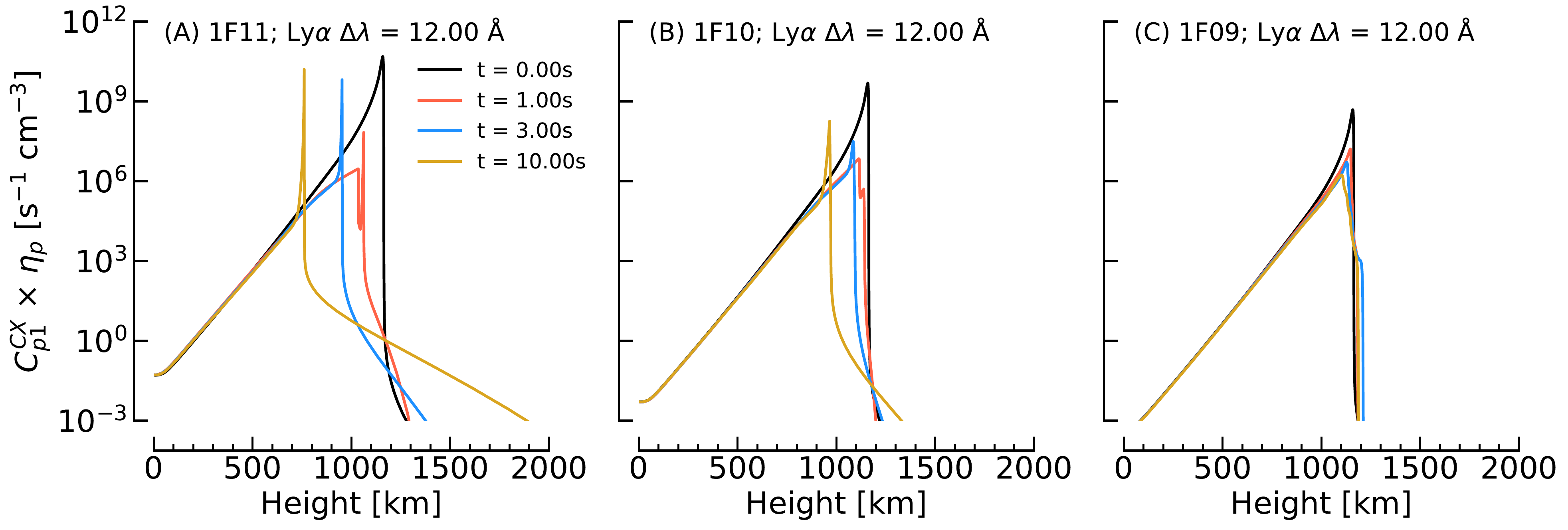}}	
	\caption{\textsl{The rate of charge exchange interactions to the ground state, at the energy responsible for producing a Doppler shift of $\Delta\lambda = 12$~\AA\ from \lya, as a function of height, for 4 times (black is $t=0$~s, red is $t=1$~s, blue is $t=3$~s, and gold is $t=10$~s) in each simulation. Panel (A) is the 1F11 simulation, (B) is 1F10, and (C) is 1F9.}}
	\label{fig:OZSpectra_CXrate}
\end{figure*}

Finally, to confirm that the rising ionization fraction is met by a reduction in the number of charge exchange interactions (and hence emissivity of non-thermal emission) we show, in Figure~\ref{fig:OZSpectra_CXrate}, the product of the charge exchange rate to the ground state and the non-thermal proton population at an energy equivalent to a Doppler shift of $\Delta\lambda = 12$~\AA. Each panel represents a different simulation and the color of each line is a different snapshot within that simulation. As expected, we see the same patterns as found for the emissivities.

In summary, although the non-thermal emission is initially very strong, with intensity directly related to the strength of the flare (due to the number of non-thermal protons injected), the increase in ionization fraction rapidly reduces the intensity. In the case of the stronger flares, the compression of the atmosphere results in a slight rebounding of intensity as the beam encounters new neutrals, but since this happens in a very narrow extent the emergent intensity never reaches that of the onset. 

\subsection{Comparison to Ambient Solar Flare Ly Emission}
\begin{figure*}
	\centering 
	\hbox{
	\subfloat{\includegraphics[width = .5\textwidth, clip = true, trim = 0.cm 0.cm 0.cm 0.cm]{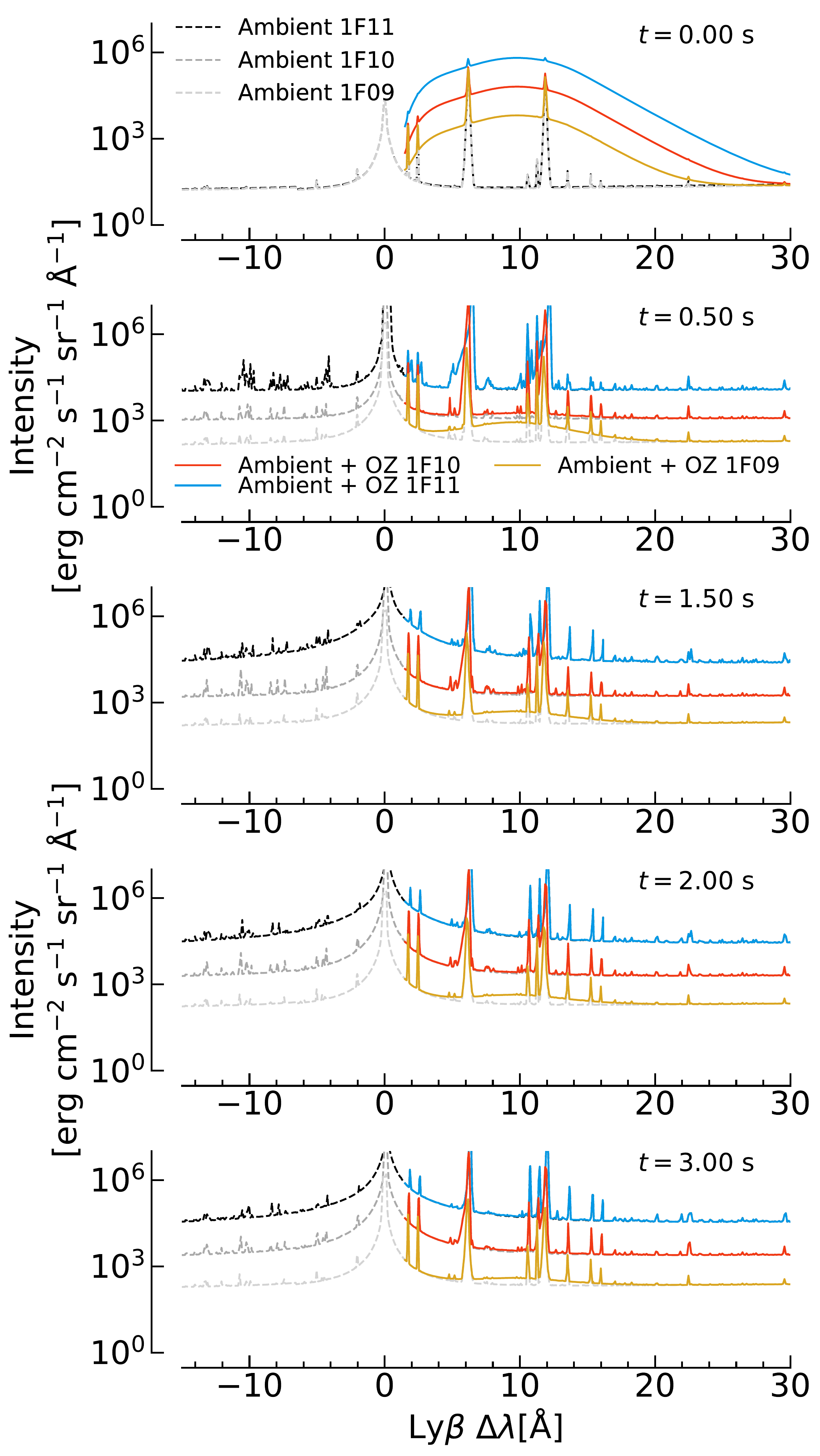}}	
	\subfloat{\includegraphics[width = .5\textwidth, clip = true, trim = 0.cm 0.cm 0.cm 0.cm]{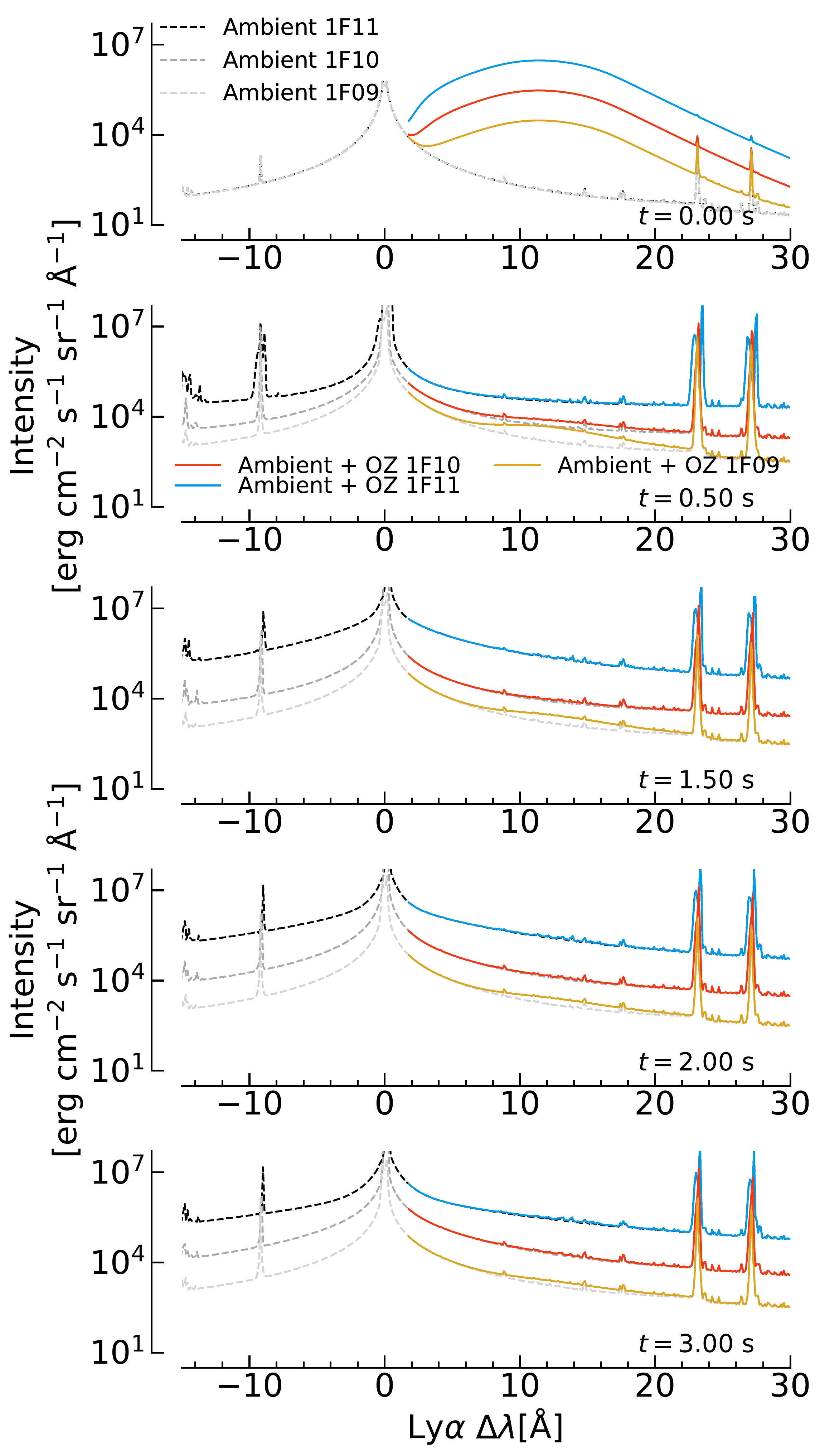}}	
	}
	\caption{\textsl{Ambient flare emission (greyscale; darkest is 1F11, lightest is 1F9) surrounding the \lyb\ (left column) and \lya\ (right column) lines, compared to the total emission (ambient summed with the non-thermal Lyman lines from the OZ effect), shown in colored lines (blue is 1F11, red is 1F10, and gold is 1F9). Each row shows a snapshot during the flare. }}
	\label{fig:OZSpectra_summedemission}
\end{figure*}

\begin{figure}
	\centering 
	\vbox{
	\subfloat{\includegraphics[width = .5\textwidth, clip = true, trim = 0.cm 0.cm 0.cm 0.cm]{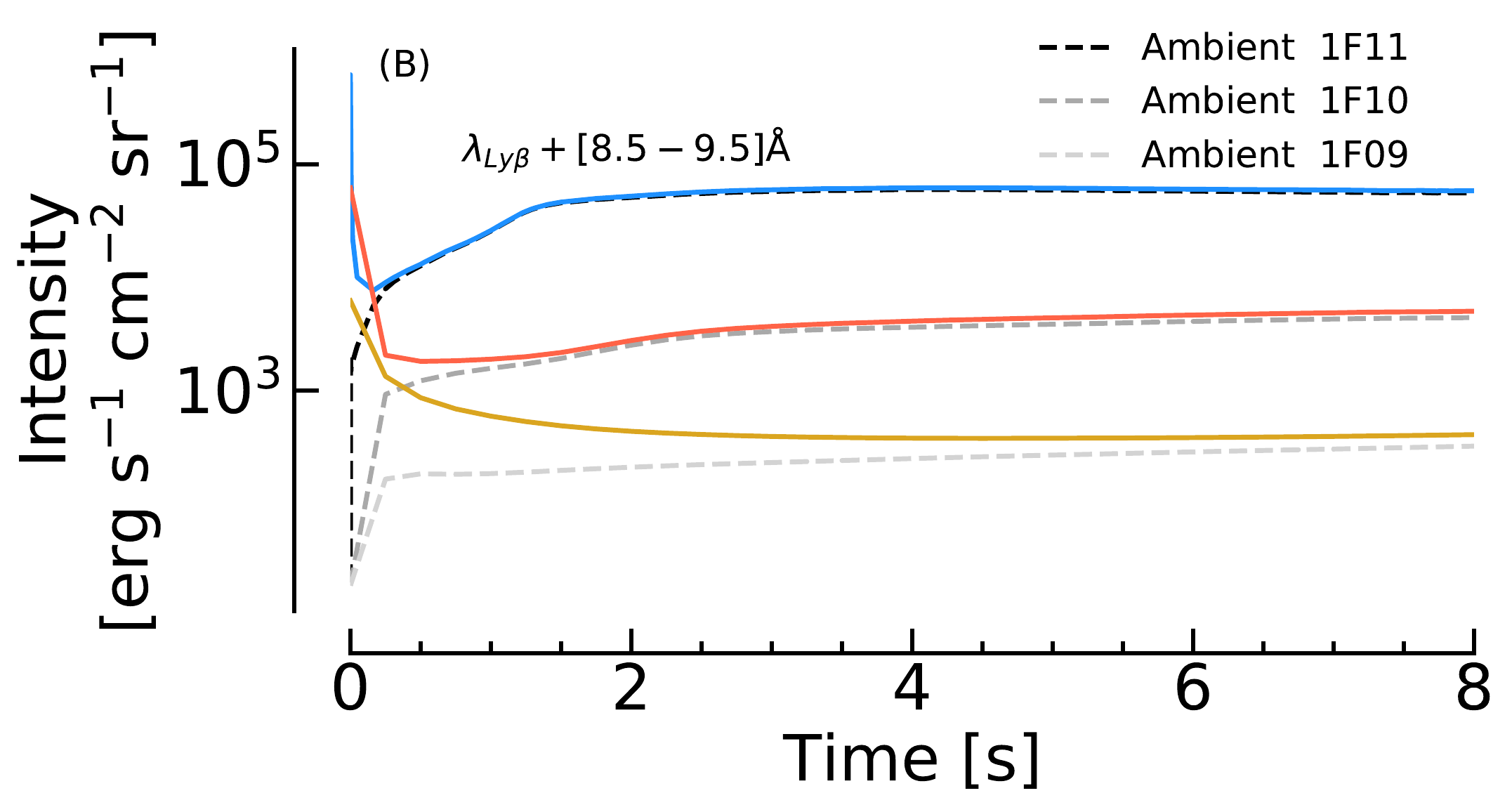}}	
	}
	\vbox{
	\subfloat{\includegraphics[width = .5\textwidth, clip = true, trim = 0.cm 0.cm 0.cm 0.cm]{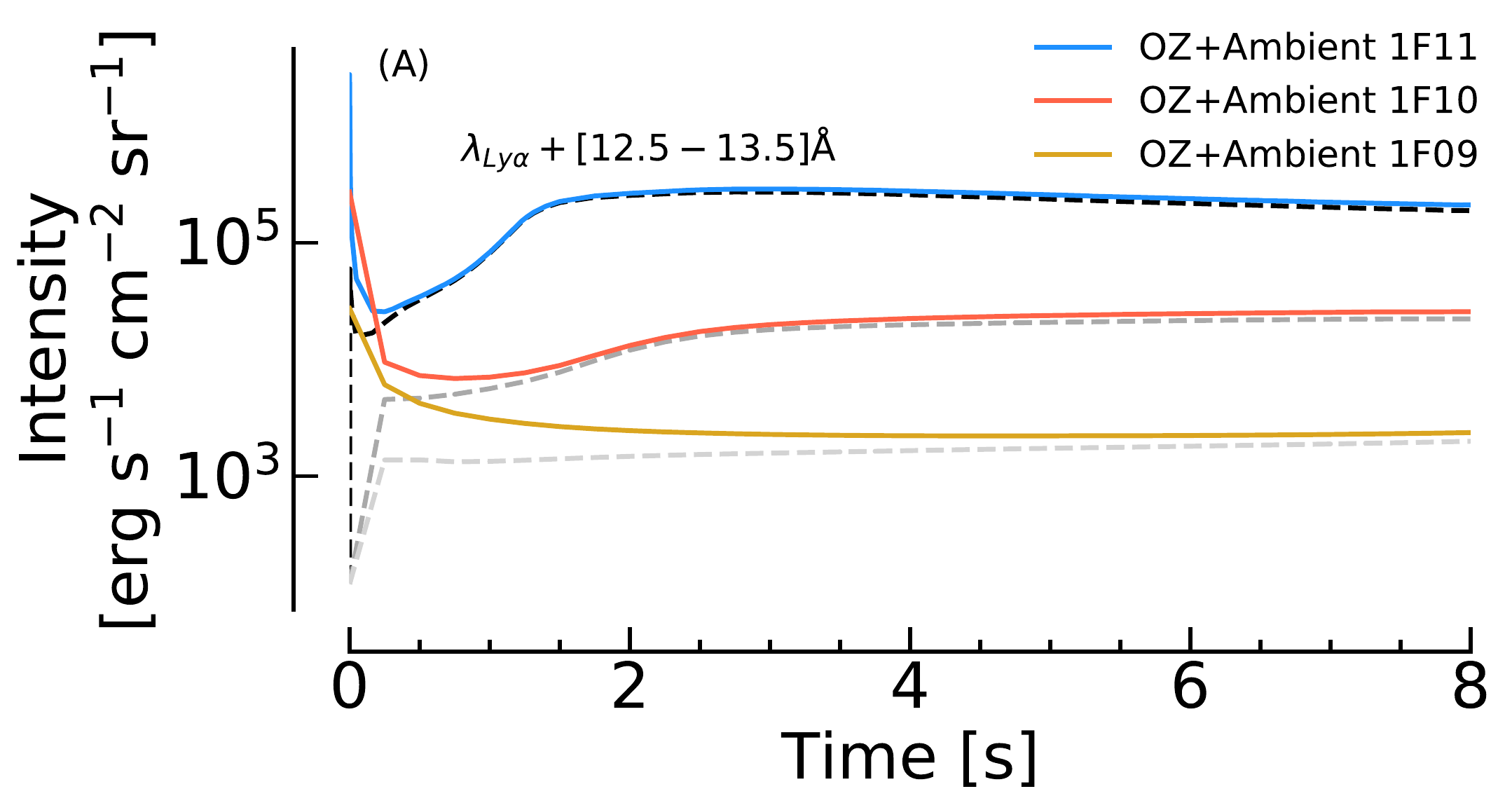}}	
	}
 \vbox{
	\subfloat{\includegraphics[width = .5\textwidth, clip = true, trim = 0.cm 0.cm 0.cm 0.cm]{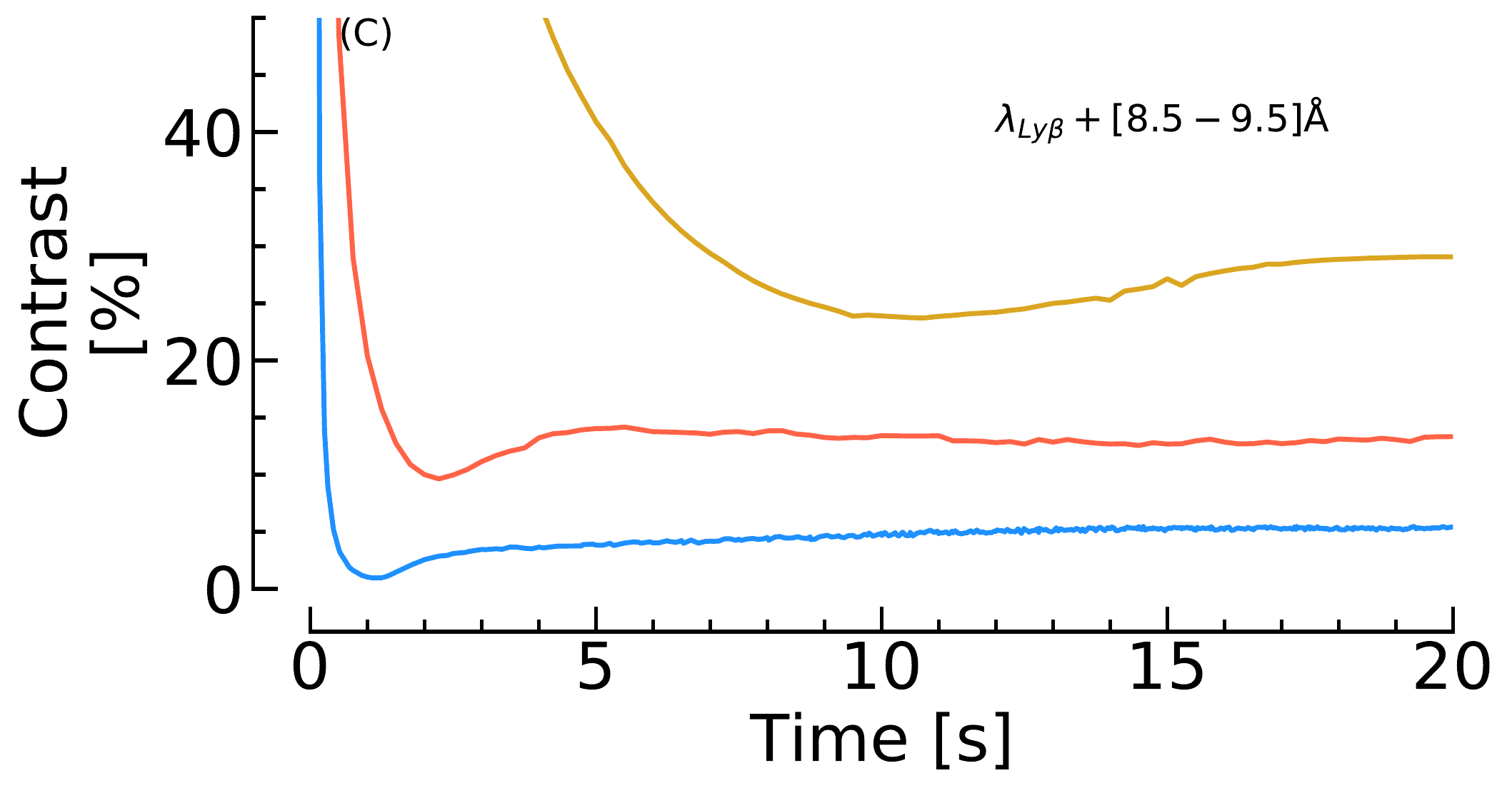}}	
	}
	\caption{\textsl{The temporal evolution of non-thermal OZ emission + thermal emission (colored lines; blue is 1F11, red is 1F10, and gold is 1F9), compared to the evolution of thermal emission alone (greyscale; darkest is 1F11, palest is 1F9). Clearly, strong OZ enhancements are short-lived, persisting only a fraction of a second in the strongest flare (1F11), and only a few seconds in the weakest simulation (1F9). Panel (A) shows \lya\ integrated between $[12.5-13.5]$~\AA, and panel (B) shows \lyb\ integrated between $[8.5-9.5]$~\AA. Panel (C) shows the contrast between non-thermal and thermal emission for \lyb\ integrated between $[8.5-9.5]$~\AA, scaled to focus on the latter part of the simulations, illustrating that after the initial onset, the differences drop substantially.}}
	\label{fig:OZSpectra_summedemission_lcurve}
\end{figure}

Now that we understand the temporal characteristics of the non-thermal Lyman line emission, the natural question is can we observe it over and above the ambient Lyman line emission? Summing thermal and non-thermal Lyman line emission, and comparing to the thermal by itself, shows us that this is likely going to be a very difficult detection to make. Figure~\ref{fig:OZSpectra_summedemission} shows those comparisons, where colored lines show the sum (gold is 1F9, red is 1F10, and blue is 1F11) and greyscale shows the ambient emission. The left column is \lyb, and the right is \lya, with each row showing a different time. At flare onset the emission is orders of magnitude above the background, but this is only true for a very short time. By $t=0.5$~s in the 1F11 simulations the combination of the strong enhancement of the background emission and the reduction in non-thermal emission makes it difficult to distinguish the non-thermal feature, which is now only a small factor larger than the thermal emission, not an order of magnitude. Non-thermal emission can still be distinguished in the 1F10 simulation and in the 1F9 simulation. It is only in the weakest flare simulation (1F9), however, that the non-thermal emission can be seen easily against the background by $t=3-5$~s. It is easier to do so for the \lyb\ line owing to the smaller wings and less intense nearby emission. At later times the difference between thermal and non-thermal emission in the 1F9 simulation is also likely too small to easily detect. 

Integrating over a region of the spectrum where non-thermal emission appears, which is free of strong lines, and comparing to the same region in the simulations without non-thermal OZ emission illustrates these timescales. Figure~\ref{fig:OZSpectra_summedemission_lcurve} shows these lightcurves, with the same color scales as Figure~\ref{fig:OZSpectra_summedemission}, where panel (A) shows \lyb\ and panel (B) shows \lya. In those figures compare the following pairs of lines: blue solid with black dashed (F11), red solid with dark grey dashed (1F10), and gold solid with light grey dashed (1F9). After a short time each solid line is only some small factor above their greyscale counterpart. In all cases the \lyb\ non-thermal emission shows a larger contrast to the thermal emission alone. Figure~\ref{fig:OZSpectra_summedemission_lcurve}(C) shows the contrast introduced by the non-thermal emission. That is, ($I_{total} - I_{ambient})/I_{ambient}$. The scale on that figure has been chosen to illustrate the contrasts present at later times in the simulations ($>10$~s), where the 1F11 is only around $\sim2.5$\%, compared to $\sim15$\% for 1F10 and $30$~\% for 1F9.


\begin{figure*}
	\centering 
    \vbox{
    \hbox{
	{\includegraphics[width = 0.5\textwidth, clip = true, trim = 0.cm 0.cm 0.cm 0.cm]{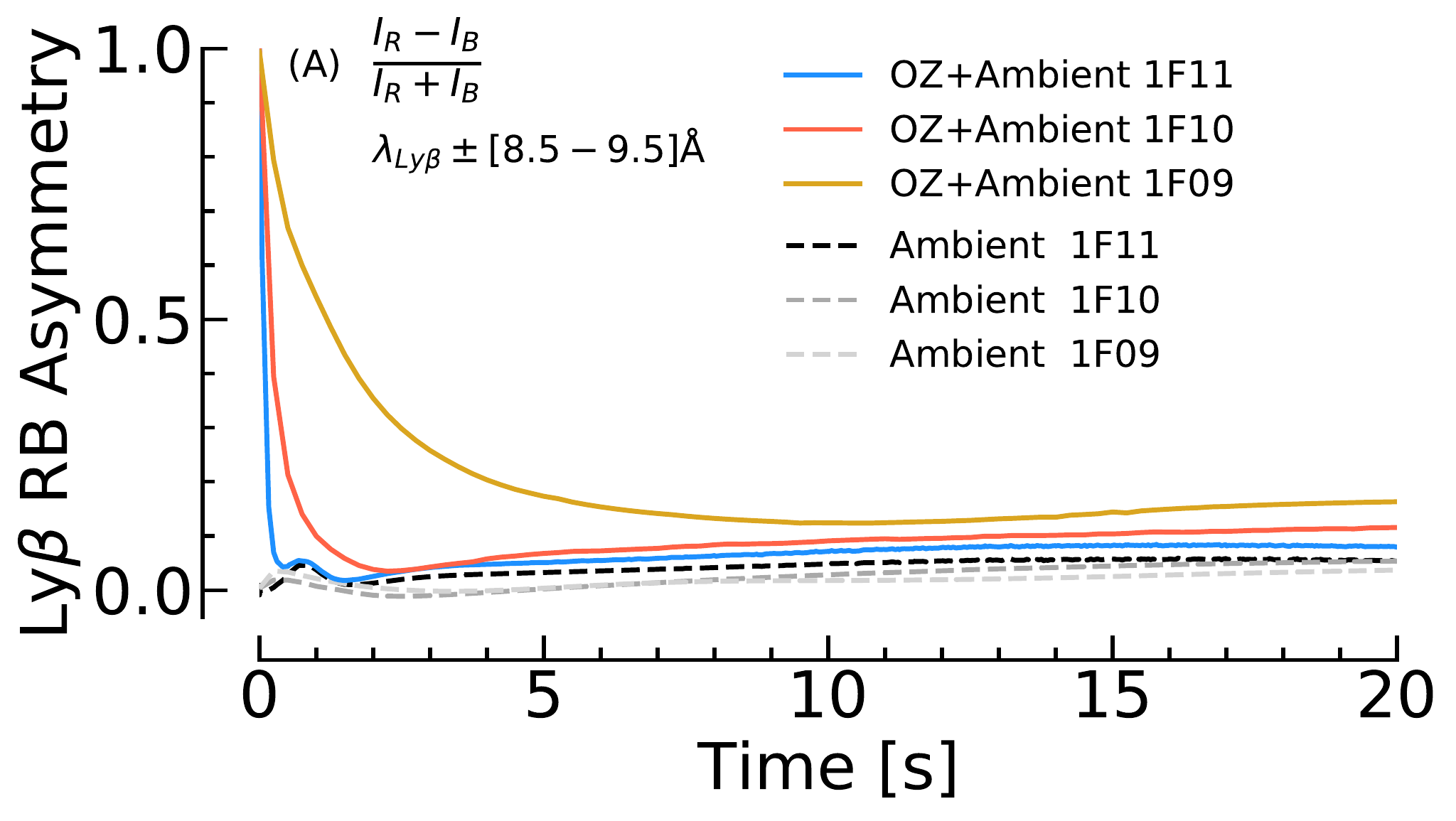}}
    {\includegraphics[width = 0.5\textwidth, clip = true, trim = 0.cm 0.cm 0.cm 0.cm]{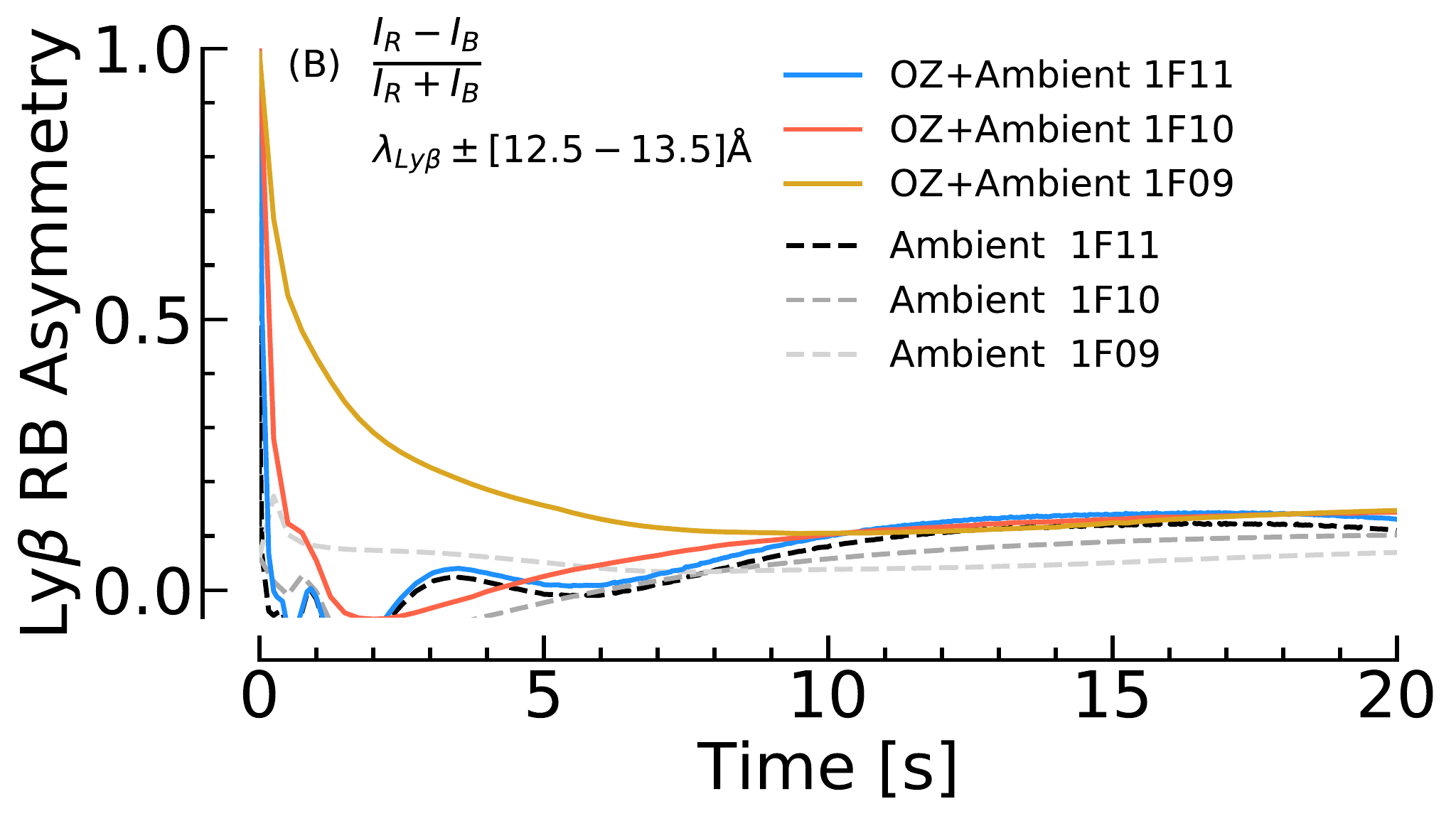}}
    }
    }
    \vbox{
    {\includegraphics[width = 0.65\textwidth, clip = true, trim = 0.cm 0.cm 0.cm 0.cm]{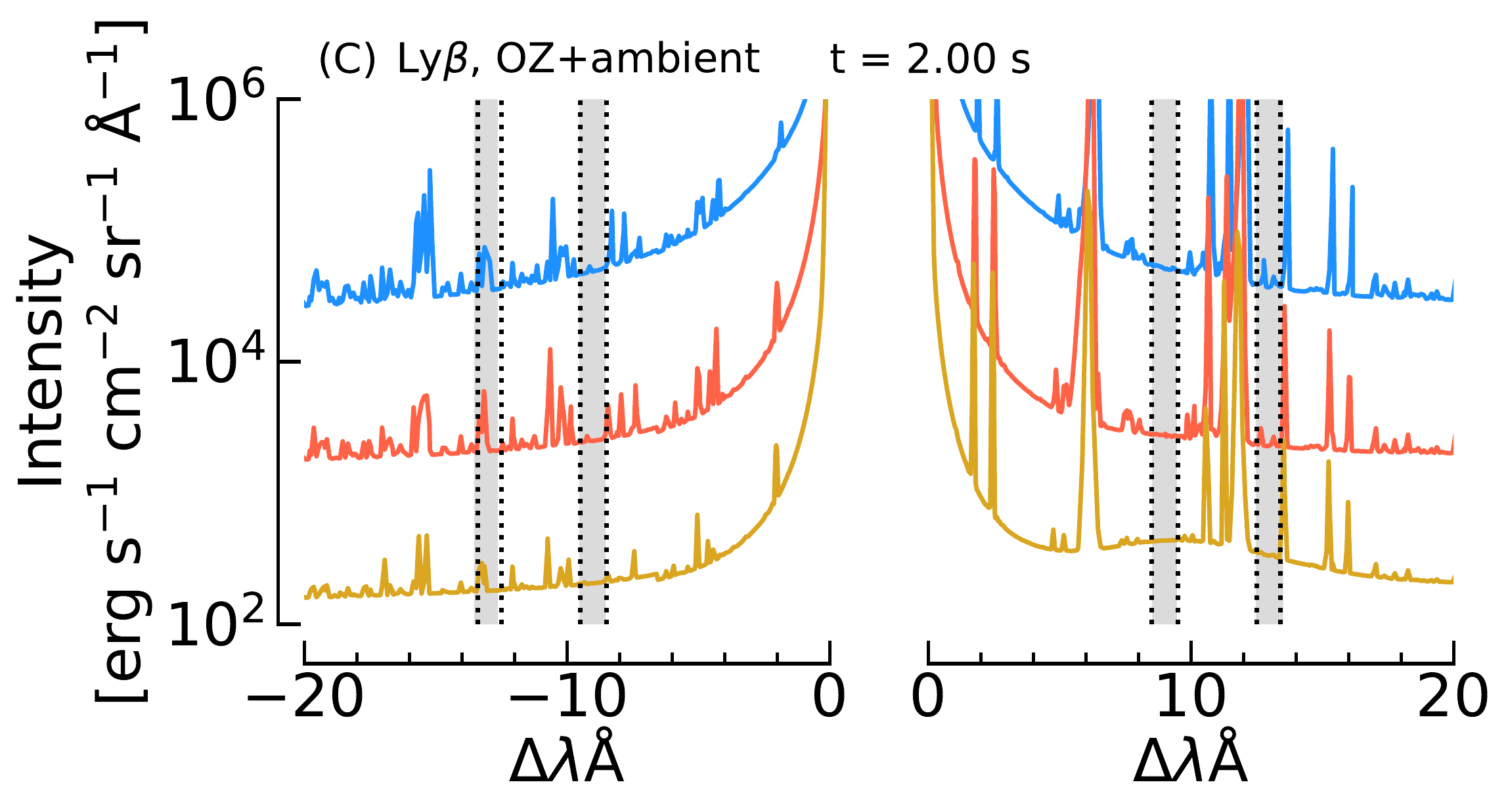}}
    }
	\caption{\textsl{The red-blue (RB) wing asymmetries, $A$, of \lyb\ with (solid colored lines; gold is the 1F9 flare, red is the 1F10 flare, and blue is the 1F11 flare) and without (dashed greyscale; darkest is 1F11, palest is 1F9) non-thermal emission. Panel (A) shows $A$ determined from $\Delta\lambda \pm [8.5-9.5]$~\AA, and panel (B) shows $A$ from $\Delta\lambda \pm [12-13]$~\AA. Panel (C) shows the line emission at $t=2$~s, indicating the regions selected (shaded grey regions bounded by dotted lines) for the asymmetry calculation.}}
	\label{fig:asymms}
\end{figure*}

If the intensity in the region near the Lyman lines is only a small factor larger than it would otherwise be due to the presence of non-thermal emission then it is problematic to unambiguously observe, especially if the signal-to-noise of the photometry is small. Asymmetries present between the red and blue line wings, caused by the presence of non-thermal emission only redward of the line cores, may help us more clearly identify the non-thermal emission beyond the first few seconds of heating in each the footpoint. To quantify these asymmetries we define the asymmetry as 

\begin{equation}
A = \frac{I_{R} - I_{B}}{I_{R} + I_{B}},
\end{equation}

\noindent where $I_{R}$ and $I_{B}$ are the intensities integrated over some $\Delta \lambda$ from line center in the red and blue wings, respectively. This is illustrated for \lyb\ in Figure~\ref{fig:asymms}, for two values of $\Delta \lambda$. Those two regions are mostly free of strong lines: (1) $\Delta\lambda\pm [12.5-13.5]$~\AA\ and (2) $\Delta\lambda\pm [8.5-9.5]$~\AA. 

In the $\Delta\lambda\pm [8.5-9.5]$~\AA\ region, very quickly ($t>0.5$~s) the large asymmetry present in strongest flare footpoint (F11) disappears, and approaches the value to be expected from ambient \lyb\ emission. Similarly, the 1F10 asymmetry decreases sharply after $t\sim1-2$~s, but does remain somewhat above the asymmetry present in the ambient emission. The asymmetry of the 1F9 flare shows a more gradual decrease out to $t\sim5-6$~s, and does lie above the background throughout the heating phase.  The maximum asymmetry in the ambient flare emission around $\Delta\lambda\pm [8.5-9.5]$~\AA\ was $A\sim[0.02-0.04]$. After 10s or so the asymmetries caused by the addition of non-thermal emission were on the order $A\sim[0.14, 0.08, 0.06]$ for 1F9, 1F10, and 1F11 simulations, respectively. For the two weaker flares these do seem comfortably above the typical ambient asymmetries. Careful study of the typical asymmetries in observations is needed to indicate if these small excess asymmetries caused by the non-thermal emission at later times is indeed detectable.  

In the $\Delta\lambda\pm [12.5-13.5]$~\AA\ region, the initial very large asymmetries are also present, but some small lines in the region makes the asymmetry more influenced by ambient flare emission. A more rigorous study when using actual observations could remove these lines, and we show our simple asymmetry calculation just to indicate that the very large asymmetry at the onset of particle precipitation into each footpoint should appear in multiple locations along the spectrum, even with the influence of these other features.

\section{Solar Orbiter / SPICE Predictions}\label{sec:spice}

\begin{figure*}
	\centering 
	{\includegraphics[width = \textwidth, clip = true, trim = 0.cm 0.cm 0.cm 0.cm]{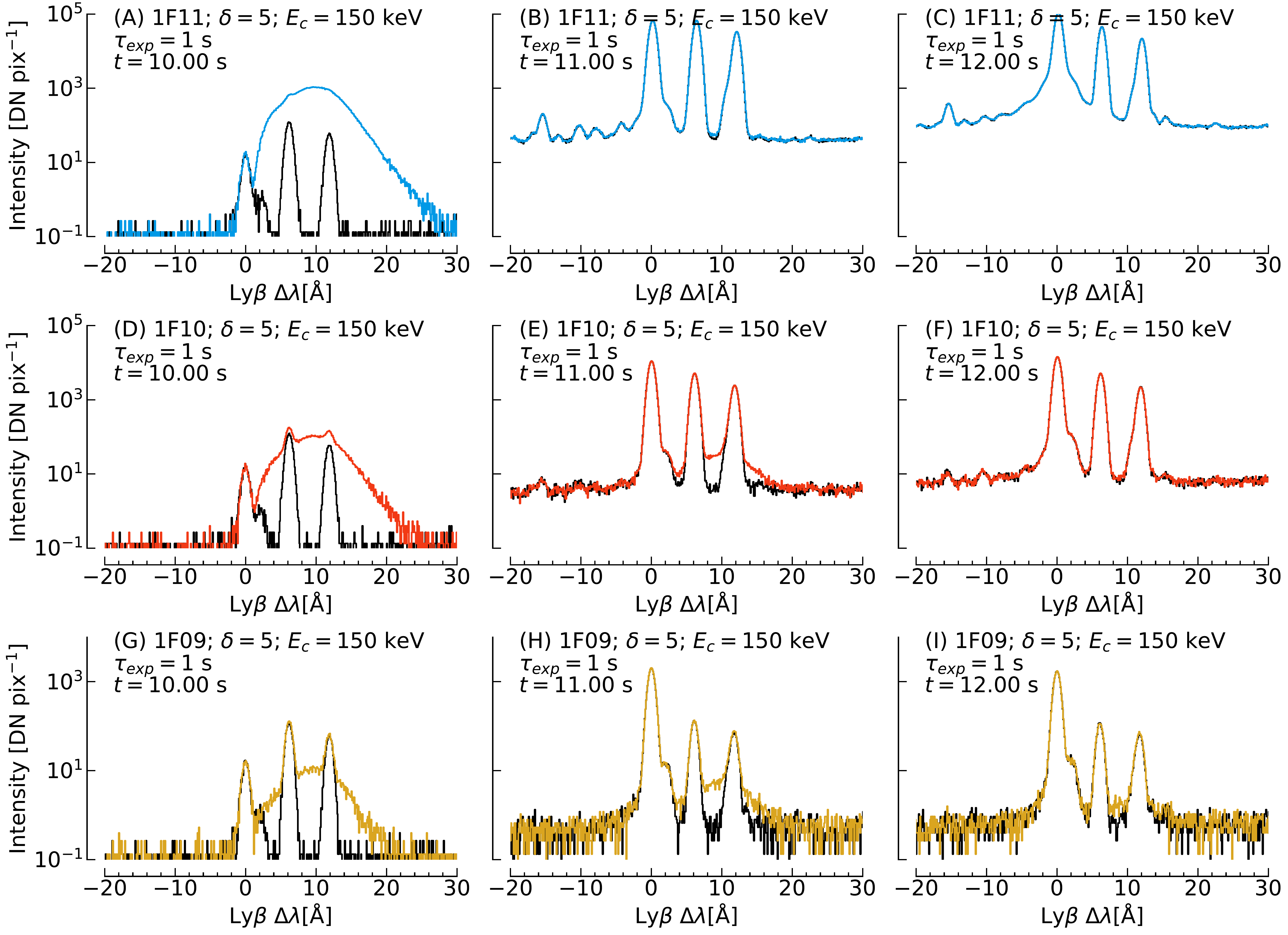}}	
	\caption{\textsl{Comparing the level of non-thermal \lyb\ emission to ambient emission around the  \lyb\ line for each of our three simulations, degraded to SPICE resolution assuming $\tau_{exp} = 1$~s, and slit width of $2^{\prime\prime}$. In each panel black is the ambient emission (i.e. without OZ emission), and the colored lines are the sum of ambient plus OZ emission. The top row shows 1F11 (blue), the middle show shows 1F10 (red), and the bottom row shows 1F9 (gold). As discussed in the text, in these synthetic observations there is $10$~s of pre-flare to facilitate averaging over exposure time, so that the flare starts at $t=10$~s here. Three consecutive exposures, starting at flare onset, are shown demonstrating the very transient nature of non-thermal emission, that does outshine the ambient emission for a brief time around and between the \ion{O}{6} doublet at 1032~\AA\ \& 1038~\AA.}}
	\label{fig:OZSpectra_SPICEres_1s}
\end{figure*}

\begin{figure*}
	\centering 
	{\includegraphics[width = \textwidth, clip = true, trim = 0.cm 0.cm 0.cm 0.cm]{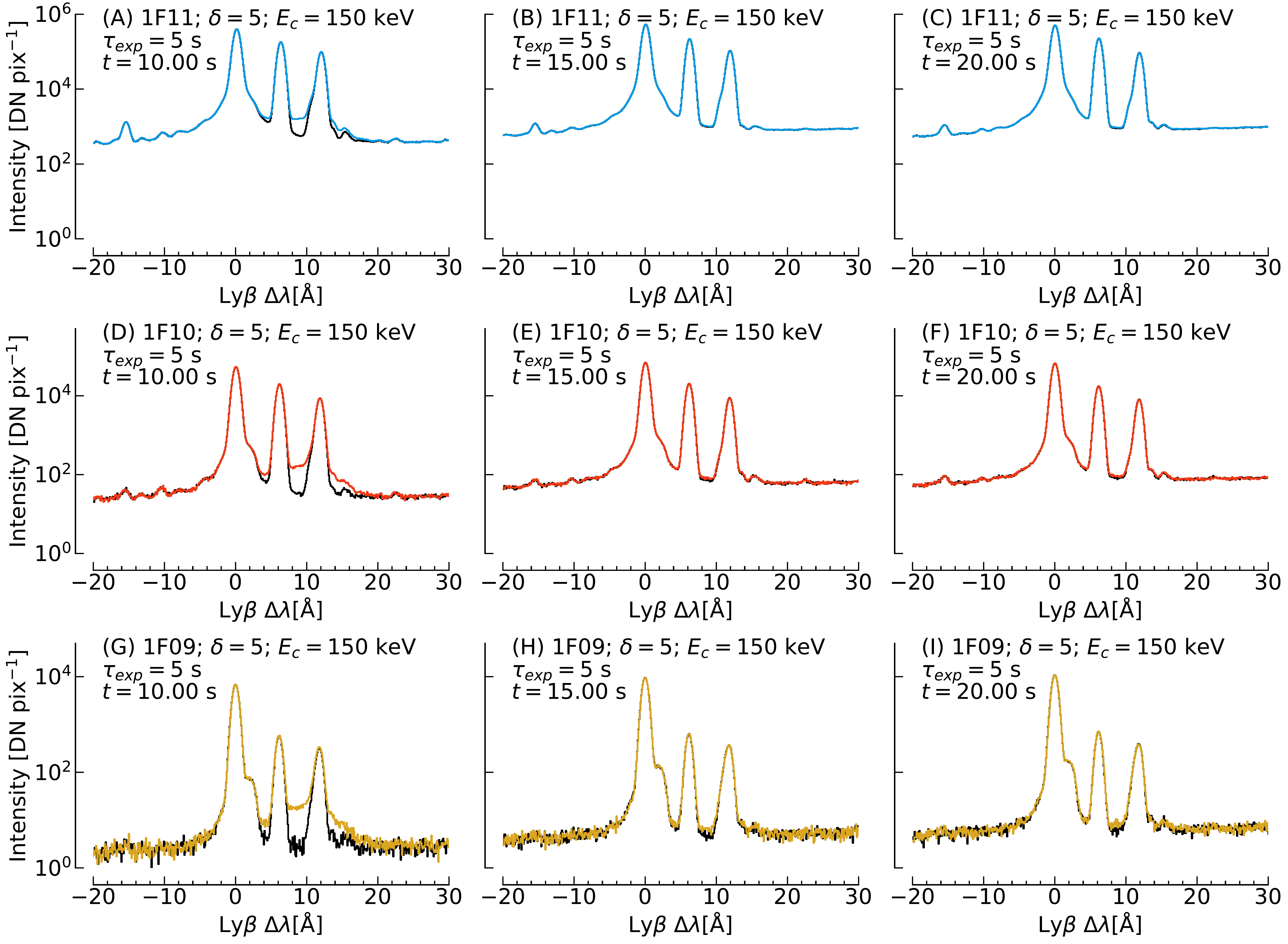}}	
	\caption{\textsl{Same as Figure~\ref{fig:OZSpectra_SPICEres_1s}, but for SPICE exposures with $\tau_{exp} = 5$~s.}}
	\label{fig:OZSpectra_SPICEres_5s}
\end{figure*}

\subsection{The Spectral Imaging of the Coronal Environment (SPICE) Instrument}
The SPICE instrument \citep{2020A&A...642A..14S} on board Solar Orbiter observes the Sun spectroscopically in the EUV in two passbands, the short-wavelength channel (SW; $\lambda = [700-792]$~\AA) and long-wavelength channel (LW; $\lambda = [970-1053]$~\AA), the latter of which includes the \lyb\ and \ion{O}{6} doublet. Though flare ribbons have, at the time of writing, not yet been observed by SPICE, this instrument does offer the best hope of detecting the OZ effect. SPICE offers relatively high temporal resolution, with typical exposure times for dynamic studies of $\tau_{exp} = [1,5]$~s, and can also provide spatial information within the flare ribbons, unlike the Sun-as-a-star \lyb\ observations from the EUV Variability Experiment \citep[EVE;][]{2012SoPh..275..115W}, onboard SDO. Solar Orbiter has tight telemetry and observing constraints, owing to the uniqueness of its orbit, and most remote sensing observations take place during 30 day windows during each perihelion. These constraints limit the time available for flare observations. However, Solar Orbiter's observing program does include targeting active region and flares, making it worth determining if SPICE would be able to detect the transient OZ effect.

\subsection{Degrading to SPICE Resolution}
The ambient plus non-thermal emission in the region around the \lyb\ line in each simulation was degraded to SPICE level-2 count rates, assuming two exposure times of $\tau_{exp} = [1,5]$~s, and that a sit-and-stare observing mode was used (i.e. continuous observation of a single source). Here, level-2 refers to the science-ready data provided to the community by the SPICE team, in which a common wavelength dispersion, spectral resolution and other properties are applied. The data for the following steps were obtained from both \cite{2020A&A...642A..14S} and from the information provided by the SPICE consortium following the second data release \citep{SPICE_DR2}.

The pre-flare ambient \lya\ emission was repeated for $10$~s at a cadence of $\delta t = 0.25$~s, and added to the original simulation time series, so that the flare footpoint heating phase in our synthetic SPICE observations was between $t_{spice}=[10-30]$~s. This new time series was degraded to SPICE resolution as follows. For each $0.25$~s snapshot, the wavelength scale was recast to the SPICE spectral plate scale ($0.09623$~\AA~pixel$^{-1}$), and the emission smeared by a spectral point spread function (PSF) assumed to be a Gaussian with full width at half maximum (FWHM) of 9.4 SPICE wavelength pixels. This was converted from an energy flux to a photon flux (using $hc/\lambda$ as the energy per photon for each SPICE wavelength bin). Our simulated intensities are per solid angle, which was removed by multiplying out the solid angle subtended by a pixel along the SPICE slit, where we assumed the $2^{\prime\prime}$ wide slit was being used (SPICE also has $4^{\prime\prime}$, $6^{\prime\prime}$, and $30^{\prime\prime}$ slits). The spatial plate scale along the slit is $1.098^{\prime\prime}$. We did not apply any spatial PSF at this stage as this would require making a number of assumptions about the ribbon width and elongation. The emission was multiplied by the effective area of the LW channel, which can be seen in Figure 24 of \cite{2020A&A...642A..14S}. From this output in photons~s$^{-1}$~pixel$^{-1}$ we then integrated the emission in time for each our modelled exposure times (ignoring the readout time, which is up to $0.42$~s if the full detector is read, but shorter if only a subset of lines and slit pixels are read). Poisson noise was added to the intensities in photons~pixel$^{-1}$, which were then converted to DN~pixel$^{-1}$ assuming 7.5~DN~photon$^{-1}$.

\subsection{Can SPICE Detect OZ Emission?}
Performing the same exercise as we did earlier to compare the sum of thermal to non-thermal emission to just the ambient emission, but this time degrading each to SPICE resolution, indicates that SPICE does offer the potential for a detection of the OZ effect. Figure~\ref{fig:OZSpectra_SPICEres_1s} shows our experiment assuming $\tau_{exp} = 1$~s. The top row is the 1F11 flare footpoint, the middle row is the 1F10 flare footpoint, and the bottom row is the 1F9 flare footpoint. Each column shows a different snapshot (recall that here the flare starts at $t=10$~s). Since the non-thermal emission at flare onset is so much larger than the ambient emission, the red-shifted feature is quite evident. It is only clearly present for one frame in the strongest flare, but persists for two frames in the 1F10 and 1F9 flares. Even if we use a longer exposure time of $\tau_{exp} = 5$~s (Figure~\ref{fig:OZSpectra_SPICEres_5s}), the emission is clearly visible in each flare for one frame. Whilst acknowledging the very transient nature of the OZ features, these results are encouraging. Depending on the orientation of the slit to the flare ribbon propagation direction we might expect to see a propagation of the redshifted feature in consecutive frames, in effect tracking the ribbon front (where energy is first deposited when new field lines reconnect).

For this experiment we varied the exposure time, but there are other considerations such as the slit width. Increasing the slit width will gather more photons, increasing the signal-to-noise, which might be necessary as the instrument response degrades over the course of the mission. The obvious downside is reduction of spatial resolution but this might become a necessary tradeoff. Similarly, we did not bin in the spectral direction, but since we are looking for a broad feature we can likely safely increase count rates that way.


\section{Summary \& Conclusions}\label{sec:conc}
New and upcoming observatories are set to provide Lyman line observations in flares, for example from Solar Orbiter/EUI (\lya; though imaging without spectra), Solar Orbiter/SPICE (\lyb), Solar-C/EUVST (\lya), ASO-S \lya, and the SNIFS sounding rocket (\lya). Further, \lya\ profiles from line scans observed by SORCE/SOLTICE are now available, and SDO/EVE sun-as-a-star observations cover the \lyb\ lines but have been relatively little studied.

Given the potential of new solar flare Lyman line observations, and the existence of untapped datasets, alongside recent important improvements to flare radiation hydrodynamic modelling (namely proton beam-driven flares including warm-target effects), we have revisited the possibility of using the \lya\ and \lyb\ lines to detect the presence of deka-keV non-thermal protons in the flaring chromosphere. To achieve this revisit of the Orrall-Zirker effect, we employ modern atomic cross-sections, and state-of-the-art numerical flare models, initially focussing on the impact of varying the magnitude of energy deposited into each footpoint. Our results have demonstrated that while a very difficult detection, accelerated protons present in the flaring chromosphere could produce transient non-thermal Lyman line emission lying above the ambient flare emission, following neutralisation of some fraction of the proton beam via charge exchange. 

In our model, a precipitating population of non-thermal protons is injected at the apex of a flare loop, assumed to be vertical and near disk-center. The transport and thermalisation of these particles is modelled using the \fpc\ code, with energy lost through Coulomb collisions heating the plasma, the evolution of which was modelled using the radiation hydrodynamics code \radyn. Together \radynfp\ provided the non-equilibrium ionization stratification and non-thermal proton distribution function over time during our flare simulations, vital improvements over previous attempts to model the OZ effect. These were used as input to a new package developed by us, \ozpy, that models the effects of the neutralisation of some fraction of the non-thermal proton distribution via charge exchange interactions between the beam and ambient plasma. This new population of energetic neutral atoms subsequently emits extremely redshifted photons of \lya\ and \lyb. Our model is open-source and freely available, and can be used to study OZ hydrogen emission given any ambient plasma stratification and non-thermal proton distribution. That is, it is flexible and can be used with inputs resulting from codes other than \radynfp, or from toy models, and could potentially be adapted to investigate similar processes on Martian aurorae as observed by MAVEN \citep[e.g.][]{2019JGRA..12410533H}.

It was found that, contrary to expectations based on the first calculations of \cite{1976ApJ...208..618O} and \cite{1985ApJ...295..275C}, this broad non-thermal `bump' in the red wings of \lya\ and \lyb\ will be noteable only for an extremely short time in strongly heated flare footpoints (sub-second), but noteable emission will persist for somewhat longer ($t\sim3-5$~s) in weaker flare footpoints. This is due to the rapid ionization of the atmosphere, quenching charge exchange interactions. Though in the strongly heated flare footpoints the precipitating protons push ever-deeper (due to warm-target effects), thus encountering a fresh supply of ambient neutrals, the compression of the atmosphere means that the emitting layer is narrow and the emergent intensity weak compared to the initial burst of non-thermal emission. Coupled to the fact that emission from the thermal \lya\ and \lyb\ wings and nearby continuum becomes enhanced, the non-thermal emission does not stand out from the background after a short time. That said, the asymmetries caused by non-thermal emission did persist, and for the 1F9 and 1F10 simulations they are $\gtrsim2$ that of the largest ambient asymmetries. These could potentially be observable by high-cadence instruments with high signal-to-noise. 

Degrading our synthetic emission to the resolution of the SPICE instrument onboard Solar Orbiter indicates that for short exposure times ($\tau_{exp} \le 5$~s, though ideally $\tau_{exp} \le 1$~s) it should be possible to observe the presence of non-thermal \lyb\ emission, albeit for only one or two consecutive frames. Perhaps focusing on the leading edge of flare ribbons in small-to-moderate flares offers the best chance of detection. These are the locations of initial energy deposition, representing a `pristine', not yet ionized target, and recent modelling results indicate that flare ribbon fronts likely undergo injection by a relatively weak energy flux \citep[e.g.][]{2022arXiv221105333P}. Since the initial onset of non-thermal emission is so large compared to the pre-flare, even a stronger flare source should be detectable in one SPICE exposure at each ribbon location (though obviously this presents a difficult observation to obtain). Very high cadence imaging, if a sufficiently broad passband is used, might also reveal this non-thermal emission as a transient flash, followed by a decay and further brightening. In general, to detect and fully exploit these signatures as diagnostics during flares we need an instrument capable of high signal-to-noise, exceptional cadence (0.1~s) and very high spatial resolution (certainly sub-arcsecond).

There are a number of fruitful avenues for investigation beyond this exploratory initial study. Here we have compared the magnitude of injected energy, but not the other parameters that define the non-thermal proton distribution. As indicated by prior investigations \citep[][]{1985ApJ...295..275C, 1999ApJ...514..430B} varying $E_{c}$ and $\delta$ should result in changes to the appearance of the broad non-thermal feature. Therefore using \ozpy\ and a large number of \radynfp\ flare models, the influence of those parameters can be investigated. Extension to stellar flares, including modelling synthetic Hubble Space Telescope observables is also an exciting direction, especially given that the only confirmed detection of the OZ effect thus far has been on a dMe star \citep{1992ApJ...397L..95W} and that follow on searches produced null detections \citep[e.g.][]{1993ApJ...414..872R,2022AJ....164..110F}. There are of course model limitations and idealisations that can be improved upon, such as modelling multi-species particle beams\footnote{If both energetic electrons and protons bombard a chromsopheric footpoint then ionization may be more rapid. Multi-species beams are are novel experiments that have yet to be performed, but which are now possible with very recent updates to \radynfp.}, and moving away from the assumptions of a near disk-center vertical beam. 

Finally, we speculate that the ratio of the \ion{O}{6} doublet may vary in the presence of highly redshifted \lyb\ emission. These lines are formed by both collisional processes as well as resonant scattering of chromospheric radiation \citep[e.g. see discussions in][]{2000JGR...105.2345S,2004ApJ...606L.159R}. Pumping of the \ion{O}{6} 1032~\AA\ line by \lyb\ during fast coronal mass ejections (CMEs), and the resulting change to the $I_{1032}:I_{1038}$ ratio has been exploited to estimate the electron density during a CME \citep[][]{2004ApJ...606L.159R}. A similar effect may occur in our OZ scenario, with pumping of \ion{O}{6} 1032~\AA\ by the proton beam-produced \lyb\ emission observable as a transient variation of the \ion{O}{6} doublet ratio. Characteristics of this change to the doublet ratio could also be diagnostically useful, since it presumably would vary depending on properties of the non-thermal component, which itself varies with properties of the injected non-thermal proton distribution.  Further study of this would require more advanced modelling of the \ion{O}{6} doublet than we have performed here, but would be a very interesting endeavour. 


 \textsc{Acknowledgments:} \small{We thank the anonymous referee for their insightful comments. GSK acknowledges the financial support from a NASA Early Career Investigator Program award (Grant\# 80NSSC21K0460).  GSK and ROM acknowledge financial support from the Heliophysics Supporting Research program (Grant\# 80NSSC21K0010). JCA, NZP, TAK and JWB acknowledge NASA funding for the SPICE instrument team at GSFC. NZP and JWB were funded through cooperative agreement 80NSSC21M0180.  JCA acknowledges funding through NASA's Heliophysics Supporting Research and Heliophysics Innovation Fund programs. HSH is grateful for the hospitality of the School of Physics and Astronomy at the University of Glasgow. This manuscript benefited from discussions held at a meeting of International Space Science Institute team: ``Interrogating Field-Aligned Solar Flare Models: Comparing, Contrasting and Improving,'' led by Dr. G. S. Kerr and Dr. V. Polito. The authors thank Dr. William Thompson for helpful discussions regarding SPICE. Solar Orbiter is a space mission of international collaboration between ESA and NASA, operated by ESA. The development of SPICE has been funded by ESA member states and ESA. It was built and is operated by a multi-national consortium of research institutes supported by their respective funding agencies: STFC RAL (UKSA, hardware lead), IAS (CNES, operations lead), GSFC (NASA), MPS (DLR), PMOD/WRC (Swiss Space Office), SwRI (NASA), UiO (Norwegian Space Agency). Resources supporting this work were provided by the NASA High-End Computing (HEC) Program through the NASA Advanced Supercomputing (NAS) Division at Ames Research Center.}

\bibliographystyle{aasjournal}
\bibliography{Kerr_2022_OrrallZirkerSpice}

\end{document}